\documentclass[a4paper,11pt]{article}
\pdfoutput=1 
\usepackage{jcappub}
\usepackage[T1]{fontenc} 

\author[a,b,c,d]{Pierre~Mourier} 
\author[e,d,f]{Asta~Heinesen} 

\affiliation[a]{Departament de F\'{i}sica, Universitat de les Illes Balears, IAC3 - IEEC, Cra. Valldemossa km 7.5, E-07122 Palma, Spain}
\affiliation[b]{Max-Planck-Institut f\"ur Gravitationsphysik (Albert
  Einstein Institute), Callinstr. 38, 30167 Hannover, Germany}
\affiliation[c]{Leibniz Universit\"at Hannover, 30167 Hannover, Germany}
\affiliation[d]{Univ Lyon, Ens de Lyon, Univ Lyon1, CNRS, Centre de Recherche Astrophysique de Lyon UMR5574, F--69007, Lyon, France} 
\affiliation[e]{Niels Bohr Institute, Blegdamsvej 17, DK-2100 Copenhagen, Denmark}
\affiliation[f]{School of Physical \& Chemical Sciences, University of Canterbury,
Private Bag 4800, Christchurch 8140, New Zealand} 

\emailAdd{pierre.mourier@uib.es}
\emailAdd{asta.heinesen@nbi.ku.dk}

\usepackage{amssymb}
\usepackage{braket}
\usepackage{cancel}
\usepackage{xcolor}
\usepackage[titletoc]{appendix}

\usepackage{relsize}
\usepackage{bm}
\usepackage{mathrsfs} 
\usepackage[normalem]{ulem}

\usepackage{footmisc} 
\usepackage{hyperref}
{
 \definecolor{BLACK}{gray}{0}
 \definecolor{WHITE}{gray}{1}
 \definecolor{RED}{rgb}{1,0,0}
 \definecolor{GREEN}{rgb}{0,1,0}
\definecolor{dgreen}{rgb}{.1,.6,.1}
\definecolor{BLUE}{rgb}{0,0,1}
 \definecolor{CYAN}{cmyk}{1,0,0,0}
 \definecolor{MAGENTA}{cmyk}{0,1,0,0}
 \definecolor{YELLOW}{cmyk}{0,0,1,0}
 \definecolor{aw}{rgb}{0.2,0.5,0.75}
  }
  \hypersetup{
    colorlinks,%
    citecolor=blue,%
    filecolor=blue,%
    linkcolor=magenta,%
    urlcolor=blue
}

\usepackage{amsthm}

\usepackage{geometry}
\definecolor{MyB}{rgb}{0.25,0.25,1.0}
\definecolor{MyB2}{rgb}{0.0,0.0,0.7}

\definecolor{MyGreen}{rgb}{0.0,0.5,0.0}
\definecolor{MyGreen2}{rgb}{0.0,0.3,0.0}

\definecolor{MyDarkRed}{rgb}{0.7,0,0}

\definecolor{MyGrey}{rgb}{0.5,0.5,0.5}
\definecolor{MyYellowish}{rgb}{0.65,0.6,0.0}

\def\beq{\begin{equation}} \def\eeq{\end{equation}}
\def\bea{\begin{eqnarray}} \def\eea{\end{eqnarray}}

\def \Scalar {S}
\def \Sc {A}
\def \Vec {V}
\def \VecField {\bm V}
\def \Bound {B}
\def \Vol {\mathcal{V}}
\def \norm {\mathcal{N}}

\def \heavi {\mathcal{H}}
\def \deltafun {\delta_D}

\def \ScZ {\Sc_0}
\def \BZ {\Bound_0}
\def \BoundH {\heavi (\BZ - \Bound)}

\def \Tube {\mathcal{T}_{\BZ}}
\def \extendedTube {\overline{\mathcal{T}}_{\!\!\BZ}}

\def \Scp {\Sc'}
\def \ScpZ {\Sc'_0}

\def \gvn {\gamma_{\VecField, \bm n}}
\def \gvnp {\gamma_{\VecField, \bm {n'}}}
\def \gnnp {\gamma_{\bm n, \bm {n'}}}

\def \CF {\mathcal{F}}
\def \CM {\mathcal{M}}
\def \Rcube {\mathbb{R}^3}
\def \SC {\mathscr{C}}

\def \functau {\tau_0}

\def \length {\mathcal{L}}
\def \lengthg {\bar{\mathcal{L}}}

\def \boundTheta {\bar{H}}
\def \bounda {\bar{a}}

\def \dl {\mathrm{d}\lambda}

\newcommand{\average}[1]{{\left\langle {#1} \right\rangle^{\Sc}_{\Sc_0}}}
\newcommand{\averagep}[1]{{\left\langle {#1} \right\rangle^{\Scp}_{\Sc_0}}}

\newcommand{\spatint}[1]{{I \! \left( {#1} \right)^{\Sc}_{\Sc_0}}}
\newcommand{\spatintp}[1]{{I \! \left( {#1} \right)^{\Scp}_{\Sc_0}}}

\title{Splitting the spacetime: A systematic analysis of foliation dependence in cosmic averaging} 

\abstract{%
It is a fundamental unsolved question in general relativity how to unambiguously characterize the effective collective dynamics of an ensemble of fluid elements sourcing the local geometry, in the absence of exact symmetries.
In a cosmological context this is sometimes referred to as the averaging problem. 
At the heart of this problem in relativity is the non-uniqueness of the choice of foliation within which the statistical properties of the local spacetime are quantified, 
which can lead to ambiguity in the formulated average theory. This has led to debate in the literature on how to best construct and view such a coarse-grained hydrodynamic theory.
Here, we address this ambiguity by performing the first quantitative investigation of foliation dependence in cosmological spatial averaging. 
Starting from the aim of constructing slicing-independent integral functionals (volume, mass, entropy, \emph{etc}.) as well as average functionals (mean density, average curvature, \emph{etc}.) defined on spatial volume sections, we investigate infinitesimal foliation variations and derive results on the foliation dependence of functionals and on extremal leaves. 
Our results show that one may only identify fully foliation-independent integral functionals in special scenarios, requiring the existence of associated conserved currents.  
We then derive bounds on the foliation dependence of integral functionals for general scalar quantities under finite variations within physically motivated classes of foliations.
Our findings provide tools that are useful for quantifying, eliminating or constraining the foliation dependence in cosmological averaging.%
}
\catcode`\@=11
\catcode`\@=12
\begin{document}
\maketitle


\tableofcontents
\newpage

\section{Introduction} 
The formulation of the theory of relativity came with the remarkable insight that proper time varies between observers, and, consequently, the Newtonian notion of a unique time parameterization of physical phenomena was abandoned. 
In concrete applications of relativity to the modelling of physical systems, it is nevertheless practical to introduce a time parametrization.
Foliations of spacetime into a set of spatial leaves that are labelled by a time coordinate, also known as `$3+1$' decompositions, appear very commonly in applications of general relativity (and alternative theories of gravity), and especially in the cosmological context.
Such foliations allow for an initial value formulation of general relativity \cite{PhysRev.116.1322,Arnowitt:1962hi}.

Due to the non-uniqueness of time parameterization in general relativity, there is a broad freedom in choosing the foliation of a given spacetime.   
The non-uniqueness of the foliation is not a problem \emph{per se}: it may be seen as an advantage that there is the freedom to consider a foliation where the physical phenomena of interest are more easily described. 
Once one applies \emph{operations} that are tied to a particular slicing, such as the integration or averaging of a field over the leaves of the foliation, the choice of foliation is important. 
In this paper, it will be our goal to systematically examine and quantify the impact of the choice of foliation in spatial averaging with a focus on applications for cosmology, and we thus give a brief review of the cosmological \emph{averaging problem} here.  

Cosmology typically aims at describing overall, statistical, or average properties of the Universe as a whole as a function of \emph{cosmic time}. Thus, the concept of $3+1$ foliations is 
inherent in the questions posed in cosmology.  
The formulation of a large-scale effective cosmological theory can be approached either (i) by making an \emph{ansatz} for the large-scale metric (that must  be assumed to be a meaningful object) and for the large-scale matter content and seeking solutions within this ansatz --- possibly also allowing for smaller-scale perturbations thereof; or (ii) by explicitly averaging over small-scale dynamics to \emph{derive} the large-scale evolution laws of the locally defined spacetime. 

The route in (i) is most often applied. It works well when the small and intermediate scales of gravitational phenomena can effectively be ignored in the large-scale evolution of the Universe, and it becomes simple when there is a notion of large-scale symmetries.  
When assuming homogeneity and isotropy over spatial sections of the Universe at its largest scales, one arrives at a  Friedman-Lema\^\i tre-Robertson-Walker (FLRW) metric description by this route. 
The Lambda Cold Dark Matter ($\Lambda$CDM) paradigm of cosmology is founded on this approach, where structures are described as perturbations around a large-scale (and background) FLRW model. 

The alternative route in (ii) is more involved, since it starts with the local spacetime description --- which must necessarily be complicated by accounting for the hierarchy of scales and nonlinearity of gravitational phenomena in the Universe --- as the basis for deriving the dynamics of the largest scales. This step also involves the construction of an appropriately defined \emph{coarse-graining} or averaging operation. 
If the assumptions of decoupling of the physics at scales comparable to the size of the visible Universe from physics at smaller scales hold, and if large-scale homogeneity and isotropy apply, then one should, by such a procedure, arrive at the same FLRW metric description as for the approach in (i). However, if these conditions are not satisfied, then the results of the route in (ii) is expected to differ from the usual applications of the procedure in (i) --- such a difference is called a \emph{backreaction} effect of the dynamics of inhomogeneous structures. 
Despite of its empirical success, the $\Lambda$CDM paradigm is subject to model anomalies on a wide range of scales~\cite{Freedman:2017yms,Bullock:2017xww,Riess:2019cxk,Perivolaropoulos:2021jda,Peebles:2022akh}, and continues to face the fundamental challenges of the nature of the dark energy and dark matter sources. 
There is a debate regarding whether exploring the route in (ii) could help resolve the anomalies and interpretational challenges of the $\Lambda$CDM paradigm; see for instance~\cite{greenwald,Isthereproof}. 

The problem of how to formally approach the route in (ii) may be denoted as the \emph{averaging problem} or the \emph{fitting problem}, and was first discussed in detail in early works by \cite{Wald:1977up,Ellis:1984bqf,Ellis:1987zz}. 
The most widely studied procedure for cosmological averaging is the Buchert averaging scheme for scalar quantities~\cite{Buchert:1999er,Buchert:2001sa,Buchert:2019mvq,Buchert:2022zaa}, but there have been numerous contributions on complementary/alternative procedures, \emph{e.g.}, \cite{Zalaletdinov:1992cg,Zalaletdinov:1996aj,Green:2010qy,Gasperini:2009wp,Gasperini:2009mu,Heinesen:2018vjp}. A variety of applications, mainly following the Buchert averaging scheme, and generalisations thereof, have directly addressed the impact of the choice of foliation in these schemes, either quantitatively or in qualitative discussions   \cite{Li:2007ci,Clarkson:2010uz,Brown:2012fx,Adamek:2017mzb,Bolejko:2017wfy,Buchert:2018yhd,Verweg:2023fov}. Very briefly summarized, these papers point towards a variation of the cosmological averages performed in a given spacetime with the choice of foliation, which is indeed natural: the choice of foliation is defining the averaging domains and therefore also the final values of the computed averages. Such a variation may be worsened by additional dependences of the averaging scheme in the choice of slices, \emph{e.g.} if the quantities to be averaged or the spatial boundaries of the domain depend on the foliation too. Without such extra dependences, it may be possible to select from physical constraints a broad class of foliations within which averaged observables remain approximately invariant, as argued qualitatively in~\cite{Buchert:2018yhd}. The foliation dependence in cosmological averaging is not in itself a problem, but if left uncontrolled, it can introduce ambiguity in the average cosmological theory.

Foliation dependence in cosmological averaging has not yet been quantified systematically in the literature.
In this paper, we present covariant and broadly applicable methods for quantifying, eliminating and/or constraining the foliation dependence in cosmological averaging. 
Although the methods are investigated with the cosmological averaging problem in mind, they  have broad applications to foliation studies in general relativity and differential geometry. 

In section~\ref{sec:scheme}, we present the general scalar averaging formalism that we use to quantify foliation dependence throughout this paper. We then consider foliation (in)dependence in an exact way, in section~\ref{sec:variation}, through calculus of variation, for integral and average functionals. We also investigate how foliations might be singled out uniquely from their extremal properties. In section~\ref{sec:foliationdependence_restricted}, we consider bounds on foliation dependence for finite variations, which are relevant when considering physically motivated restricted classes of cosmic foliations. Finally, in section~\ref{sec:discussion}, we summarize and discuss our results and their potential application to the foliation dependence of large-scale backreaction terms and effects in cosmological averaging.

\section{Covariant averaging over spatial foliations} 
\label{sec:scheme}

We now define an averaging scheme relevant for averaging scalar functions over 3-dimensional (spatial or null) slices as embedded within the four-dimensional spacetime manifold $\mathcal{M}$. 
We use the $3+1$ averaging formalism presented in~\cite{Heinesen:2018vjp}, building upon~\cite{Gasperini:2009wp,Gasperini:2009mu,Gasperini:2011us} --- see also the related development in~\cite{fanizza2020}. 
This scheme is very general for $3+1$ averaging of scalars\footnote{%
There exists also a covariant formalism for smoothing (coarse-graining) tensorial fields, known as Macroscopic Gravity~\cite{Zalaletdinov:1992cg,Zalaletdinov:1996aj}. It differs in scope from the approach used here however, due to a focus on coarse-graining fields within the 4-dimensional bulk, rather than on defining averages over given bounded regions of hypersurfaces.%
}, as it allows for the selection of an arbitrary averaging domain within an arbitrary slicing, and --- as an extension to~\cite{Gasperini:2009wp,Gasperini:2009mu,Gasperini:2011us} --- it further allows for any volume measure and weighting function for the integrand.

\begin{figure}
    \centering
    \includegraphics[width=0.55\textwidth]{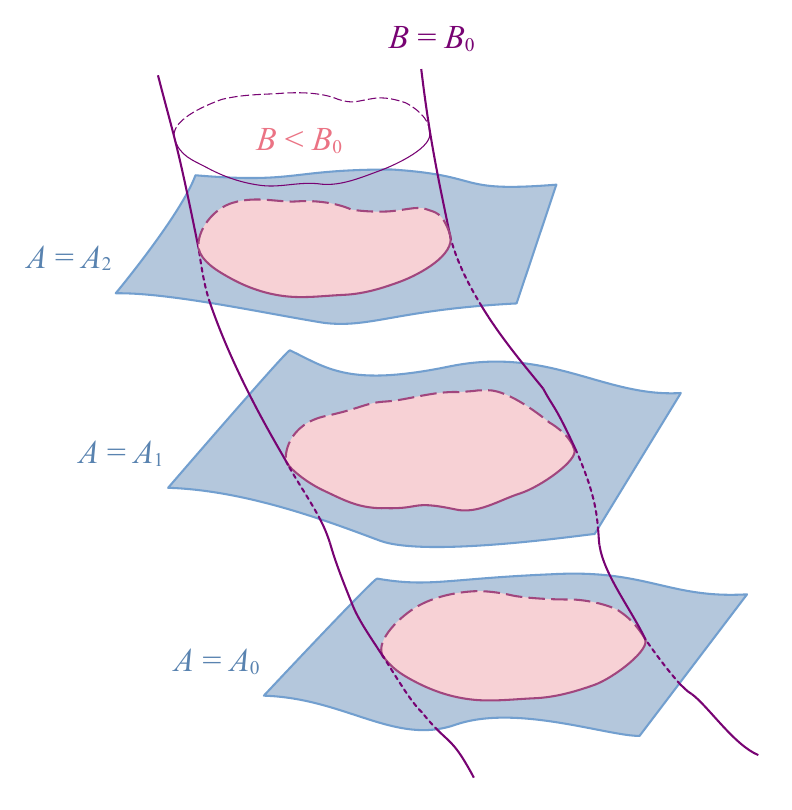}
    \caption{Representation of the foliation and integration domain setup defined by the scalar functions $\Sc$ and $\Bound$ in the window function~\eqref{eq:window}. The figure shows several hypersurfaces from the foliation defined by the level sets of $\Sc$, the spacetime tube specified by $\{ \Bound \leq \Bound_0 \}$ bounding the spatial extent of the region of interest, and the compact integration domain defined within each $\{ \Sc = cst. \}$ slice from the intersection of the slice with the above tube. Note that we here illustrate the case of space-like slices, but a function $\Sc$ will light-like level sets could be chosen as well.}
    \label{fig:slicing_setup}
\end{figure}

Following~\cite{Heinesen:2018vjp}, we consider spacetime domains that are selected by a window function of the form:
\begin{equation}
\label{eq:window}
W_{\Sc, \Sc_0, \Bound, \Bound_0, \bm \Vec}
{} = \Vec^{\mu} \nabla_{\mu} \Sc \; \deltafun (\Sc - \Sc_0) \, \heavi (\Bound_0 - \Bound)   \; ,
\end{equation} 
where $\heavi$ is the Heaviside step-function and $\deltafun$ the Dirac delta function.
Above, the \emph{foliation scalar} $\Sc$, with an everywhere nonzero gradient, defines the foliation, each level set $\{ \Sc = cst. \}$ defining a hypersurface. The gradient of $\Sc$ is assumed to be either time-like or null, making the hypersurfaces space-like or light-like, accordingly. The constant $\Sc_0$ parametrizes the hypersurface within the foliation. The \emph{boundary scalar} $\Bound$ is set to have a space-like, outwards-pointing gradient, hence one may think of $\Bound$ as a radial coordinate in a suitable coordinate system. Together with the (fixed) constant $\Bound_0$, it defines the spatial boundaries of the finite domain of averaging within each hypersurface. 
We shall restrict ourselves to the case where $\Bound$ is independent of $\Sc$ and its derivatives, \emph{i.e.}, by selecting the region where $\Bound \leq \Bound_0$ we consider a fixed spatially bounded ($3+1$)--dimensional tube in spacetime. We illustrate this setup and these notations on Fig.~\ref{fig:slicing_setup}.

The \emph{volume measure vector} $\bm \Vec$, on the other hand, determines the integration measure on the selected 3-dimensional surfaces. For instance, for space-like hypersurfaces, taking $\VecField$ to be their unit normal vector $\bm n$, \emph{i.e.}, $\Vec^\mu \! = n^\mu \! \equiv - \nabla^\mu \Sc \, (- \nabla^\nu \Sc \nabla_\nu \Sc)^{-1/2}$, is the choice that would let $\sqrt{g} \,  W_{\Sc, \Sc_0, \Bound, \Bound_0, \bm \Vec} \, d^4 x$ reduce to the Riemannian volume element within the hypersurfaces, inside the integration domain, where $g$ is the \emph{modulus} of the determinant of the spacetime metric $\mathbf{g}$ (the components of which are noted $g_{\mu \nu}$), $g \equiv  | \det(g_{\mu \nu}) | = - \det(g_{\mu \nu})$. We shall, however, allow $\bm \Vec$ to be any time-like vector field, and the integration measure will thus generally not coincide with the Riemannian volume measure on the leaves. Allowing for a non-Riemannian volume measure is convenient for certain types of fluid-intrinsic averaging~\cite{Buchert:2018yhd,Buchert:2019mvq}. Those formalisms are recovered by setting $\VecField$ to be the $4$-velocity $\bm u$ of a certain fluid flow, independently of the foliation set by $\Sc$, and the volume measure then differs from the Riemannian one on the slices (if $\bm{\nabla} \Sc$ is time-like) by the Lorenz factor $\gamma \equiv - \bm u \cdot \bm n$. The generality of $\VecField$ shall also be convenient in the present paper for making explicit which properties of the averaged expressions are related to the domain selection and which are related to the volume measure. We finally note that $\VecField$ is not necessarily requested to be unitary. This allows for weighted averages --- where the norm $| \VecField |$ of $\VecField$ can be considered as the local weight and the normalized $\VecField$ as defining the volume measure ---, such as mass-weighted averages when $| \VecField |$ is a mass density~\cite{massweighted,Heinesen:2018vjp}.

For the examples of applications considered in this paper we mainly have in mind spatial foliations ($\bm \nabla \Sc \cdot \bm \nabla \Sc < 0$), which are relevant for formulating averaged evolution equations in time, i.e., viewing the averaging problem as an initial value problem. 
We shall however also be interested in considering null foliations ($\bm \nabla \Sc \cdot \bm \nabla \Sc = 0$) in some cases. 
In such situations, we will sometimes be interested in allowing $\Bound$ to rather have a time-like gradient. Then, $\{ \Bound= \Bound_0 \}$ would become a space-like hypersurface rather than a time-like tube boundary, and the regions selected by $W_{\Sc, \Sc_0, \Bound, \Bound_0, \bm \Vec}$ for a range of values of $\Sc_0$ would then rather be interpreted as a set of light-cones truncated at that fixed hypersurface.

For the unit Heaviside step-function $\heavi$, we use the right-continuous convention $\heavi(0) = 1$ throughout. 
In the following we shall omit the subscripts on $W_{\Sc, \Sc_0, \Bound, \Bound_0, \bm \Vec}$ except for those relating to the foliation and refer to the window function (\ref{eq:window}) as $W^{\Sc}_{\Sc_0}$, leaving the choice of boundaries and volume measure implicit. 

We define the integral over a scalar $\Scalar$ over the spacetime domain singled out by $W^{\Sc}_{\Sc_0}$ in the following way
\begin{equation} 
\label{eq:integration}
I (\Scalar)_{\Sc_0}^{\Sc}  \equiv  \int_{\mathcal{M}} d^4 x \sqrt{g} \, \Scalar \, W^{\Sc}_{\Sc_0} \; .
\end{equation} 
We then define the volume $\Vol_{\Sc_0}^{\Sc}$ of the domain and 
the volume average $\braket{\Scalar}_{\Sc_0}^{\Sc}$ of an arbitrary scalar $\Scalar$, respectively, as
\begin{equation} 
\label{eq:average}
\Vol_{\Sc_0}^{\Sc}  = I(1)_{\Sc_0}^{\Sc}    \qquad  ;  \qquad   \braket{\Scalar}_{\Sc_0}^{\Sc}   \equiv   \frac{I(\Scalar)_{\Sc_0}^{\Sc} }{\Vol_{\Sc_0}^{\Sc} }  \; .
\end{equation}

The derivative of the integral (\ref{eq:integration}) with respect to $\Sc_0$ is given by (see~\cite{Heinesen:2018vjp}):
\begin{equation} 
\label{eq:commutation_rule}
 \frac{\partial I (\Scalar)_{\Sc_0}^{\Sc} }{\partial \Sc_0} = \int_{\mathcal{M}} d^4 x \sqrt{g} \,  \Vec^{\nu} \nabla_{\nu} \Sc \; \deltafun (\Sc - \Sc_0) \,\frac{\nabla_{\mu} ( \Scalar \Vec^{\mu} \heavi (\Bound_0 - \Bound)   ) }{\Vec^{\sigma} \nabla_{\sigma} \Sc}   \; ,
\end{equation} 
and the analogous derivative of the average (\ref{eq:average}) is given by:
\begin{align}
    \label{eq:commutation_rule_average}
 \frac{\partial \braket{\Scalar}_{\Sc_0}^{\Sc} }{\partial \Sc_0} & = \frac{1}{I (1)_{\Sc_0}^{\Sc}}   \frac{\partial I (\Scalar)_{\Sc_0}^{\Sc} }{\partial \Sc_0}  - \frac{\braket{\Scalar}_{\Sc_0}^{\Sc}}{I (1)_{\Sc_0}^{\Sc}} \frac{\partial I (1)_{\Sc_0}^{\Sc} }{\partial \Sc_0}  \nonumber \\
 & {} =   - \frac{1}{I (1)_{\Sc_0}^{\Sc}}  \! \int_{\mathcal{M}} {\! d^4 x \sqrt{g} \,  \Vec^{\nu} \nabla_{\nu} \Sc \; \deltafun (\Sc - \Sc_0) \,\frac{\nabla_{\mu} \left( \Scalar \Vec^{\mu} \heavi (\Bound_0 - \Bound)    \right) -  \braket{\Scalar}_{\Sc_0}^{\Sc} \nabla_{\mu} ( \Vec^{\mu} \heavi (\Bound_0 - \Bound)    )   }{\Vec^{\sigma} \nabla_{\sigma} \Sc}  }   \, . 
\end{align} 
When $\Sc$ is a time-function, meaning that $\bm \nabla \Sc \cdot \bm \nabla \Sc < 0$, these derivatives may be thought of as describing the time-evolution of the integral/average.

In the above expressions, we re-introduced the volume weighting factor $\VecField \cdot \bm \nabla \Sc$ to explicitly separate the window function and the scalar to be integrated (generally foliation dependent due to its $\Vec^\sigma \nabla_\sigma \Sc$ denominator). For instance, given our $\heavi(0) = 1$ convention, Eq.~\eqref{eq:commutation_rule} may be rewritten as,
\begin{equation}
    \frac{\partial \spatint{\Scalar}}{\partial \Sc_0} = \spatint{ \frac{\nabla_{\mu} ( \Scalar \Vec^{\mu} \heavi (\Bound_0 - \Bound)) }{\Vec^{\sigma} \nabla_{\sigma} \Sc}}  = \spatint{\frac{\nabla_{\mu} ( \Scalar \Vec^{\mu} ) - \Scalar \Vec^{\mu} \nabla_\mu \Bound \; \deltafun (\Bound - \Bound_0)) }{\Vec^{\sigma} \nabla_{\sigma} \Sc}} \; .
\end{equation}

We do require that $\Bound$ be parametrized in such a way that space-like or null sections of $\{ \Bound \leq cst. \}$ tubes are compact. In the case where the spatial sections of $\mathcal{M}$ are infinite, integrals or averages over the entire $\{ \Sc = cst. \}$ slices may be obtained from a $\Bound_0 \rightarrow \infty$ limit, the existence of which would however set some convergence conditions that $\Scalar$ must obey; we will not directly consider such a limit in this work. In the case of a spatially closed topology, on the other hand, the above formalism can directly encompass the special case of integrating over the entire compact slices. This is obtained by choosing a large enough $\Bound_0$; in that case, the corresponding equations may as well be simplified by dropping the Heaviside factors $\heavi(\Bound_0 - \Bound)$. Boundary conditions arising from derivatives of this factor, \emph{i.e.} $\deltafun(\Bound - \Bound_0)$ terms, then disappear as $\Bound < \Bound_0$ everywhere within $\mathcal{M}$.

\section{Infinitesimal variation of the foliation}
\label{sec:variation} 

In this section we consider local extrema of the integrals and averages defined above, when they are viewed as functionals of the foliation. 
We thus compute stationarity conditions of the integrals/averages under variation of the foliation scalar $\Sc$. 
Such stationarity conditions have at least two useful applications. 
Firstly, we are interested in identifying foliation independent quantities. 
Foliation-independence is equivalent to stationarity of the given functional for \emph{all} foliations. 
Secondly, we may be interested in finding a foliation (or leaf of a foliation) that extremise a specific integral or average functional. Extremal leaves of such functionals can be thought of as generalisations of paths of shortest distance, and they may be thought of defining natural hypersurfaces in specific contexts.
Thus, such extremals may provide an interesting way of defining leaves/foliations uniquely. 
Moreover, the corresponding extremum value of the functional would be a natural foliation independent measure of this functional, generalising the shortest distance (minimised over all possible paths) as a preferred measure of distance between two points.  

We leave the foliation scalar unconstrained in the variation. Thus the leaves associated with the solutions to the resulting stationarity conditions may in principle be space-like, time-like, null, or of a varying nature (depending on the point) amid these three possibilities; and the nature of the solution will have to be checked in each case.  
Analogous constrained stationarity conditions can be derived by imposing local or global constraints of physical interest, such as constraining the foliation scalar to have an everywhere null gradient (e.g. when discussing light cones) or to be a proper-time function for a given 4-velocity field. We avoid such considerations in this paper for the sake of simplicity. 

We shall consider stationarity conditions either for a single leaf or for an entire foliation, the latter being more restrictive. 
We consider cases where the integrated scalar $\Scalar$ of interest is possibly dependent on $\bm \nabla \Sc$, but not $\Sc$ or higher-order derivatives, and where the volume measure vector $\bm \Vec$ is possibly dependent on $\Sc$ and $\bm \nabla \Sc$\footnote{%
We consider possible dependence of $\bm \Vec$ on $\bm \nabla \Sc$ since we want to allow for cases where $\bm \Vec$ is normal to the hypersurfaces defined by $\Sc$. We also consider possible dependence of $\Scalar$ on $\bm \nabla \Sc$, the reason being that factors of $\bm \Vec \cdot \bm \nabla \Sc$ arise naturally for time-derivatives as seen in the commutation rules (\ref{eq:commutation_rule}) and (\ref{eq:commutation_rule_average}). For analysing higher-order derivatives or scalar curvature of the slices, dependence on second-order derivatives $\bm \nabla \bm \nabla \Sc$ in $\Scalar$ would need to be considered in the variation.%
}.

\subsection{Freedom of parameterisation of the foliation}
Before considering the variation of the average/integral functionals in generality, we shall consider the trivial subset of variations that define a map of the foliation onto itself. We shall require that the functional is  invariant under such mappings.   
Consider a given foliation $\mathcal{F} = \{\Sigma_{\Sc = \Sc_0} \}$, where $\Sc$ is a spacetime scalar representing the foliation, $\Sc_0$ is a parameter defined over some range $\Sc_{0,1} \leq \Sc_0 \leq \Sc_{0,2}$ selecting a leaf of the foliation, and where $\Sigma_{\Sc = \Sc_0}$ represents a leaf of the foliation. 
The transformation
\begin{equation}
\label{eq:foliation_reparam}
\Sc \mapsto f(\Sc) \quad  , \quad \Sc_0 \mapsto f(\Sc_0) \; ,
\end{equation}
where $f$ is a strictly monotonic function, defines a map from the foliation onto itself $\mathcal{F} \mapsto \mathcal{F}$. 
We thus denote the choice of $f$ the \emph{freedom of parameterisation} of the foliation.   

We can define an averaging operation independent of this parameterisation by requiring that a change of parametrisation leaves the integral (\ref{eq:integration}) invariant when $\Scalar$ is itself independent of the parametrisation of the foliation. Thus, we require that transformations $\Sc \mapsto f(\Sc)$, $\Sc_0 \mapsto f(\Sc_0)$ lead to mappings of the integral onto itself $I (\Scalar)_{\Sc_0}^{\Sc} \mapsto I (\Scalar)_{f(\Sc_0)}^{f(\Sc)} = I (\Scalar)_{\Sc_0}^{\Sc}$, where $\Scalar$ is an arbitrary scalar independent of the parametrisation of the foliation $\mathcal{F}$. 
Demanding that $I(\Scalar)_{\Sc_0}^{\Sc}$ remains invariant under the gauge transformations $\Sc \mapsto f(\Sc)$, $\Sc_0 \mapsto f(\Sc_0)$ is equivalent to requiring stationarity of $I(\Scalar)_{\Sc_0}^{\Sc}$ for all representations $\Sc$ of the foliation $\mathcal{F}$ under variations $\Sc \mapsto \tilde{\Sc}= \Sc + \delta f (\Sc) \, , \, \Sc_0 \mapsto \tilde{\Sc}_0 = \Sc_0 + \delta f (\Sc_0) $, where $\delta f (\Sc)$ (and its derivatives) is an infinitesimal scalar function of $\Sc$. 

For a generic integral operation (\ref{eq:integration}) the integral evaluated at the representation $\tilde{\Sc}$ of the foliation can be expressed in terms of the integral evaluated in the representation $\Sc$ as follows
\begin{align}
\label{eq:integrationparametrisation}
I (\Scalar)_{\tilde{\Sc}_0}^{\tilde{\Sc}} & =  \int_{\mathcal{M}} d^4 x \sqrt{g} \, \Scalar  \Vec^{\mu}[\tilde{\Sc}, \bm \nabla \tilde{\Sc}] \, \nabla_{\mu}( \tilde{\Sc} )  \, \deltafun (\tilde{\Sc}_0 - \tilde{\Sc}) \heavi (\Bound_0 - \Bound) \nonumber \\
& {} = \int_{\mathcal{M}} d^4 x \sqrt{g} \, \Scalar \Vec^{\mu}[\tilde{\Sc}, \bm \nabla \tilde{\Sc}] \, \nabla_{\mu} \Sc \, \deltafun (\Sc_0 - \Sc) \heavi (\Bound_0 - \Bound)  \; ,
\end{align} 
where the second line follows from $\partial_{\Sc} \tilde{\Sc} \; \deltafun (\tilde{\Sc}_0 - \tilde{\Sc} ) = \deltafun (\Sc_0 - \Sc)$. We have made the functional dependence of $\VecField$ explicit by writing $\Vec^{\mu}[\tilde{\Sc}, \bm \nabla \tilde{\Sc}]$. 
Performing the first order functional expansion of $\Vec^{\mu}(\tilde{\Sc})$ around the representation $\Sc$ and plugging it into  (\ref{eq:integrationparametrisation}) we have 
\begin{align} 
\label{eq:integrationparametrisation2}
 \left. \delta I (\Scalar)_{\Sc_0}^{\Sc} \right|_{\mathcal{F}}  & \equiv I (\Scalar)_{\tilde{\Sc}_0}^{\tilde{\Sc}} - I (\Scalar)_{\Sc_0}^{\Sc}   \nonumber \\
&= \int_{\mathcal{M}} d^4 x \sqrt{g} \, \Scalar \left( \frac{ \partial \Vec^{\mu} }{\partial \Sc} \, \delta f(\Sc) +  \frac{ \partial \Vec^{\mu} }{\partial \nabla_{\nu} \Sc} \,   \nabla_{\nu} \delta f(\Sc)   \right) \nabla_{\mu} \Sc \, \deltafun (\Sc_0 - \Sc) \heavi (\Bound_0 - \Bound) \nonumber \\
&= \delta f(\Sc_0) \, I \! \left(\frac{\Scalar \, \frac{ \partial \Vec^{\mu}(\Sc) }{\partial \Sc} \nabla_{\mu} \Sc }{\Vec^{\kappa}(\Sc) \nabla_{\kappa} \Sc }\right)_{\Sc_0}^{\Sc} +   \frac{\partial \delta f(\Sc_0)}{\partial \Sc_0}\; I \! \left(\frac{\Scalar \frac{ \partial \Vec^{\mu}(\Sc) }{\partial \nabla_{\nu} \Sc}  \nabla_{\nu}\Sc  \nabla_{\mu}\Sc }{\Vec^{\kappa}(\Sc) \nabla_{\kappa} \Sc }\right)_{\Sc_0}^{\Sc} \; , 
\end{align} 
where $ \left. \delta I (\Scalar)_{\Sc_0}^{\Sc} \right|_{\mathcal{F}}$ denotes the first order variation of $\delta I (\Scalar)_{\Sc_0}^{\Sc}$ under variations of $\Sc, \Sc_0$ that maps the foliation $\mathcal{F}$ onto itself. 
We require (\ref{eq:integrationparametrisation2}) to vanish for any choice of $\Scalar$.
Since $\delta f(\Sc_0)$ and its derivative can be chosen arbitrarily and independently on a given leaf $\Sc = \Sc_0$, the following two conditions,
\begin{equation}
 \label{eq:integrationparametrisationResult}
\frac{ \partial \Vec^{\mu}(\Sc) }{\partial \Sc} \nabla_{\mu} \Sc = 0 \qquad  \text{and}  \qquad  \frac{ \partial \Vec^{\mu}(\Sc) }{\partial \nabla_{\nu} \Sc}  \nabla_{\nu}\Sc  \nabla_{\mu}\Sc = 0   \; ,
\end{equation}
must be independently satisfied. 
Loosely speaking, \eqref{eq:integrationparametrisationResult} states that the volume measure (as determined by $\bm V$) can depend only on the \emph{direction} defined by $\bm \nabla A$, but not on its norm or the values of $A$. Besides any foliation-independent $\VecField$, these conditions allow in particular for taking $\VecField$ as the unit normal to the hypersurfaces as in~\cite{Gasperini:2009mu,Gasperini:2009wp}, in the case where those surfaces are space-like.

\subsection{Variation of integral quantities with respect to the foliation}
\label{StationarityConditionsIntegral} 
We shall now derive the stationarity conditions for the integral functional (\ref{eq:integration}) under variation of the hypersurface scalar\footnote{%
We are here varying the \emph{domain} of integration, by varying the spatial slice. This is in contrast to most variational problems, where a field living on a fixed domain is varied.  
Note that the functional $I (\Scalar)_{\Sc_0}^{\Sc}$ is differentiable even though it contains a delta-function in $\Sc$, since the variation of the delta-function is defined through partial integration.%
} $\Sc$.
Physical integral functionals of interest may for instance be volume or mass functionals.  

We write the first order variation of the integral $I(\Scalar)_{\Sc_0}^{\Sc}$ (\ref{eq:integration}) as a function of the variation $\Sc \rightarrow \Sc + \delta \Sc$ as 
\begin{equation}
\label{eq:Iexpansion}
I (\Scalar)_{\Sc_0}^{\Sc} \rightarrow I (\Scalar)_{\Sc_0}^{\Sc + \delta \Sc} = I (\Scalar)_{\Sc_0}^{\Sc} +  \delta I (\Scalar)_{\Sc_0}^{\Sc} \; , 
\end{equation}
with
\begin{align}
\label{eq:I_inf_variation}
 \delta I (\Scalar)_{\Sc_0}^{\Sc}  &=  \int_{\mathcal{M}} d^4 x \sqrt{g} \, \delta \! \left( \Scalar W^{\Sc}_{\Sc_0} \right)    \nonumber   \\ 
{}&= - \int_{\mathcal{M}} d^4 x \, \delta \Sc \sqrt{g} \, \deltafun (\Sc_0 - \Sc)   \nabla_{\mu} ( \Scalar \Vec^{\mu}  \heavi (\Bound_0 - \Bound)    )   \nonumber \\
 & {\phantom{={}}} + \int_{\mathcal{M}} d^4 x \, \delta (\nabla_{\mu}\Sc) \sqrt{g} \, \deltafun (\Sc_0 - \Sc)  \heavi (\Bound_0 - \Bound)  \left[ \frac{\partial}{ \partial (\nabla_{\mu} \Sc)} (   \Scalar \Vec^{\nu} \nabla_{\nu} \Sc) -    \Scalar  \Vec^{\mu}   \right]  \nonumber \\
{} &= - \int_{\mathcal{M}} d^4 x \, \delta \Sc \sqrt{g} \, \deltafun (\Sc_0 - \Sc)   \nabla_{\mu} ( \Scalar \Vec^{\mu}  \heavi (\Bound_0 - \Bound)    )   \nonumber \\
&{\phantom{={}}} + \int_{\mathcal{M}} d^4 x \, \delta (\nabla_{\mu}\Sc) \sqrt{g} \, \deltafun (\Sc_0 - \Sc)  \heavi (\Bound_0 - \Bound) \nabla_{\nu} \Sc \left[ \frac{\partial}{ \partial (\nabla_{\mu} \Sc)} (   \Scalar \Vec^{\nu} )   \right]   \; . 
\end{align} 
We note that local stationarity requirements cannot yet be directly extracted from the variation as written in \eqref{eq:I_inf_variation}, because $\delta(\nabla_\mu \Sc) = \nabla_\mu (\delta \Sc)$ may not be considered fully independent of $\delta \Sc$. It is only the component of $\nabla_\mu (\delta \Sc)$ along a direction pointing away from the surface, that may be considered independent from $\delta \Sc$ as evaluated on the surface. Hence, we make the following rewriting, first using partial integration of \eqref{eq:I_inf_variation} to obtain
\begin{align} 
\label{eq:I_inf_variation2}
 \delta I (\Scalar)_{\Sc_0}^{\Sc}  =  & - \int_{\mathcal{M}} d^4 x \, \delta \Sc \sqrt{g} \frac{\partial}{\partial \Sc} ( \deltafun (\Sc_0 - \Sc) ) \heavi (\Bound_0 - \Bound)  \nabla_{\mu}\Sc \nabla_{\nu}\Sc  \, \frac{\partial}{ \partial (\nabla_{\mu} \Sc)} (   \Scalar \Vec^{\nu} )  \nonumber   \\
{} & - \int_{\mathcal{M}} d^4 x \, \delta \Sc \sqrt{g} \, \deltafun (\Sc_0 - \Sc)  \, \nabla_{\mu} \! \left( \frac{\partial}{ \partial (\nabla_{\mu} \Sc)} (   \Scalar \Vec^{\nu} \nabla_{\nu} \Sc )  \heavi (\Bound_0 - \Bound)    \right)  \; , 
\end{align} 
and then introducing an arbitrary, fixed vector field $\bm Z$ chosen such that $\bm Z \cdot  \bm \nabla \Sc \neq 0$ (this is for instance guaranteed for a time-like $\bm Z$, if $\Sc$ has an everywhere time-like or null gradient) to rewrite the first term in (\ref{eq:I_inf_variation2}). This results in the following useful form of the variation:
\begin{align}
\label{eq:I_inf_variation3}
\delta I (\Scalar)_{\Sc_0}^{\Sc} &=    \int_{\mathcal{M}} d^4 x  \, \frac{ \delta \left(  Z^\sigma \nabla_{\sigma} \Sc \right)}{Z^{\kappa} \nabla_{\kappa} \Sc} \sqrt{g} \, \deltafun (\Sc_0 - \Sc) \heavi (\Bound_0 - \Bound)   \Vec^{\nu} \nabla_{\nu}\Sc  \nabla_{\mu}\Sc  \, \frac{\partial \Scalar}{ \partial (\nabla_{\mu} \Sc)}  \nonumber    \\
 &{\quad}+  \int_{\mathcal{M}} d^4 x  \, \delta  \Sc  \sqrt{g} \, \deltafun (\Sc_0 - \Sc) \, \nabla_{\sigma} \! \left( \frac{Z^{\sigma}}{Z^{\kappa} \nabla_{\kappa} \Sc}     \heavi (\Bound_0 - \Bound)  \Vec^{\nu} \nabla_{\nu}\Sc  \nabla_{\mu}\Sc  \, \frac{\partial \Scalar}{ \partial (\nabla_{\mu} \Sc)} \right)    \nonumber   \\
 & {\quad} - \int_{\mathcal{M}} d^4 x \, \delta \Sc \sqrt{g} \, \deltafun (\Sc_0 - \Sc)   \, \nabla_{\mu} \! \left( \frac{\partial}{ \partial (\nabla_{\mu} \Sc)} (   \Scalar \Vec^{\nu} \nabla_{\nu} \Sc )  \heavi (\Bound_0 - \Bound)    \right) \;  ,
\end{align}
where the requirement of gauge-invariance of the averaging operation (\ref{eq:integrationparametrisationResult}) has been used, and with $\delta (  Z^\sigma \nabla_{\sigma} \Sc ) = Z^\sigma \delta(\nabla_\sigma \Sc)$. 

In Eq.~\eqref{eq:I_inf_variation3} above, the variation $\delta \Sc$ on a given surface can now be considered independent from its derivative as the latter is taken along the direction $\bm Z$, away from the surface. Accordingly, the first line of \eqref{eq:I_inf_variation3} gives rise to a first local constraint equation, and the last two lines give rise to a second one. Altogether, we arrive at the two independent constraints for stationarity of $I (\Scalar)_{\Sc_0}^{\Sc}$ around the surface $\Sc = \Sc_0$:
\begin{align} 
& \qquad \qquad \qquad \qquad \delta I (\Scalar)_{\Sc_0}^{\Sc} = 0  \quad  \forall  \,  \delta \Sc    \qquad  \Leftrightarrow   \nonumber \\
& \left\{
  \begin{array}{@{}ll@{}}
 & \left.   \heavi (\Bound_0 - \Bound)   \nabla_{\mu}\Sc  \frac{\partial \Scalar}{ \partial (\nabla_{\mu} \Sc)} \right|_{\Sc_0}  = 0     \\
 &  \text{and} \\
 &\left. \nabla_{\sigma} \! \left( \frac{ Z^{\sigma} }{ Z^{\kappa} \nabla_{\kappa} \Sc}     \heavi (\Bound_0 - \Bound)  \Vec^{\nu} \nabla_{\nu}\Sc   \nabla_{\mu}\Sc \frac{\partial \Scalar }{ \partial (\nabla_{\mu} \Sc)} - \frac{\partial (   \Scalar \Vec^{\nu} \nabla_{\nu} \Sc ) }{ \partial (\nabla_{\sigma} \Sc)}  \heavi (\Bound_0 - \Bound)    \right) \right|_{\Sc_0}   = 0  \, ,   \label{eq:IntegralStationarity}
\end{array}
\right.
\end{align} 
where we have made use of the assumption $\Vec^{\nu}  \nabla_{\nu}\Sc\neq 0$.
The first condition above is nothing but the requirement for $\Scalar$ to be invariant under a change of the foliation $\mathcal{F} = \{\Sigma_{\Sc = \Sc_0} \}$. 
Note that the conditions for stationarity in Eq.~\eqref{eq:IntegralStationarity} are independent on the arbitrary choice of $\bm Z$ despite the apparent dependence on $\bm Z$ in the second condition.
This can be seen by expanding $\bm Z$ in a component proportional to $\bm \nabla \Sc$ and a component orthogonal to $\bm \nabla \Sc$, and noting that all contributions orthogonal to $\bm \nabla \Sc$ vanish due to the first condition of (\ref{eq:IntegralStationarity}) and to its spatial derivatives along the $\{ \Sc = \Sc_0 \}$ hypersurface.

Requiring that $I (\Scalar)_{\Sc_0}^{\Sc}$ is stationary for all surfaces $\Sc_0$ results in the following conditions:\footnote{%
We could have considered a domain with a small width in time, by replacing the delta function in (\ref{eq:window}) by a narrow step-function. In that case we arrive at (\ref{eq:I_inf_variation2}) with the delta-function replaced by a step function. In this case the expression equivalent to (\ref{eq:I_inf_variation2}) is not singular, but the ``boundary condition'' $\nabla_{\mu}\Sc  \; \partial \Scalar / \partial (\nabla_{\mu} \Sc) = 0$ would still need to be fulfilled separately in order for there to be extrema.%
}
\begin{align}
& \delta I (\Scalar)_{\Sc_0}^{\Sc} = 0  \quad  \forall  \,  \delta \Sc, \Sc_0    \qquad  \Leftrightarrow  \qquad   \left\{
  \begin{array}{@{}ll@{}}
&   \heavi (\Bound_0 - \Bound)    \nabla_{\mu}\Sc  \frac{\partial \Scalar }{ \partial (\nabla_{\mu} \Sc)} = 0      \\
&  \text{and} \\
& \nabla_{\sigma} \! \left(  \frac{\partial (   \Scalar \Vec^{\nu} \nabla_{\nu} \Sc ) }{ \partial (\nabla_{\sigma} \Sc)}  \heavi (\Bound_0 - \Bound)    \right) = 0      \; .   \label{eq:IntegralStationarityAll}
\end{array}
\right.
\end{align}
Note that the stationarity requirements (\ref{eq:IntegralStationarity}) and (\ref{eq:IntegralStationarityAll}) for the integral $I (\Scalar)_{\Sc_0}^{\Sc} $ are \emph{local} conservation equations. 
Due to the first condition, when $\Scalar$ is dependent on the representation $\Sc$ of the foliation such that $\nabla_{\mu}\Sc  \; \partial \Scalar / \partial (\nabla_{\mu} \Sc) \neq 0$, no stationary foliations exist for $I (\Scalar)_{\Sc_0}^{\Sc}$, when allowing for all possible variations of the foliation\footnote{%
Sometimes we might be interested in considering restricted subclasses of variations, e.g. variations within a class of proper time-foliations $\tau \rightarrow \tau + \delta \tau$ of a 4-velocity $\bm u$, that by construction satisfy $\bm u \cdot \bm \nabla \delta \tau = 0$ (see section~\ref{subsec:constant-tau}). In such cases stationary foliations for certain functionals may exist for $\nabla_{\mu}\Sc  \; \partial  \Scalar / \partial (\nabla_{\mu} \Sc)  \neq 0$.%
}. Due to the derivative of the spatial boundary selection function $\heavi(\Bound_0 - \Bound)$, the second condition in either Eq.~\eqref{eq:IntegralStationarity} or Eq.~\eqref{eq:IntegralStationarityAll} may itself be split into a condition inside the $\Bound \leq \Bound_0$ domain and a condition on the $\Bound = \Bound_0$ boundary. For instance for Eq.~\eqref{eq:IntegralStationarityAll}, the second condition is expanded as
\begin{equation}
\nabla_\sigma \left(\frac{\partial (\Scalar \Vec^\nu \nabla_\nu \Sc)}{\partial(\nabla_\sigma \Sc)} \right) \heavi(\Bound_0 - \Bound) - \frac{\partial (\Scalar \Vec^\nu \nabla_\nu \Sc)}{\partial(\nabla_\sigma \Sc)} \nabla_\sigma \Bound \, \deltafun(\Bound - \Bound_0) = 0 \; ,
\end{equation}
and is thus equivalent to:
\begin{align} 
& \left\{
  \begin{array}{@{}ll@{}}
 & \heavi(\Bound_0 - \Bound) \, \nabla_\sigma \left(\frac{\partial (\Scalar \Vec^\nu \nabla_\nu \Sc)}{\partial(\nabla_\sigma \Sc)} \right)  = 0     \\
 &  \text{and} \\
 &\left. \left(\frac{\partial (\Scalar \Vec^\nu \nabla_\nu \Sc)}{\partial(\nabla_\sigma \Sc)} \nabla_\sigma \Bound \right) \right|_{\Bound = \Bound_0}   = 0  \; .   
\end{array}
\right.
\end{align}

It can be shown from (\ref{eq:commutation_rule}) and the conditions (\ref{eq:IntegralStationarity}) that stationarity of $I (\Scalar)_{\Sc_0}^{\Sc}$ implies the vanishing of its derivative with respect to $\Sc_0$:
\begin{equation}
\delta I (\Scalar)_{\Sc_0}^{\Sc} = 0  \quad  \forall  \,  \delta \Sc    \qquad  \Rightarrow  \qquad  \left. \frac{ \partial I (\Scalar)_{\Sc_0}^{\Sc}}{\partial \Sc_0}  \right|_{\Sc_0}  = 0 \; .  \label{eq:IntegralGlobal}  
\end{equation} 
Similarly (\ref{eq:IntegralStationarityAll}) implies
\begin{equation}
\delta I (\Scalar)_{\Sc_0}^{\Sc} = 0  \quad  \forall  \,  \delta \Sc, \Sc_0   \qquad  \Rightarrow  \qquad   \frac{ \partial I (\Scalar)_{\Sc_0}^{\Sc}}{\partial \Sc_0}   = 0  \; .  \label{eq:IntegralGlobalAll}   
\end{equation} 
These results are expected since (infinitesimal) constant translations of $\Sc$ at fixed $\ScZ$, which are part of the class of variations $\delta \Sc$, can also be seen as a translation in time within the original foliation. In fact, the results (\ref{eq:IntegralGlobal}) and (\ref{eq:IntegralGlobalAll}) are general and apply to integral functionals with arbitrary functional dependence on $\bm \nabla \Sc, \bm \nabla \bm \nabla \Sc, .. ,\bm \nabla^{(n)} \Sc$. 

\paragraph{We now consider the special case where $\Scalar$ and $\VecField$ are independent of the foliation.} 
In particular, we let $\Scalar$ and $\VecField$ be independent on the direction vector $\bm \nabla \Sc$ of the foliation, and it follows that
\begin{align}
\label{eq:specialcaseSV}
&\left.  \nabla_{\nu}\Sc  \frac{\partial \Vec^{\nu}  }{ \partial (\nabla_{\mu} \Sc)}  \right|_{\Sc_0}  = 0  \quad  \text{and} \quad   \left.   \frac{\partial \Scalar }{ \partial (\nabla_{\mu} \Sc)}  \right|_{\Sc_0}  = 0  \quad  \text{and} \quad    \\
&  \left. \nabla_{\sigma} \left( \nabla_{\nu}\Sc  \frac{\partial \Vec^{\nu}}{ \partial (\nabla_{\mu} \Sc)}    \right) \right|_{\Sc_0}  = 0       \quad  \text{and} \quad          \left. \nabla_{\sigma} \left(  \frac{\partial \Scalar }{ \partial (\nabla_{\mu} \Sc)}    \right) \right|_{\Sc_0}  = 0 \; ,   \nonumber
\end{align}
in the case of extremisation for a single leaf; or, 
\begin{equation} 
\label{eq:specialcaseSVAll}
\nabla_{\nu}\Sc  \frac{\partial  \Vec^{\nu} }{ \partial (\nabla_{\mu} \Sc)}  = 0    \quad  \text{and} \quad      \frac{\partial  \Scalar }{ \partial (\nabla_{\mu} \Sc)}  = 0  \; , 
\end{equation}
for extremisation for the entire foliation. In these cases, 
the first condition of (\ref{eq:IntegralStationarity}) and of (\ref{eq:IntegralStationarityAll}) respectively are automatically satisfied. Moreover, the condition for stationarity for a single hypersurface simplifies to 
\begin{equation} 
\delta I (\Scalar)_{\Sc_0}^{\Sc} = 0  \quad  \forall  \,  \delta \Sc    \qquad  \Leftrightarrow  \qquad  \left. \nabla_{\sigma} \left(  \Scalar \Vec^{\sigma}  \heavi (\Bound_0 - \Bound)    \right) \right|_{\Sc_0}   = 0  \; ,  \label{eq:IntegralConstraintsReduced}  
\end{equation}
and the stationarity condition for all surfaces reduces to 
\begin{equation} 
\delta I (\Scalar)_{\Sc_0}^{\Sc} = 0  \quad  \forall  \,  \delta \Sc, \Sc_0   \qquad  \Leftrightarrow  \qquad   \nabla_{\sigma} \left(  \Scalar \Vec^{\sigma}  \heavi (\Bound_0 - \Bound)    \right)  = 0 \; .  \label{eq:IntegralConstraintsAllReduced}  
\end{equation}
The requirements (\ref{eq:specialcaseSV}) or (\ref{eq:specialcaseSVAll}) are indeed natural in many cases\footnote{%
Note that the case of $\bm \Vec$ being the unit normal to the hypersurfaces defined by $\Sc$ (when those are space-like) automatically satisfies (\ref{eq:specialcaseSV}) and (\ref{eq:specialcaseSVAll}) when $\Scalar$ is independent of the foliation.%
}%
.
In the case where $\bm \Vec$ is independent of the foliation, the stationarity requirement (\ref{eq:IntegralConstraintsReduced}) for a single surface is only dependent on the foliation through the surface of evaluation, and the requirement (\ref{eq:IntegralConstraintsAllReduced}) for stationarity for an entire foliation does not depend on the foliation at all. 
In this case, if (\ref{eq:IntegralConstraintsAllReduced}) is satisfied for a particular foliation, then it is satisfied for any possible foliation. 

As in the general case of Eqs.~\eqref{eq:IntegralStationarity}--\eqref{eq:IntegralStationarityAll}, the above stationarity conditions may be split into a bulk and a boundary condition,
\begin{equation}
    \heavi(\Bound_0 - \Bound) \, \nabla_\sigma (\Scalar \Vec^\sigma) = 0 \quad \text{and} \quad \left. \left(\Scalar \Vec^\sigma \nabla_\sigma \Bound \right) \right|_{\Bound = \Bound_0} = 0 \; ,
    \label{eq:IntegralConstraintsAllReducedSplit}
\end{equation}
to be simultaneously satisfied either on the single $\{ \Sc = \Sc_0 \}$ slice or on all slices.

\paragraph{We remark that the stationarity conditions for integral functionals are restrictive.} 
In particular, the stationarity requirement (\ref{eq:IntegralConstraintsAllReduced}) for an entire foliation is extremely restrictive. The foliation is stationary only when $\Scalar \VecField$ is a conserved current.
The stationarity requirement (\ref{eq:IntegralConstraintsReduced}) for a single leaf is more flexible: obtaining stationarity in a given foliation amounts to being able to collect points for which $\bm \nabla ( \Scalar \bm  \Vec ) = 0$ is satisfied to construct a space-like (or null-like) leaf.
We emphasise that the stationarity conditions (\ref{eq:IntegralConstraintsReduced}) and (\ref{eq:IntegralConstraintsAllReduced}) are derived under assumptions. For instance, they are derived under the assumption of no functional dependence of $I (\Scalar)_{\Sc_0}^{\Sc}$ on second or higher order derivatives of $\Sc$ (thus, for integrals over foliation-adapted curvature degrees of freedom the results derived in the present section do in general not apply). 
We also emphasise, that extrema which are not stationary points can exist in the form of infimums or in the form of local extrema introduced by a ``boundary'' in the solution space.

\subsubsection{Example: Rest mass}
\label{example_integrated_mass}
Consider a conserved local rest mass current 
\begin{equation} 
\label{eq:restmass_current}
M^{\mu} = \varrho u^{\mu} ,   \qquad  \nabla_{\mu}M^{\mu} = 0 \; ,
\end{equation} 
where $\varrho$ is a rest mass density of a fluid with 4-velocity $\bm u$. 
We might seek to define a total rest mass of spatial domains as volume integrals over $\varrho$. 

Let us consider the case where we take $\VecField$ to coincide with the fluid 4-velocity: $\VecField = \bm u$.
This is a natural choice when averaging local quantities intrinsic to the fluid \cite{Buchert:2001sa,Buchert:2018yhd}, and it allows for the integrated $\varrho$, $I (\varrho)_{\Sc_0}^{\Sc}$, to properly define a total rest mass \cite{Heinesen:2018vjp}. 
We will assume that we can define boundaries $\Bound$ intrinsic to the fluid flow through parallel transport, $\bm u \cdot \bm \nabla \Bound = 0$, such that the spatial boundary is comoving with the fluid flow. We may, alternatively, consider the case of a boundary-free domain, \emph{i.e.}, for a spatially closed manifold and an integration domain coinciding with the whole hypersurfaces. Either way, this ensures the preservation of the fluid content of the averaging domain over time, and will accordingly allow for the preservation of its total rest mass. 

Accordingly, we define a total rest-mass associated with the (fluid-comoving) averaging domain in a given foliation $\Sc$ as
\begin{equation} 
\label{eq:integration_mass}
M_{\Sc_0}^{\Sc }= I (\varrho)_{\Sc_0}^{\Sc}  =  \int_{\mathcal{M}} d^4 x \sqrt{g} \, \varrho \,  W^{\Sc}_{\Sc_0}  \; ,
\end{equation}
with 
\begin{equation} 
\label{eq:window_mass}
W^{\Sc}_{\Sc_0} = u^{\mu} \nabla_{\mu} \Sc \,  \deltafun(\Sc - \Sc_0) \, \heavi (\Bound_0 - \Bound) \; .
\end{equation}
The mass (\ref{eq:integration_mass}) can be shown to be conserved over time ($\partial M_{\Sc_0}^{\Sc } / \partial \Sc_0 = 0$) for any choice of foliation $\Sc$ \cite{Heinesen:2018vjp}.
Furthermore, we have from (\ref{eq:I_inf_variation3}) that 
\begin{equation} 
\label{eq:I_inf_variation_mass}
\delta M_{\Sc_0}^{\Sc } \equiv \delta I (\varrho)_{\Sc_0}^{\Sc}  =  - \int_{\mathcal{M}} d^4 x \, \delta \Sc \, \sqrt{g} \, \deltafun (\Sc_0 - \Sc)   \nabla_{\mu} ( \varrho u^{\mu} \heavi (\Bound_0 - \Bound)    )    \; ,
\end{equation}
and we recover the condition (\ref{eq:IntegralConstraintsAllReduced}), $\nabla_\mu (\varrho u^\mu \heavi(\Bound_0 - \Bound)) = 0$, for the stationarity of $M_{\Sc_0}^{\Sc }$ with respect to the foliation. Using the conservation (\ref{eq:restmass_current}) of the rest mass current and the comoving boundary (or no-boundary) assumption, $\nabla_\mu (\varrho u^\mu \heavi(\Bound_0 - \Bound)) = \nabla_\mu (\varrho u^\mu) \heavi(\Bound_0 - \Bound) - \varrho u^\mu \nabla_\mu \Bound \,\deltafun(\Bound_0 - \Bound)$ always vanishes, implying that the mass is stationary for all foliations and thus foliation independent.   

The rest mass definition (\ref{eq:integration_mass}) and its stationarity still hold if $\bm \nabla \Sc$ is light-like\footnote{%
The induced volume measure $(\bm \Vec \cdot \bm \nabla \Sc) \, \sqrt{g}$ is in this case the generalisation of the induced volume measure on the null-cone considered in \cite{Gasperini:2011us}. The present integration measure reduces to that of \cite{Gasperini:2011us} when $\bm \Vec$ is taken to be a time-like unit vector field which is hypersurface-forming.%
}.
Thus, this mass is not only invariant with respect to a choice of spatial hypersurface; it is also invariant under a change from spatial to light-like hypersurfaces, providing a potentially interesting correspondence between light-cone and spatial hypersurface averaging. 

\subsubsection{Example: Volume}
\label{example_integrated_volume}
Suppose we want to consider extremal foliations for the volume $\Vol_{\Sc_0}^{\Sc} \equiv I(1)_{\Sc_0}^{\Sc}$ of a domain lying within the hypersurfaces. 
We consider variations $\Sc \rightarrow \Sc + \delta \Sc$ that vanish ($\delta \Sc = 0$) on the boundary $\Bound = \Bound_0$ (or, as above, that the domain is boundary-free, \emph{i.e.} $\Bound < \Bound_0$ everywhere on $\mathcal{M}$). 
This implies slightly weaker stationarity requirements, removing the conditions imposed at the domain boundary since only the interior region has to be constrained. 

The problem of finding the extremal volume enclosed by a fixed spatial boundary, can be thought of as a higher-dimensional generalisation of finding the shortest path between two fixed spacetime events.
This is a well-studied problem in the literature, at least for the Riemannian volume measure (see e.g. \cite{Yau1982} and references therein for the case of Riemannian geometry and \cite{ASNSP_1976_4_3_3_361_0} and references therein for Lorentzian manifolds). It is nevertheless worth recalling here as an illustrative example for the present discussion, as the volume functional is of great importance in cosmology.

\paragraph{We first consider the case where the hypersurfaces are space-like and $\bm \Vec$ is their unit normal.}
In this case, we recover the Riemannian volume measure on the surfaces. We write 
\begin{equation}
\label{eq:vec_unitnormal}
\Vec_{\mu} = n_\mu =  \frac{-  \nabla_{\mu} \Sc}{ \norm }  , \qquad \norm \equiv (- g^{\alpha \beta}  \nabla_{\alpha} \Sc \nabla_{\beta} \Sc)^{1/2} \; , 
\end{equation}
where $\bm n$ is the future-pointing unit normal to the hypersurfaces, and $\norm^{-1}$ is the associated lapse function. 
In this case the condition (\ref{eq:specialcaseSVAll}) is satisfied, and the stationarity requirement reduces to (\ref{eq:IntegralConstraintsAllReduced}). This simplifies further due to the fixed value of $\Sc$ at the boundary (or the lack of a boundary), removing the condition at $\Bound = \Bound_0$, $(\bm V \cdot \bm \nabla \Bound) \, \deltafun (\Bound_0 - \Bound) = 0$. It follows that a foliation extremises $I(1)^{\Sc}_{\Sc_0}$ if and only if the extrinsic curvature scalar of each hypersurface vanishes inside the domain, $\bm \nabla \cdot \bm \Vec = \bm \nabla \cdot \bm n = 0$. We thus recover the well-known condition for the stationarity of this volume functional.

Suppose that we can find a foliation for which this condition is satisfied. 
We want to determine whether such an extremal foliation is a maximum, minimum, or a saddle-point.  
For this purpose we use the identities 
\begin{equation}
\label{eq:derNorm}
\frac{\partial \norm}{ \partial \nabla_{\mu} \Sc} = V^{\mu}  , \qquad   \frac{\partial \Vec^{\nu}}{ \partial \nabla_{\mu} \Sc} = -  \frac{1}{\norm} h^{\mu \nu} , \qquad h^{\mu \nu} \equiv \Vec^{\mu} \Vec^{\nu} + g^{\mu \nu} \; , 
\end{equation}
to compute the second variation of the volume
\begin{align}
\label{eq:second_var_volume}
 \delta^2 \Vol_{\Sc_0}^{\Sc}  \equiv \delta^2 I(1)_{\Sc_0}^{\Sc}   &=  \int_{\mathcal{M}} d^4 x \sqrt{g} \heavi(\Bound_0 - \Bound)  \delta^2 \left(\deltafun( \Sc_0 - \Sc)  \, \norm  \right) \nonumber \\
 &{} =  \int_{\mathcal{M}} d^4 x \sqrt{g} \heavi(\Bound_0 - \Bound)  \left( \frac{\partial^2 \deltafun( \Sc_0 - \Sc) }{ \partial \Sc ^2}     \norm  \delta \Sc^{2}   \right.      \nonumber   \\
 & \left. \quad \; \; + \, 2 \, \frac{\partial \deltafun( \Sc_0 - \Sc) }{ \partial \Sc }   \frac{ \partial  \norm }{ \partial \nabla_{\mu} \Sc } \delta \Sc \nabla_{\mu} (\delta \Sc) + \deltafun( \Sc_0 - \Sc)    \frac{ \partial^2  \norm }{ \partial \nabla_{\nu} \Sc \partial \nabla_{\mu} \Sc } \nabla_{\nu} (\delta \Sc)    \nabla_{\mu} (\delta \Sc)      \right)   \nonumber  \\
  &{} =  \int_{\mathcal{M}} d^4 x \sqrt{g} \heavi(\Bound_0 - \Bound)  \deltafun( \Sc_0 - \Sc)    \frac{ \partial^2  \norm }{ \partial \nabla_{\nu} \Sc \partial \nabla_{\mu} \Sc } \nabla_{\nu} (\delta \Sc)    \nabla_{\mu} (\delta \Sc)        \nonumber \\
  &{} =  - \int_{\mathcal{M}} d^4 x \sqrt{g} \heavi(\Bound_0 - \Bound)   \deltafun( \Sc_0 - \Sc)    \frac{1}{\norm} h^{\mu \nu}  \nabla_{\nu} (\delta \Sc)    \nabla_{\mu} (\delta \Sc)         \; , 
\end{align}
where the second last equality follows from partial integration of the term involving $\norm  \delta \Sc^{2} \, (\partial^2 \deltafun( \Sc_0 - \Sc) / \partial \Sc ^2 ) $, the condition for stationarity $\bm \nabla \cdot \bm V = 0$, and (\ref{eq:derNorm}). The last equality follows from (\ref{eq:derNorm}). 
Since $h^{\mu \nu}$ is positive semi-definite, we have $\delta^2 \Vol_{\Sc_0}^{\Sc}   \leq 0$, with equality when the perturbation $\delta \Sc$ depends solely on $\Sc$ (with $\bm \nabla \delta \Sc \propto \bm \nabla \Sc$). Such perturbations simply map the foliation onto itself --- and in fact, they have to vanish when the domain does have a boundary, due to the boundary condition $\left. \delta \Sc \right|_{\Bound = \Bound_0} = 0$ that we set in this case for the current example.     
Thus $\delta^2 \Vol_{\Sc_0}^{\Sc} < 0$ for any infinitesimal actual change of foliation, and hence the extremal foliation \emph{maximises} the volume. This is an expected consequence of the Lorentzian signature of the metric, while a Riemannian signature would induce a minimisation of the volume. 

\paragraph{We next consider the case where $\bm \Vec$ is independent of the foliation.}
For variations of $\Sc$ that are zero on the boundary $\Bound = \Bound_0$ and in the boundary-free case, all foliations are extrema of the volume if and only if $\bm \nabla \cdot \VecField = 0$, and it follows that the volume is foliation independent if this condition is satisfied. Furthermore, we trivially have $\delta^2 \Vol_{\Sc_0}^{\Sc} = 0$ for all foliations in this class. 
The volume has no local extremum, on the other hand, if $\bm \nabla \cdot \bm \Vec \neq 0$. This is the case for a fluid proper volume measure, i.e. $\bm V = \bm u$ where $\bm u$ is the 4-velocity of a fluid source \cite{Heinesen:2018vjp,Buchert:2018yhd}, if any expansion or contraction of the fluid occurs within the integration domain. 

Choosing $\bm \Vec$ as a conserved current defines a foliation-independent ``volume''. 
We can consider again the example of a conserved rest mass current $\bm M = \varrho \bm u$ (\ref{eq:restmass_current}) from the above subsection, and set $\bm \Vec = \varrho \bm u$. 
In this example, the conserved ``volume'' is simply the total rest mass as defined in (\ref{eq:integration_mass}), and averages with such a window function are mass-weighted \cite{massweighted,Heinesen:2018vjp}. The foliation-independence is still, in principle, restricted to the case $\left. \delta \Sc \right|_{\Bound = \Bound_0} = 0$ or that of a boundary-free domain, but is recovered for any deformation and more generic domains with the additional requirement of a domain boundary comoving with the rest mass current, $\bm u \cdot \bm \nabla \Bound = 0$. 

\paragraph{A few remarks are in order on extremal leaves. } 
In this section we have commented on volume extremising foliations. However, we remark that \emph{a single leaf} that extremises volume -- or another integral functional of physical interest -- could be used as a preferred surface for the initial value problem in cosmology. A preferred initial surface may in turn be extrapolated to form a foliation, by for instance propagating the initial surface along a physically motivated 4-velocity field. 
Since stationarity requirements for a single leaf are easier to satisfy than for a full foliation, single leaf extrema may be explored in cases where it is not possible to identify a full extremal foliation.

\subsection{Variation of averaged quantities with respect to the foliation}
\label{StationarityConditionsAverage} 
We will now derive stationaity conditions for the average functional (\ref{eq:average}) under variations of the hypersurface scalar $\Sc$, analogous to the above results for the integral functional. 
Examples of physical average functionals of interest in cosmology are average density, expansion rate, and spatial curvature degrees. 

We write the first order variation of the average (\ref{eq:average}) under the   variation $\Sc \rightarrow \Sc + \delta \Sc$ as 
\begin{equation}
\label{eq:Aexpansion}
\braket{\Scalar}_{\Sc_0}^{\Sc} \rightarrow \braket{\Scalar}_{\Sc_0}^{\Sc + \delta \Sc} =\braket{\Scalar}_{\Sc_0}^{\Sc} +  \delta \braket{\Scalar}_{\Sc_0}^{\Sc} .
\end{equation}
The variation of the average can be expressed through the variation of integral quantities in the following way
\begin{equation} 
\label{eq:average_inf_variation_def}
\delta \braket{\Scalar}_{\Sc_0}^{\Sc}   = \frac{ \delta I (\Scalar)_{\Sc_0}^{\Sc}}{ I (1)_{\Sc_0}^{\Sc} } - \braket{\Scalar}_{\Sc_0}^{\Sc} \frac{ \delta I (1)_{\Sc_0}^{\Sc}}{ I (1)_{\Sc_0}^{\Sc} }  ,
\end{equation}
and we can plug in (\ref{eq:I_inf_variation3}) to obtain stationarity conditions for averaged quantities.
The conditions for demanding stationarity for the entire foliation are
\begin{align} 
& \qquad \qquad \qquad \qquad \delta \Braket{\Scalar}_{\Sc_0}^{\Sc} = 0   \quad  \forall  \,  \delta \Sc, \Sc_0   \qquad  \Leftrightarrow   \nonumber   \\
& \left\{
  \begin{array}{@{}ll@{}}
&\heavi (\Bound_0 - \Bound)   \Vec^{\nu}  \nabla_{\nu}\Sc  \nabla_{\mu}\Sc \left(  \frac{\partial}{ \partial (\nabla_{\mu} \Sc)} (   \Scalar )     \right)   = 0    \\
& \text{and} \\
&    \nabla_{\sigma}  \left(    \frac{\partial (  \Scalar \Vec^{\nu} \nabla_{\nu} \Sc ) }{ \partial (\nabla_{\sigma} \Sc)}    \, \heavi (\Bound_0 - \Bound)  \right) -  \Braket{\Scalar}_{\Sc_0}^{\Sc}        \nabla_{\sigma}  \left(    \frac{\partial (  \Vec^{\nu} \nabla_{\nu} \Sc ) }{ \partial (\nabla_{\sigma} \Sc)} \,   \heavi (\Bound_0 - \Bound)   \right)  =  0 \; .  \label{eq:AverageConstraintsAll}
\end{array}\right.
\end{align}
The condition for stationarity of an average functional for a single leaf can be similarly derived from (\ref{eq:I_inf_variation3}), providing an analogous criterion to (\ref{eq:IntegralStationarity}) for integral functionals. We do not include this condition for simplicity
given that such a situation will not be of interest for the investigations discussed below. We also note that, as for stationary integrals above, the second condition in Eq.~\eqref{eq:AverageConstraintsAll} --- or its single-slice equivalent --- may be further split into a bulk ($\Bound \leq \Bound_0$) and a boundary ($\Bound = \Bound_0$) conditions stemming from the domain selection factor $\heavi(\Bound - \Bound_0)$ and its derivative.

The global constraint equation
\begin{equation} 
\delta \Braket{\Scalar}_{\Sc_0}^{\Sc} = 0  \quad  \forall  \,  \delta \Sc, \Sc_0    \qquad  \Rightarrow  \qquad  \frac{ \partial \Braket{\Scalar}_{\Sc_0}^{\Sc} }{\partial \Sc_0}  = 0     \label{eq:AverageGlobalAll}  
\end{equation} 
must be satisfied in order to obtain stationarity for a single slice of a foliation selected by $\Sc=\Sc_0$ and for stationarity for the entire foliation respectively. 
The necessary condition (\ref{eq:AverageGlobalAll}) is analogous to the integral condition (\ref{eq:IntegralGlobalAll}).
As for (\ref{eq:IntegralGlobalAll}), the condition (\ref{eq:AverageGlobalAll}) can be shown to hold for averages arising from a general window function with arbitrary functional dependence on $\bm \nabla \Sc, \bm \nabla \bm \nabla \Sc, \dots ,\bm \nabla^{(n)} \Sc$. 

\paragraph{We now consider the special case where $\Scalar$ and $\VecField$ are independent of the foliation.} 
In this case, the constraints (\ref{eq:specialcaseSVAll}) are satisfied. 
Stationarity conditions for the entire foliation, Eq.~\eqref{eq:AverageConstraintsAll}, accordingly reduce to
\begin{equation}
\delta \Braket{\Scalar}_{\Sc_0}^{\Sc} = 0  \quad  \forall  \,  \delta \Sc, \Sc_0    \qquad  \Leftrightarrow  \qquad     \nabla_{\sigma} \left(  \Scalar   \Vec^{\sigma}  \heavi (\Bound_0 - \Bound)    \right)            -  \Braket{\Scalar}_{\Sc_0}^{\Sc}   \nabla_{\sigma} \left(  \Vec^{\sigma}  \heavi (\Bound_0 - \Bound)    \right)           = 0  \; . \label{eq:AverageConstraintsAllReduced}  
\end{equation}
Using (\ref{eq:AverageGlobalAll}), we see that the above condition is equivalent to $\Braket{\Scalar}_{\Sc_0}^{\Sc}$ being constant (independent of $\Sc_0$) together with $\big( \Scalar -  \Braket{\Scalar}_{\Sc_0}^{\Sc} \big)  \Vec^{\sigma}$ being a conserved current comoving with the boundaries of the domain, such that $ \nabla_{\sigma} \big( \big( \Scalar -  \Braket{\Scalar}_{\Sc_0}^{\Sc} \big)  \Vec^{\sigma}  \heavi (\Bound_0 - \Bound)    \big) = 0$ is satisfied  --- \emph{i.e.}, $ \nabla_{\sigma} \big( \big( \Scalar -  \Braket{\Scalar}_{\Sc_0}^{\Sc} \big)  \Vec^{\sigma} \big) = 0$ for $\Bound \leq \Bound_0$ and $\big( \Scalar -  \Braket{\Scalar}_{\Sc_0}^{\Sc} \big)  \big( \Vec^{\sigma} \nabla_{\sigma} \Bound \big) = 0$ on $\Bound = \Bound_0$. The latter boundary condition can be neglected if we consider variations that are fixed on the boundary, or a global averaging domain on a spatially closed manifold.

\paragraph{We remark that the conditions for stationarity of average functionals are very restrictive.}  
The existence of extremal foliations are conditioned on the existence of a locally-conserved current. 
Considering the case $\bm V = \bm u$, where $\bm u$ is a fluid four velocity field with an associated rest mass density $\varrho$, we note that a natural conserved current is $\varrho \, \bm u$: $\nabla_\mu (\varrho u^\mu) = 0$. However, since $\Braket{\varrho}_{\Sc_0}^{\Sc} \geq 0$ with equality only when $\varrho = 0$ everywhere, it follows that setting $\big(\Scalar - \average{\Scalar} \big) \propto \varrho$ cannot generate stationary solutions to (\ref{eq:AverageConstraintsAllReduced}), except for the trivial case of taking $\Scalar$ as a constant of spacetime. We note that $\varrho$ might itself be such a constant, but in addition to a homogeneous rest-mass distribution, this would imply a non-expanding fluid flow, $\nabla_\mu u^\mu = 0$, unless $\varrho = 0$, due to the conservation equation.

We note that solutions found in this section are valid only for the specified functional dependence of the average functional on the foliation. 
As discussed for integral functionals in section \ref{StationarityConditionsIntegral}, there might exist extremals that are not stationary points, occuring as infimums or as local extrema on the boundary of a set of allowed foliations.

\subsubsection{Example: Entropy}
The study of entropy is a rich topic in gravitational physics, and in cosmology it has found various applications, for instance in the characterisation of initial conditions and inflationary scenarios \cite{Brahma:2020zpk}, cyclic universe models \cite{Ijjas:2021zwv}, and structure formation \cite{Clifton:2013dha}. 
Here, we focus on the following entropy measure, inspired by the Kullback-Leibler relative information entropy \cite{10.1214/aoms/1177729694,Hosoya:2004nh}: 
\begin{equation} 
\label{eq:entropy_S}
\mathcal{S}_{\Sc_0}^{\Sc}  \equiv   I\left(\Scalar \ln\left(\frac{\Scalar}{ \braket{\Scalar}_{\Sc_0}^{\Sc}  }\right)\right)_{\Sc_0}^{\Sc}  . 
\end{equation}
where $\Scalar$ must be a field in which gravity induces an increased clustering or inhomogeneity. 
Physically relevant substitutions for $\Scalar$ include rest mass densities $\varrho$ as in \cite{Hosoya:2004nh}, expansion scalars $\theta$ (in case of positive expansion everywhere), and proper time measures $\tau$. 
The variation of $\mathcal{S}$ with the foliation reads
\begin{align}
\label{eq:entropy_S_variation}
\delta \mathcal{S}_{\Sc_0}^{\Sc} &=     \delta I\left(\Scalar \ln\left( \Scalar \right)  \right)_{\Sc_0}^{\Sc}   - \delta \left( I\left(\Scalar \right)_{\Sc_0}^{\Sc}  \ln\left( \braket{\Scalar}_{\Sc_0}^{\Sc}\right)   \right) \nonumber  \\
&{}=  \delta I\left(\Scalar \ln\left( \Scalar \right)  \right)_{\Sc_0}^{\Sc}  -  \left(  \ln\left( \braket{\Scalar}_{\Sc_0}^{\Sc}\right) +1   \right) \delta  I\left(\Scalar \right)_{\Sc_0}^{\Sc} + \braket{\Scalar}_{\Sc_0}^{\Sc} \delta I\left( 1 \right)_{\Sc_0}^{\Sc} \; . 
\end{align}

\paragraph{We restrict our investigation to the case where $\VecField$ is independent of the foliation.} 
In this case, the condition (\ref{eq:specialcaseSVAll}) applies, and using (\ref{eq:I_inf_variation3}), the stationarity condition becomes
\begin{equation}
\delta \mathcal{S}_{\Sc_0}^{\Sc}   = 0   \quad   \forall  \quad  \delta A, \Sc_0      \qquad  \Leftrightarrow \qquad  \left\{
  \begin{array}{@{}ll@{}}
& \left. \left( \Scalar \ln\left(\frac{\Scalar}{ \braket{\Scalar}_{\Sc_0}^{\Sc}  }\right) + \braket{\Scalar}_{\Sc_0}^{\Sc} - \Scalar \right) \Vec^{\mu}  \nabla_{\mu} B  \right|_{\Bound = \Bound_0}  = 0   \\
&\text{and} \\
&  \ln\left(\frac{\Scalar}{ \braket{\Scalar}_{\Sc_0}^{\Sc}  }\right)  \nabla_{\mu} \left( \Scalar  \Vec^{\mu} \right)  -  \left(\Scalar - \braket{\Scalar}_{\Sc_0}^{\Sc}\right)  \nabla_{\mu} \Vec^{\mu} = 0 \; .  \label{eq:Av_current_entropy} 
\end{array}\right.
\end{equation}
We have here explicitly split the condition into its boundary and bulk components (respectively from $\deltafun(\Bound - \Bound_0)$ and $\heavi(\Bound_0 - \Bound)$ terms). Accordingly, the second requirement above is to be satisfied everywhere inside the $\Bound \leq \Bound_0$ tube; the first requirement, as a boundary condition, is dropped in the case of a boundary-free averaging domain.
Assuming that $\Scalar$ is hypersurface-forming, the foliation into constant-$\Scalar$ hypersurfaces, $\Sc = \Scalar$, satisfies (\ref{eq:Av_current_entropy}) since $\Scalar$ is then by construction homogeneous over each slice, and $\Scalar = \braket{\Scalar}_{\Sc_0}^{\Sc}$.
It follows that $\mathcal{S}$ is stationary with value $\mathcal{S} = 0$ for this foliation. 
This point of stationarity is guaranteed to be a unique global minimum when $\Scalar$ is a strictly positive function (as is the case for a rest mass density) since
\begin{equation}
\label{eq:entropy_S_Jensen}
\frac{\mathcal{S}_{\Sc_0}^{\Sc}}{ I\left( 1 \right)_{\Sc_0}^{\Sc} }  =  \Braket{ \Scalar \ln\left(\frac{\Scalar}{ \braket{\Scalar}_{\Sc_0}^{\Sc}  }\right)}_{\Sc_0}^{\Sc}  = \Braket{\frac{\Scalar}{ \braket{\Scalar}_{\Sc_0}^{\Sc}  } \ln\left(\frac{\Scalar}{ \braket{\Scalar}_{\Sc_0}^{\Sc}  }\right)}_{\Sc_0}^{\Sc}  \braket{\Scalar}_{\Sc_0}^{\Sc}   \geq  0 \; ,
\end{equation}
where the inequality follows from Jensen's inequality for the function $x \mapsto x\ln(x)$, with equality only if $\Scalar$ is constant over the $\Sc = \Sc_0$ hypersurface. 
For a hypersurface-forming $\Scalar$, such an extremal-foliation exists irrespective of how $\bm \Vec$ and $\Bound$ are defined. However, to satisfy non-singularity of the integration measure ($\bm \Vec \cdot \bm \nabla \Sc \neq 0$) for this foliation, we would have to demand $\bm \Vec \cdot \bm \nabla \Scalar \neq 0$. 

Note that in case $\bm \nabla \Scalar \cdot \bm \nabla \Scalar < 0$ is not fulfilled everywhere, the constant-$\Scalar$ solution does not define a spatial foliation. 
This situation may occur for the choices of $\Scalar$ suggested above, $\Scalar \in \{ \varrho$, $\theta$, $\tau \}$, for which the hypersurface-forming property of $\Scalar$ must always be checked. 
Moreover, in general, the constant-$\Scalar$ foliation needs not be the unique solution to (\ref{eq:Av_current_entropy}), \emph{i.e.}, there may be other local extrema. 

\paragraph{We now consider the special case where $\Scalar \VecField$ is a conserved current.} 
In this case, we have $\nabla_\mu ( \Scalar \Vec^\mu ) = 0$. This  includes the physical example where $\VecField$ is a 4-velocity field and $\Scalar$ is the associated rest mass density. 
In this case, the stationarity conditions (\ref{eq:Av_current_entropy}) reduce to 
\begin{equation} 
\delta \mathcal{S}_{\Sc_0}^{\Sc}   = 0   \quad   \forall  \quad  \delta A, \Sc_0      \qquad  \Leftrightarrow  \qquad   \left\{
  \begin{array}{@{}ll@{}}
& \left. \left( \Scalar \ln\left(\frac{\Scalar}{ \braket{\Scalar}_{\Sc_0}^{\Sc}  }\right) + \braket{\Scalar}_{\Sc_0}^{\Sc} - \Scalar \right) \Vec^{\mu}  \nabla_{\mu} B  \right|_{\Bound = \Bound_0}  = 0   \\
& \text{and} \\
&( \braket{\Scalar}_{\Sc_0}^{\Sc} - \Scalar)   \nabla_{\mu} \Vec^{\mu} = 0 \; .  \label{eq:Av_current_entropySpecial} 
\end{array}\right.
\end{equation}
For $\bm \nabla \cdot \bm \Vec = 0$ the second condition of (\ref{eq:Av_current_entropySpecial}) is automatically satisfied for any foliation. 
Hence, in this case, the entropy is foliation independent and has a constant value (which is zero if and only if $\Scalar$ is a constant of spacetime), up to the boundary condition, i.e. the first condition of (\ref{eq:Av_current_entropySpecial}). The latter can be accounted for by either imposing $\bm \Vec \cdot  \bm \nabla  \Bound = 0$, by considering the particular case of a global boundary-free integration domain, or by simply keeping the foliation fixed at the domain's boundaries, $\left. \delta \Sc \right|_{\Bound=\Bound_0} =  0$. 
Note that this case of $\bm \nabla \cdot \VecField = 0$ also implies $\bm \Vec \cdot \bm \nabla \Scalar = 0$, and thus a foliation defined by $\Sc = \Scalar$ (where $\mathcal S$ would realise its global minimum at $\mathcal S = 0$) would result in a singular volume element. 

For $\bm \nabla \cdot \bm \Vec \neq 0$ everywhere, (\ref{eq:Av_current_entropySpecial}) is equivalent to $\Scalar = \braket{\Scalar}_{\Sc_0}^{\Sc}$, i.e., the global minimum of $\mathcal S$ corresponding to the constant-$\Scalar$ foliation is the only local extremum. It follows for instance that, for an everywhere expanding fluid with 4-velocity $\bm \Vec = \bm u$ and rest mass density $\Scalar = \varrho$, the constant-$\varrho$ foliation is the unique minimiser for the entropy.

\subsubsection{Example: Minimally differing frames}
Suppose that we have a physical time-like vector field $\bm u$ in our cosmological theory in the frame of which averaged quantities would be desirable. 
This would for instance mean averaging in the rest frame of a fluid source if $\bm u$ represents its 4-velocity. 
This 4-velocity field can have vorticity, which will prevent defining hypersurfaces that are orthogonal to its flow lines. In this case we may ask whether there is a unique space-like foliation (defined by a scalar $\Sc$ with time-like gradient), or a family of foliations, such that their normal vector field $\bm n$, given by 
\begin{equation}
\label{eq:normal}
n^{\mu} = \frac{- \nabla^{\mu} \Sc}{( - g^{\nu \kappa} \nabla_{\nu} \Sc \nabla_{\kappa} \Sc )^{1/2} } \; , 
\end{equation}
is maximally close to $\bm u$ by some measure. 
In cases where such a foliation could be defined, this would provide a natural frame for definining averages as close to the frame of $\bm u$ as possible. 
The tilt between the two normalised time-like vector fields 
\begin{equation}
\label{eq:tilt}
\gamma = - n^{\mu} u_{\mu} , \qquad \gamma \geq 1 \; , 
\end{equation}
is a natural local scalar measure of their closeness, 
where $\gamma = 1$ if and only if $\bm n = \bm u$. 
We define the measure of ``statistical closeness'' of the vector fields $\bm n$ and $\bm u$ over the domain defined by $\{ \Sc = \Sc_0 \, , \, \Bound \leq \Bound_0 \}$, as
\begin{equation} 
\label{eq:averaged_tilt}
\Braket{\gamma}_{\Sc_0}^{\Sc }= \frac{ \int_{\mathcal{M}} d^4 x \sqrt{g} \, W_{\Sc_0, \bm n}^{\Sc}  \gamma }{ \int_{\mathcal{M}} d^4 x \sqrt{g} \, W_{\Sc_0, \bm n}^{\Sc}  } =  \frac{ \int_{\mathcal{M}} d^4 x \sqrt{g} \, W_{\Sc_0, \bm u}^{\Sc}  }{ \int_{\mathcal{M}} d^4 x \sqrt{g} \, W_{\Sc_0, \bm n}^{\Sc}  }  \; \geq 1 \; ,
\end{equation}
with 
\begin{equation}
\label{eq:window_tilt}
 W_{\Sc_0, \bm n}^{\Sc} = n^{\mu} \nabla_{\mu} \Sc \; \deltafun (\Sc - \Sc_0)  \, \heavi (\Bound_0 - \Bound) \; ; \qquad W_{\Sc_0, \bm u}^{\Sc} = u^{\mu} \nabla_{\mu} \Sc \; \deltafun (\Sc - \Sc_0)  \, \heavi (\Bound_0 - \Bound)  \; ,
\end{equation}
being the window functions with volume measure defined with respect to the normal vector $\bm n$ (\ref{eq:normal}) and the vector field $\bm u$, respectively. 
Plugging in $\bm \Vec = \bm n$ and $\Scalar = \gamma$ in (\ref{eq:AverageConstraintsAll}), we have that the first condition of (\ref{eq:AverageConstraintsAll}) is automatically satisfied while the second condition of (\ref{eq:AverageConstraintsAll}), split into its bulk and boundary components, becomes 
\begin{equation}
 \delta \Braket{\gamma}_{\Sc_0}^{\Sc} = 0   \quad  \forall  \,  \delta \Sc, \Sc_0      \qquad  \Leftrightarrow  \qquad   
 \left\{
  \begin{array}{@{}ll@{}}
&\Braket{\gamma}_{\Sc_0}^{\Sc } \nabla_{\sigma}  n^{\sigma}  -  \nabla_{\sigma} u^{\sigma}   =  0      \\
& \text{and} \\
& \left.  \left( \Braket{\gamma}_{\Sc_0}^{\Sc }  n^{\sigma} - u^{\sigma}    \right) \nabla_{\sigma} \Bound   \right|_{\Bound = \Bound_0}   = 0 \; .  \label{eq:VariationTilt}   
\end{array}\right.
\end{equation} 
The second condition in (\ref{eq:VariationTilt}), \emph{i.e.}, the boundary condition, needs to be satisfied only if the domain has a boundary and we include variations of the foliation at that boundary as well as on the domain's interior. 
A necessary condition for (\ref{eq:VariationTilt}) is $\partial_{\Sc_0} \Braket{\gamma}_{\Sc_0}^{\Sc} = 0$. In solving (\ref{eq:VariationTilt}), we can thus consider $\average{\gamma}$ as a constant parameter. 

Investigations of the general solution to Eq.~\eqref{eq:VariationTilt} is beyond the scope of this paper. However, we consider here the special case where $\bm u$ is divergence-free, $\bm \nabla \cdot \bm u = 0$. The first condition of (\ref{eq:VariationTilt}) then reduces to $\bm \nabla \cdot \bm n = 0$. 
Finding a solution to the first condition of Eq.~\eqref{eq:VariationTilt} in this case amounts to examining the existence of zero extrinsic scalar curvature foliations.
The problem of extremising the averaged tilt (\ref{eq:averaged_tilt}) thus becomes equivalent to extremising the volume as in section \ref{example_integrated_volume}.
This is because $\int_{\mathcal M} \mathrm{d}^4 x \sqrt{g} \, W^\Sc_{\Sc_0, \bm u}$ becomes foliation independent, so that we are extremising the inverse of the Riemannian volume (as defined by $\int_{\mathcal M} \mathrm{d}^4 x \sqrt{g} \, W^\Sc_{\Sc_0, \bm n}$).
From the results in section \ref{example_integrated_volume} we know that stationary points for the Riemannian volume are maximal, meaning that the stationary points for the averaged tilt (\ref{eq:averaged_tilt}) are minimal in this case.

\section{Finite foliation changes and quantitative bounds on foliation dependence}
\label{sec:foliationdependence_restricted} 

The results of section~\ref{sec:variation} show that, while foliation independent statements can be made for special cases, most integral or average functionals are foliation dependent. 
Nevertheless, it may be possible in many cases of interest to quantify the level of foliation dependence. 
The aim of this section is to determine quantitative bounds on the foliation dependence in scenarios relevant for cosmological models.

\subsection{Correspondence between hypersurfaces of different foliations}
\label{corr_between_foliations}

In the following, we will consider two different foliations $\mathcal{F}$ and $\mathcal{F}'$ corresponding to the respective level sets of two scalars with past-pointing time-like or null gradients\footnote{%
Due to the signature of the Lorentzian metric tensor, a function with causal gradient that increases towards the future has a past-pointing gradient.%
}, $\Sc$ and $\Scp$, where the transformation $\Sc \mapsto \Scp$ needs not be infinitesimal.

In order to make comparisons of leaves $\Sigma_0 \equiv \Sigma_{\Sc = \ScZ}$ and $\Sigma_0' \equiv \Sigma_{\Scp = \ScpZ}$ of the two foliations in a meaningful way, we must ensure that the two slices correspond to the ``same time'' in some sense. 
We shall specifically require that $\Sigma_0$ and $\Sigma_0'$ intersect at at least one event within the bounded region determined by the tube $\Tube \equiv \{ x \in \mathcal{M} \, / \, \Bound(x) \leq \Bound_0 \}$. This ensures a notion of synchronisation at at least one point within the domain of interest.
It prevents in particular artificial differences due to the comparison of two different parametrisations  of the same foliation.

We can always choose a parametrization such that $\ScpZ \equiv \ScZ$, such that $\Sigma_0' = \Sigma_{\Scp = \ScpZ} = \Sigma_{\Scp = \ScZ}$ for all $\ScZ$.
This can be achieved by using the freedom of reparametrization of $\mathcal{F}'$ as per transformations of $\Scp$ of the class (\ref{eq:foliation_reparam}). 
The requirement of intersection of the pairs of corresponding slices from the foliations  
will suffice in what follows, even though it does not always uniquely specify the parametrization\footnote{%
To obtain, if necessary, a unique determination of the parametrization $\Scp$ given $\mathcal{F}$, $\mathcal{F}'$, and $\Sc$, one could, for instance, specify the intersection point of each pair of slices $\Sigma_{\Sc = \ScZ}$, $\Sigma_{\Scp = \ScZ}$) by requiring that $\Scp = \Sc$ everywhere along a given time-like curve within the domain $\Tube$. Such a curve could correspond to the worldline of a given (\emph{e.g.} geocentric) observer.%
} $\Scp$ for $\mathcal{F}'$ from the parametrization choice $\Sc$ for $\mathcal{F}$. It already ensures, in particular, that $\Scp$ must be chosen as equal to $\Sc$ (\emph{i.e.}, the transformation $\Sc \mapsto \Scp = f(\Sc)$ reduces to the identity) in the case $\mathcal{F} = \mathcal{F}'$ mentioned above.

\subsection{Simplifying assumptions and notations}
\label{sec:assumptions}

In this section, we shall only consider cases where $\Scalar$ and $\VecField$ are invariant under deformations $\Sc \mapsto \Scp$ of the foliation.
As has been argued in the above sections, defining $\VecField$ as a physical vector field independent of the foliation is natural for many purposes --- a natural choice of $\VecField$ could for instance be the 4-velocity field of a physical matter source.  
In most cases we will also be interested in averaging physical scalars that are independent of the foliation\footnote{%
In some cases, factors of $\VecField \cdot \bm \nabla \Sc$ appear naturally, when computing derivatives of global quantities, as seen in (\ref{eq:commutation_rule}) and (\ref{eq:commutation_rule_average}). 
We can still consider scalars that have dependence on $\VecField \cdot \bm \nabla \Sc$ in this framework, when we restrict the deformation of the foliation $\Sc$ to have gradient orthogonal to $\VecField$. Such a class of deformations is in fact of physical interest, \emph{e.g.} in the context of proper time foliations (see Sec.~\ref{subsec:constant-tau} below).%
}, as for instance the rest mass density or expansion rate of a physical matter source. 

The boundaries of the domain and their propagation between slices, as determined by the scalar $\Bound$, are already considered to be set independently of the foliation as part of our averaging scheme. Here, we shall make the simplifying assumption that the domain propagation follows the flow of the volume-measure vector: $\VecField \cdot \bm \nabla \Bound = 0$ --- unless a boundary-free domain is being considered so that boundary terms can already be discarded, with $\heavi(\Bound_0 - \Bound) = 1$ and $\deltafun(\Bound - \Bound_0) = 0$ everywhere, and $\Tube = \mathcal{M}$. We set $\VecField$ to be unitary, $\VecField \cdot \VecField = -1$; a non-normalized vector field (corresponding to weighted volume averages) could formally be absorbed into the scalar $\Scalar$ to be averaged, $\Scalar \VecField = \left[ \Scalar (\VecField \cdot \VecField)^{1/2} \right] \times \left[ (\VecField \cdot \VecField)^{-1/2} \, \VecField \right]$. The above assumption and normalization convention are again compatible with the choice of $\VecField$ as a source fluid's $4$-velocity; in this case with a fluid-comoving domain propagation. This is indeed one of the main applications we have in mind for this section (see, \emph{e.g.}, \cite{Buchert:2018yhd,Buchert:2001sa}).

\subsection{The difference of integral functionals between leaves of two foliations}
\label{sec:dif}  

We now consider the difference between integrals of a given scalar $S $ over leaves of two foliations $\mathcal{F}$ and $\mathcal{F}'$: $\Delta I(\Scalar) \equiv \spatintp{\Scalar} - \spatint{\Scalar}$, where $\Scalar$ is any foliation-independent scalar. Note that we are using the short hand notation for the difference, $\Delta I(\Scalar)$, where the dependence on the foliations and the leaves selected by $\Sc_0$ is implicit. 
We shall rewrite this difference in a way that will be convenient for defining upper bounds for its norm. 

\subsubsection{$\Delta I(\Scalar)$ in terms of covariant 4-integration}

We consider a given $\ScZ$ and the corresponding pair of intersecting slices $\Sigma_0 = \Sigma_{\Sc = \ScZ}$, $\Sigma_0' = \Sigma_{\Scp = \ScZ}$ from the two foliations $\mathcal{F}$, $\mathcal{F}'$, obeying the above assumptions. While the roles played by $\mathcal{F}$ and $\mathcal{F}'$ are formally symmetric, we will consider $\mathcal{F}$ as the reference foliation in which the integrals $\spatint{\xi}$ or averages $\average{\xi}$ of various scalars $\xi$ are known. 
$\mathcal{F}'$, on the other hand, will be considered as an arbitrary other space-like or null foliation that may be subjected to certain conditions, such as having a small tilt everywhere with respect to $\mathcal{F}$ in the case where both foliations are space-like. We will keep $\mathcal{F}'$ fully general in the present subsection.
We assume that $\mathcal{M}$ is a path-connected manifold, hence so are $\Sigma_0$ and $\Sigma_0'$ according to the global hyperbolicity assumption. 

In the following,
we will make use of the flowlines of $\VecField$ as a diffeomorphism between the domains of $\Sigma_0$ and $\Sigma_0'$ that are within the tube $\Tube$. 
While this mapping is covariantly defined, it will be convenient to introduce an associated set of spatial coordinates. 
We do so by arbitrarily choosing a coordinate basis $(X^i,\, i = 1,2,3)$ on one of the slices $\Sigma_{\Sc= \Sc_1}$ of $\CF$ (or some open subset of $\Sigma_{\Sc = \Sc_1}$ containing $\Sigma_{\Sc = \Sc_1} \cap \Tube$). The three coordinates $X^i$, assumed to span $\Rcube$ without loss of generality, can then be extended into an incomplete (spatial) set of coordinates in spacetime by requiring them to be constant along the flow lines of $\VecField$: $\VecField \cdot \bm \nabla (X^i) = 0, \, \forall i \in \{1,2,3 \}$. The spacetime tube $\Tube = \{ x \in \mathcal{M} \, / \, \Bound(x) \leq \BZ \}$ with $\VecField \cdot \bm \nabla \Bound = 0$, then corresponds to a given compact domain in the space of the spatial coordinates $(X^i)$; in other words $\BoundH$ is a function of $(X^i)$. (This also holds in the case of a spatially closed manifold with a global, boundary-free averaging domain.)
One may then introduce any time coordinate to complete $(X^i)$ into a spacetime coordinate set --- again for convenience in the below calculations. We shall complete it into a synchronous--comoving coordinate set adapted to the $4-$vector field $\VecField$ by introducing a proper time $\tau$ of $\VecField$, i.e., a function satisfying $\VecField \cdot \bm \nabla \tau = 1$ (see section~2.4.2 of \cite{Buchert:2001sa}). We can use the residual freedom in the definition of $\tau$ to demand that $\tau = 0$ on $\Sigma_0$ --- even if $\Sigma_0$ is light-like. This then uniquely specifies\footnote{%
Although we still simply denote it as $\tau$ for convenience, note that the scalar function uniquely defined in this way is specific to $\Sigma_0$; it will define a different function if another $\ScZ$ is considered.%
} $\tau$ among the proper-time parametrizations of the family of worldlines of $\VecField$.  

With the simplifying assumptions from Sec.~\ref{sec:assumptions}, the difference $\Delta I(\Scalar) = \spatintp{\Scalar} - \spatint{\Scalar}$  can be expressed covariantly as:
\begin{equation}
\label{eq:delta_I_basic}
\!\!\!\! \Delta I (\Scalar) = \int_{\mathcal{M}} d^4x \, \sqrt{g} \, \Scalar \, \BoundH  \left[ \Vec^\mu \nabla_\mu \Scp \; \deltafun(\Scp - \ScZ)  - \Vec^\mu \nabla_\mu \Sc \;  \deltafun(\Sc - \ScZ)  \right] \; . 
\end{equation}

The spacetime functions $\Sc$, $\Scp$ and $\tau$ are all nondecreasing along each of the worldlines of $\VecField$. Hence, these three scalars are nondecreasing functions of each other for any constant value of the $\VecField$-comoving spatial coordinates $X^i$. 
This implies that slices of both foliations $\CF$ and $\CF'$ can be parametrized by $\tau$. In particular, $\Sigma_0'$ can be characterized as: $P \in \Sigma_0' \Leftrightarrow \tau(P) =  \functau(X^i(P))$ (for any spacetime point $P$), where the values taken by the scalar field $\tau$ on $\Sigma_0'$ define a smooth function of the spatial coordinates\footnote{%
The Cauchy hypersurface $\Sigma_0'$ from $\CF'$, like $\Sigma_{\Sc = \Sc_1}$ from $\CF$, intersects each of the (time-like) flow lines of $\VecField$ exactly once. Hence, the flow of $\VecField$ defines a diffeomorphic parametrization of the points of $\Sigma_0'$ by their spatial coordinates $(X^i)$ within the domain of interest.%
} that we denote as $\functau(X^i)$.  In turn, $\Sigma_0$ is simply parametrized as $P \in \Sigma_0 \Leftrightarrow \tau(P) = 0$.
Using these parametrizations, the weighted slice-selecting terms $\Vec^\mu \nabla_\mu \Scp \; \deltafun(\Scp - \ScZ)$ and $\Vec^\mu \nabla_\mu \Sc \;  \deltafun(\Sc - \ScZ)$ in Eq.~(\ref{eq:delta_I_basic}) can be rewritten as, respectively: 
\begin{equation}
\Vec^\mu \nabla_\mu \Scp \; \deltafun(\Scp - \ScZ) = \Vec^\mu \nabla_\mu \tau \; \deltafun(\tau - \functau(X^i)) \quad \mathrm{and} \quad \Vec^\mu \nabla_\mu \Sc \;  \deltafun(\Sc - \ScZ) = \Vec^\mu \nabla_\mu \tau \; \deltafun(\tau) \; .
\end{equation}
Eq.~(\ref{eq:delta_I_basic}) then becomes
\begin{equation}
\Delta I (\Scalar) = \int_{\mathcal{M}} d^4x \, \sqrt{g} \,  \Scalar \, \BoundH (\Vec^\mu \nabla_\mu \tau) \left[ \deltafun(\tau - \functau(X^i))  -  \deltafun(\tau )  \right] .
\label{eq:delta_I_Znablatau}
\end{equation} 

\subsubsection{$\Delta I(\Scalar)$ in terms of 3-integration in adapted coordinates}
Using the definition $\Vec^\mu \nabla_\mu \tau = 1$ and choosing the spacetime coordinate system $x^\mu = (\tau,X^i)$ as the coordinates of integration in (\ref{eq:delta_I_Znablatau}), we can integrate this expression over $\tau$: 
\begin{equation}
 \Delta I (\Scalar) = \int_{\Rcube} d^3 X \, \BoundH \left[ \big( \sqrt{g} \, \Scalar \big)_{(\tau = \functau(X^i), X^i)} - \big( \sqrt{g} \, \Scalar \big)_{(\tau=0, X^i)} \right] \, , 
\label{eq:delta_I_as_sqrtb_S_difference}
\end{equation}
where the subscript coordinates refer to an evaluation point, that is, $f_{(\tau,X^i)}$ would be a function $f$ evaluated at the event of coordinates $(\tau, X^i)$.

The metric determinant modulus $g$ appearing in Eq.~(\ref{eq:delta_I_as_sqrtb_S_difference}) above is the one obtained in the coordinate system $(\tau,X^i)$. To recover a more coordinate-independent expression, one may introduce the (positive) determinant $b = \det(b_{ij})$ of the local spatial projector orthogonal to $\VecField$, $\mathbf{b} = {b_{\mu \nu} \, dx^\mu \otimes dx^\nu} = b_{ij} \, dX^i \otimes dX^j$, $i,j = 1,2,3$, with $b_{\mu \nu} \equiv g_{\mu \nu} + \Vec_\mu \Vec_\nu$. This determinant remains invariant under a change of the time coordinate, and it coincides with the value taken by $g$ in the coordinate system $(\tau,X^i)$. (We generalize the choice of time coordinate and show the relation between the two determinants in Appendix~\ref{app:volumeelement}.) The occurrences of $\sqrt{g}$ in Eq.~(\ref{eq:delta_I_as_sqrtb_S_difference}) may thus equivalently be replaced by $\sqrt{b}$.  The associated volume $3-$form $\sqrt{b} \, d^3 X$ is then invariant both under a change of time coordinate and under a time-independent change of the spatial coordinates $X^i$ (\emph{i.e.}, under a relabelling of the flow lines of $\VecField$). This volume $3$-form corresponds to the (manifestly covariant) Hodge dual $\star \underline{\VecField}$ to the $1$-form $\underline{\VecField} \equiv \Vec_\mu \, dx^\mu$ associated to $\VecField$, and may be interpreted as the infinitesimal spatial volume element in the local $\VecField$-orthogonal frames (see \cite[sec.~4 {and Appendix~D}]{Buchert:2001sa} and \cite{Heinesen:2018vjp}). In cases where $\VecField$ is hypersurface-forming, $\mathbf{b}$ would correspond to the Riemannian spatial metric tensor induced by $\mathbf{g}$ on the $\VecField$-orthogonal hypersurfaces, and $\sqrt{b} \, d^3 X$ would then simply be the associated spatial volume form.

To compute the difference in $\sqrt{b} \, \Scalar$ between two points along a given flow line of $\VecField$ that appears in Eq.~(\ref{eq:delta_I_as_sqrtb_S_difference}), let us first write the evolution equation along $\VecField$ of the $\sqrt{b}$ factor, in the adapted coordinates $(\tau,X^i)$: 
\begin{equation}
\label{eq:evol_sqrt_b_0}
\Vec^\mu \partial_\mu \! \left(\sqrt{b} \right)  = \partial_\mu \! \left(\sqrt{b} \, \Vec^\mu \right) =  \partial_\mu \! \left(\sqrt{g} \, \Vec^\mu \right) = \sqrt{g} \,\nabla_\mu \Vec^\mu = \sqrt{b} \,\nabla_\mu \Vec^\mu \, ,
\end{equation}
where we again made use of two relations holding in these coordinates: $\sqrt{b} = \sqrt{g}$ and $\Vec^\mu = (1,0,0,0)$. The first and last sides of Eq.~(\ref{eq:evol_sqrt_b_0}), and hence their equality, are nevertheless independent of the choice of the time coordinate. 
The evolution rate of the volume measure $\sqrt{b}$, which is additionally invariant under a change of $\VecField$-comoving spatial coordinates $(X^i)$, is thus given by
\begin{equation}
\label{eq:evol_sqrt_b}
\frac{1}{\sqrt{b}} \, \Vec^\mu \partial_\mu \! \left( \sqrt{b} \right) = \frac{1}{\sqrt{b}} \, \left. \frac{\mathrm{d}}{\mathrm{d} \tau} \right|_{X^i} \! \left( \sqrt{b} \right) = \nabla_\mu \Vec^\mu \, ,
\end{equation}
where the operator $\left. ( \mathrm{d}/\mathrm{d}\tau ) \right|_{X^i}$ corresponds to a derivative along $\VecField$ with respect to $\tau$.
Using the coordinates $(\tau,X^i)$, we can now  explicitly integrate Eq.~(\ref{eq:evol_sqrt_b}) into
\begin{equation}
\big( \sqrt{b} \big)_{(\tau=\tau_1,X^i)} = \big( \sqrt{b} \big)_{(\tau=0,X^i)} \, \exp \left( \int_{\tau=0}^{\tau_1} (\nabla_\mu \Vec^\mu)_{(\tau,X^i)} \; d\tau \right) \, ,
\label{eq:solution_sqrt_b}
\end{equation}
for any value $\tau_1$ of $\tau$. The main integrand in expression~(\ref{eq:delta_I_as_sqrtb_S_difference}) for $\Delta I(\Scalar)$ is then rewritten as:
\begin{equation} 
\big( \sqrt{b} \, \Scalar \big)_{(\tau = \functau(X^i), X^i)} - \big( \sqrt{b} \, \Scalar \big)_{(\tau=0, X^i)} 
= \big( \sqrt{b} \big)_{(\tau=0,X^i)} \; \psi(X^i) \; ,
\label{eq:diff_sqrt_b_S_1}
\end{equation}
with
\begin{align}
& \psi(X^i) \equiv \left[ \exp \left( \int_{\tau=0}^{\functau(X^i)} (\nabla_\mu \Vec^\mu)_{(\tau,X^i)} \; d\tau \right) - 1 \right] \Scalar_{(\tau=0,X^i)} \nonumber \\
{} & \quad +  \exp \left( \int_{\tau=0}^{\functau(X^i)} (\nabla_\mu \Vec^\mu)_{(\tau,X^i)} \; d\tau \right) \; \int_{\tau=0}^{\functau(X^i)} (\Vec^\mu \partial_\mu \Scalar)_{(\tau,X^i)} \; d \tau \; .
\label{eq:psi_result}
\end{align}

With this rewriting, the expression for $\Delta I(\Scalar)$ finally reduces to
\begin{equation}
\Delta I(\Scalar) = \int_{\Rcube} d^3 X \, \BoundH \, \big( \sqrt{b} \big)_{(\tau=0,X^i)} \,  \psi(X^i) = \spatint{\bar{\psi}} \, ,
\label{eq:delta_I_as_I_psi}
\end{equation}
where $\psi(X^i)$ is extended into a spacetime scalar $\bar{\psi}$ by defining the latter as equalling $\psi(X^i)$ on some Cauchy hypersurface (say $\Sigma_0$) and satisfying $\VecField \cdot \bm \nabla \bar{\psi} = 0$; \emph{i.e.}, $\bar{\psi}_{(\tau,X^i)} = \psi(X^i) \; \forall \tau$. With a slight abuse of notation, one may simply write $\Delta I(\Scalar) = \spatint{\psi(X^i)}$.

We give in Appendix~\ref{app:psi_form2} an alternative form of $\psi(X^i)$ re-expressed in terms of the local current $\nabla_\mu \left( \Scalar \Vec^\mu \right)$. This form allows for a more direct connection with our results on exact foliation-(in)dependence of Sec.~\ref{StationarityConditionsIntegral} and for alternative bounds on finite variations of spatial integrals and averages to the ones presented below.

\subsection{Bounds for foliations with small relative tilts}
\label{sec:smalltiltsbounds} 

Cosmology is typically studied under the assumption of small (nonrelativistic) relative velocities between relevant observers in spacetime. In this section, we shall therefore consider bounds on integrals relevant for a class of space-like foliations which are close to the reference field $\VecField$ (and to each other) in terms of their relative Lorentz factors. 

\subsubsection{The small tilts assumption}
\label{smalltiltsassumption}

In this subsection, we restrict our attention to space-like foliations.
We denote as $\bm n$ the future-pointing unit time-like normal vector field to the foliation $\mathcal{F}$, \emph{i.e.} satisfying
\begin{equation}
\bm n = -N \, \bm \nabla \Sc \quad ; \quad N = (-\nabla_\mu \Sc \, \nabla^\mu \Sc)^{-1/2} \; ,
\end{equation}
where $N$ is the lapse function associated with $\Sc$. We similarly define $\bm {n'}$ as the unit time-like normal associated with $\mathcal{F}'$ and $N'$ as the lapse associated with $\Scp$, with $\bm{n'} = - N' \, \bm{\nabla} \Scp$. 

We can then define the local Lorentz factors between $\VecField$, $\bm n$ and $\bm{n'}$, as follows:
\begin{equation}
\gvn \equiv  - \VecField \cdot \bm n \geq 1 \quad ; \qquad \gvnp \equiv  - \VecField \cdot \bm{n'} \geq 1 \quad ; \qquad \gnnp \equiv  - \bm n \cdot \bm{n'} \geq 1 \; , 
\end{equation} 
and introduce the local decomposition of $\bm{n'}$ with respect to $\VecField$: 
\begin{equation}
\bm{n'} = \gvnp \left(  \VecField + \bm{v'} \right) \; , \qquad v'^\mu \Vec_\mu = 0 \; , 
\label{eq:decomp_np}
\end{equation}
where $\bm{v'}$ automatically satisfies $v'^\mu v'_\mu = 1 - \gvnp^{-2}$.

The key assumption that we shall use in this subsection is that both $\bm n$ and $\bm {n'}$ are close to $\VecField$, that is, that the tilt velocities $\sqrt{1-\gvn^{-2}}$ and $\sqrt{1-\gvnp^{-2}}$  ( $ = \sqrt{v'^\mu v'_\mu}$ ) are small, \emph{i.e.}, globally bounded by a small parameter $v_1 \ll 1$. 
This implies in particular that the relative tilt between slices of the two foliations also remain small and globally bounded: $\sqrt{1 - \gnnp^{-2}} \leq 2 v_1$. 
{We then also have $\gvn \leq (1-v_1^2)^{-1/2} \simeq 1$, and we can} introduce the global small parameter {$v_0 \equiv v_1 (1-v_1^2)^{-1/2} \simeq v_1$, satisfying everywhere $v_1 \leq \gvn \, v_1 \leq v_0 \ll 1$} as the main characteristic spatial velocity to be used in the below bounds. 

\subsubsection{Bounding the distance between two tilted slices}
\label{sec:slicedistancebound}
From the expression of $\psi$ above, Eq.~(\ref{eq:psi_result}), or its rewriting in Eq.(\ref{eq:psi_result_alt}), it is clear that in addition to assuming upper limits on the local $\Scalar$-- and $\VecField$--based variables $| \Scalar |$ (on $\Sigma_0$), $|\Vec^\mu \partial_\mu \Scalar |$, $|\nabla_\mu (\Scalar \Vec^\mu) |$ and/or $|\nabla_\mu\Vec^\mu |$, one also needs to be able to bound the proper-time parametrization function $| \functau(X^i) |$ of $\Sigma_0'$ for all points. 
This quantity provides a measure of the distance between the two slices $\Sigma_0$ and $\Sigma_0'$, and in this section we shall provide a bound of $| \functau(X^i) |$ in terms of the small velocity parameter $v_0$.

\begin{figure}[h]
    \centering
    \includegraphics[width=0.7\textwidth]{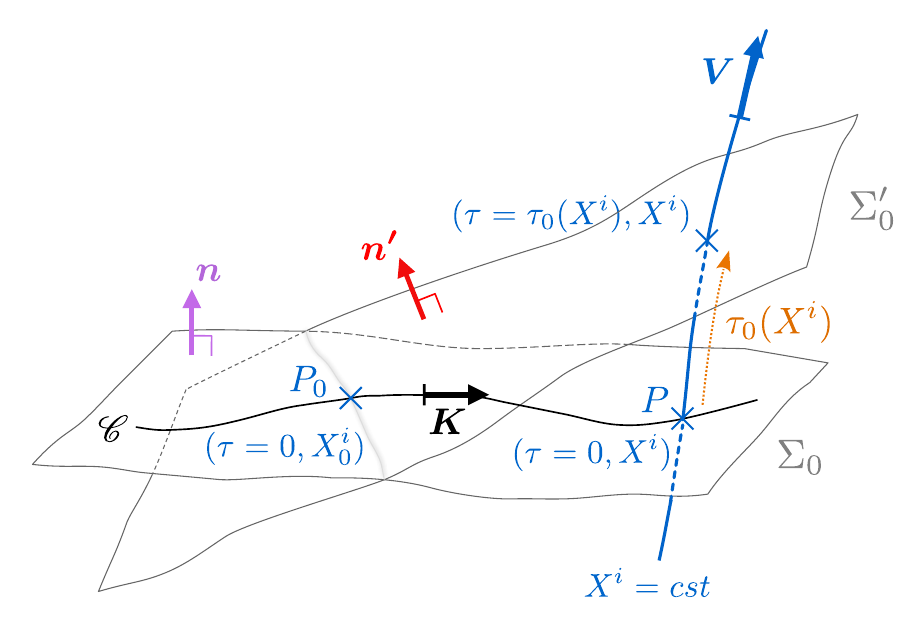}
    \caption{The spatial slices $\Sigma_0$ and $\Sigma_0'$, the spatial curve $\mathscr{C} \subset \Sigma_0$, and the main points and vector fields used for expressing and bounding the time-like distance $\functau$ between the two slices. $P$ is an arbitrary point of $\Sigma_0$, while $P_0$, at coordinates $(\tau=0, X^i_0)$, belongs to the intersection $\Sigma_0 \cap \Sigma'_0$ of both slices under consideration: $\functau(X^i_0) = 0$. Note that for this schematic representation which is not specifically concerned with causality, we use a Riemannian picture of orthogonality for easier visualisation.}
    \label{fig:dtau_dlambda_setup}
\end{figure}

Consider any given point $P$ --- of coordinates $(\tau=0,X^i)$ for a certain $(X^i)$ --- within the integration domain on $\Sigma_0$. One can draw a geodesic spatial curve $\mathscr{C}$ within $\Sigma_0$
joining $P$ to a reference point $P_0$ --- say of coordinates $(\tau=0,X^i_0)$ --- within the integration domain on $\Sigma_0$ where $\functau(X^i_0) = 0$. That is, $P_0$ is taken as an intersection point of the two slices: $P_0 \in \Sigma_0 \cap \Sigma_0' \cap \Tube$ which we assumed to be non-empty. (See Fig.~\ref{fig:dtau_dlambda_setup} for an illustration of this geometric setup.) $\mathscr{C}$ can be parametrized by its unit space-like $\bm{n}$--orthogonal tangent vector $\bm{K}$, and the associated affine parameter $\lambda$.  Setting $\lambda = 0$ at $P_0$, $\lambda$ then runs from $0$ at $P_0$ to $\length$ at $P$, where $\length$ is the total proper length of the curve within $\Sigma_0$. The coordinates of the point at parameter $\lambda$ along $\mathscr{C}$ can then be parametrized as $x^\mu (\lambda) = (\tau=0, X^i(\lambda))$.
These definitions allow us to perform spatial integrations along $\SC$, using $P_0$ as a reference point where $\functau = 0$, writing for instance
\begin{equation}
\functau(X^i) = \int_{\lambda=0}^{\length} \, \frac{\mathrm{d}}{\dl}  \! \left( \functau(X^i(\lambda)) \right) \, d\lambda \; .
\label{eq:int_along_C_trivial}
\end{equation}
In the following we will use the short-hand notation $\functau(\lambda) \equiv \functau(X^i(\lambda))$.

We may extend the tangent vector $\bm{K}$ {of $\mathscr{C}$} 
into a vector field {$\bm{\tilde L}$} on the congruence generated by $\VecField$ from $\mathscr{C}$ --- \emph{i.e.}, along all worldlines of $\VecField$ that intersect $\Sigma_0$ at a point on $\mathscr{C}$ --- by Lie dragging $\bm{K}$ along $\VecField$: $\mathscr{L}_{\VecField} \bm{\tilde{L}} \equiv \nabla_{\VecField} \bm{\tilde{L}} - \nabla_{\bm{\tilde{L}}} \VecField = 0$, \emph{i.e.}, $ (\mathrm{d}/\mathrm{d}\tau) \tilde L^\nu |_{X^i} = \tilde L^\mu \partial_\mu \Vec^\nu $, with $\bm{\tilde{L}} = \bm{K}$ at $\tau = 0$, along any given $\mathscr{C}$--intersecting worldline of $\VecField$. Along these same worldlines, we can then introduce a unit space-like vector $\bm{L}$ built from $\bm{\tilde L}$: 
\begin{equation}
    L^\mu \equiv \frac{b^\mu_{\ \nu} \tilde L^\nu}{b_{\mu \nu} \tilde L^\mu \tilde L^\nu} \quad ; \quad \bm L \cdot \bm L = b_{\mu \nu} L^\mu L^\nu = 1 \; .
\end{equation}
The $\VecField$--orthogonal projection $b^\mu_{\ \nu} \tilde L^\nu$ is indeed nonvanishing, given that $\bm{\tilde L}$ is not parallel to $\VecField$ at $\tau = 0$ (as $\bm K \cdot \bm K = 1$ while $\VecField$ is time-like) and that this property is preserved by the Lie dragging. Note that, on points of $\mathscr{C}$, $\bm K$ coincides with $\bm{\tilde L}$ (by construction), but \emph{a priori} not with the projected vector $\bm L$, since $\bm K$ needs not be orthogonal to the local $\VecField$.

The derivative of $\functau$ along $\mathscr{C}$ involved in Eq.~(\ref{eq:int_along_C_trivial}) is then obtained as  
\begin{equation}
\frac{d \functau(\lambda) }{d \lambda} = \left( \frac{\bm{\tilde L} \cdot \bm{n'} }{\gvnp}\right)_{(\tau=\functau(\lambda),X^i(\lambda))}  \quad .
\label{eq:dtau_dlambda_compact}
\end{equation}
It can also be re-expressed as follows:
\begin{equation}
\frac{d \functau(\lambda) }{d \lambda} = (\VecField \!\cdot \bm K)_{\lambda} + \sqrt{1+(\VecField \!\cdot \bm K)_{\lambda}^2}  \left[ (\bm L \cdot \bm{v'})_{(\tau=\functau(\lambda),X^i(\lambda)) }\;  F(\functau(\lambda),\lambda) + G(\functau(\lambda),\lambda) \right] \; ,
\label{eq:dtau_dlambda}
\end{equation}
where $(\VecField \!\cdot \bm K)_\lambda \equiv (\VecField \!\cdot \bm K)_{(\tau=0,X^i(\lambda))}$ is $\VecField \!\cdot \bm{K}$ evaluated at the current point on $\mathscr{C}$ and where 
\begin{align}
\label{eq:defF}
F(\tau_1,\lambda) & \equiv \exp \left[ \int_{\tau=0}^{\tau_1} {\, (\Theta_{\mu \nu} \, L^\mu L^\nu)_{(\tau,X^i(\lambda))} \; d\tau} \right] \; ; \\
\label{eq:defG}
G(\tau_1,\lambda) & \equiv \int_{\tau=0}^{\tau_1}{ \, (\bm L \cdot \bm a)_{(\tau,X^i(\lambda))} \; F(\tau,\lambda) \, d\tau} \; , 
\end{align}
for any $\tau_1$.
The derivation of these results, Eqs.~\eqref{eq:dtau_dlambda_compact}--\eqref{eq:defG}, is given in Appendix~\ref{app:derivation_dtaudlambda}. In the above expressions for $F$ and $G$, we have used the decomposition of the covariant derivative of $\underline{\VecField}$ with respect to $\VecField$ \cite{Ehlers:1961xww}:
\begin{equation}
\nabla_\mu \Vec_\nu = - \Vec_\mu \, \Vec^\rho \nabla_\rho \Vec_\nu + b^\rho_{\, \mu} b^\sigma_{\, \nu} \nabla_\rho \Vec_\sigma  = - \Vec_\mu \, a_\nu + \Theta_{\mu \nu} + b^\rho_{\, [\mu} b^\sigma_{\, \nu]} \nabla_\rho \Vec_\sigma \, .
\label{eq:decomp_grad_Vec}
\end{equation}
This defines the expansion tensor of $\VecField$, with components $\Theta_{\mu \nu} \equiv b^\rho_{\, (\mu} b^\sigma_{\, \nu)} \nabla_\rho \Vec_\sigma$; its vorticity tensor, of components $b^\rho_{\, [\mu} b^\sigma_{\, \nu]} \nabla_\rho \Vec_\sigma$; and its acceleration vector $\bm{a}$, with components $a^\mu \equiv \Vec^\nu \nabla_\nu \Vec^\mu$.

We can right away use the above global smallness assumption on the relative tilts between $\bm n$, $\bm n'$ and $\VecField$ to derive bounds on the terms $| \VecField \!\cdot \bm K |$ and $\sqrt{1 + (\VecField \cdot \bm K)^2_\lambda}   \, \left| \bm L \cdot \bm{v'} \right|$ appearing in Eq.~\eqref{eq:dtau_dlambda}. We first note that $| \VecField \!\cdot \bm K | = | h_{\mu \nu} \Vec^\mu K^\nu |  \leq \sqrt{h_{\mu \nu} \Vec^\mu \Vec^\nu} = \sqrt{\gvn^2 - 1} = \gvn \sqrt{1-\gvn^{-2}}$, where $h_{\mu \nu} \equiv g_{\mu \nu} + n_\mu n_\nu$ are the components of the spatial projector (which is also the induced metric) on the leaves of $\CF$. 
Using this, the second above term obeys the following inequality:
\begin{align}
\sqrt{1 + (\VecField \cdot \bm K)^2_\lambda}   \, \left| \bm L \cdot \bm{v'} \right|_{(\tau,X^i(\lambda))} & {}={}  \sqrt{1 + (\VecField \cdot \bm K)^2_\lambda} \, \left| b_{\mu \nu} L^\mu v'^\nu \right|_{(\tau,X^i(\lambda))} \nonumber \\
&\leq (\gvn)_{(\tau=0,X^i(\lambda))} \sqrt{(v'^\mu v'_\mu)_{(\tau,X^i(\lambda))}}  \;  .
\label{eq:v0_bound_second_term}
\end{align}
With the small tilt velocities $v_1$, $v_0$ satisfying {everywhere $\sqrt{1-\gvn^{-2}} \leq v_1$, $\sqrt{v'^\mu v'_\mu} \leq v_1$} and $\gvn v_1 \leq v_0$, both of the above terms {from Eq.~\eqref{eq:dtau_dlambda}} are everywhere smaller than $v_0$.

\paragraph{We shall now assume the existence of a global bound on the norm of the expansion tensor.} 
That is, we assume that there exists a bound on $(\Theta^\mu_{\; \nu} \, \Theta^\nu_{\; \mu})^{1/2}$ that applies throughout the part of $\CM$ under consideration (\emph{i.e.}, a certain portion of $\Tube$). 
This is equivalent to assuming a global bound $\boundTheta$ on all eigenvalues of the (diagonalizable) matrix $\Theta^\mu_{\; \nu}$ over the local tangent spaces orthogonal to $\VecField$, so that for any point $P$ in the spacetime region of interest, for each eigenvalue $\theta_k(P)$ ($k = 1,2, 3$) of $\Theta^\mu_{\; \nu}$ at $P$ over the tangent space orthogonal to $\VecField$ at $P$, $\left| \theta_k (P) \right| \leq \boundTheta$. Note that these eigenvalues are real, and covariantly defined.  This implies, in particular, that the norm of the volume expansion rate $\nabla_\mu \Vec^\mu = b^{\mu \nu} \nabla_\mu \Vec_\nu = \Theta^\mu_{\; \mu}$ is also globally bounded: $\left| \nabla_\mu \Vec^\mu \right| \leq 3 \boundTheta$. 
We also set the acceleration of $\VecField$ to zero (see the discussion below for the case of a nonzero acceleration), so that $G(\functau(\lambda),\lambda) = 0$.

We shall now use the above quantities to set bounds on $\functau(X^i)$. 
Let us first consider the case $\functau(X^i) > 0$. This correspond to $\Sigma_0'$ lying to the future of the $\VecField$-comoving observer at $P$.
If $\functau(\lambda)$ changes sign along $\SC$, it will cross $0$ again at some $\lambda = \lambda_0 > 0$, corresponding to another point in $\Sigma_0' \cap \Sigma_0 \cap \Tube$.\footnote{%
This requires the integration domain $\Sigma_0 \cap \Tube$ to be path-connected, and geodesically convex or at least star-shaped with respect to $P_0$. If this is not satisfied but $\Sigma_0$ as a whole does satisfy these conditions, spatial paths may be drawn within an extended flow-tube $\extendedTube$ of $\VecField$, encompassing $\Tube$, such that $\Sigma_0 \cap \extendedTube$ satisfies these properties while keeping its diameter as small as possible (\emph{e.g.}, $\Sigma_0 \cap \extendedTube$ could be taken as the smallest sphere containing $\Sigma_0 \cap \Tube$). The assumed global bounds on tilts, $\Theta_{\mu \nu}$, or the integration domain's diameter, then simply are understood to hold on a portion of $\extendedTube$ rather than only the corresponding one of $\Tube$, while the domain of spatial integration is unchanged ($\Sigma_0 \cap \Tube$ or $\Sigma_0' \cap \Tube$).
If $\Sigma_0$ is itself not geodesically convex \emph{e.g.} due to punctures, when referring to spatial curves within $\Sigma_0$ the meaning of `geodesic' has to be extended to a non-necessarily unique path taken as short as possible within $\Sigma_0$ (for instance, we do not need $\bm K$ to satisfy the spatial geodesic equation within $\Sigma_0$).%
} We then only need to consider the part of $\SC$ beyond this point, and we may simply replace the integration bounds $[\lambda = 0, \, \lambda = \length]$ in integrals like Eq.~(\ref{eq:int_along_C_trivial}) by $[\lambda = \lambda_0, \, \lambda = \length]$. Otherwise, we can simply set $\lambda_0 = 0$. We thus ensure in either case that $\functau(\lambda) > 0 \; \forall \lambda \in \, ] \lambda_0, \length]$, with $\functau(\lambda = \lambda_0) = 0$.

From the bound $v_0$ on $\left| (\VecField \cdot \bm{K})_\lambda \right|$ and on {$\sqrt{1 + (\VecField \cdot \bm K)^2_\lambda}   \, \left| \bm L \cdot \bm{v'} \right|_{(\tau,X^i(\lambda))}$ (for any $\tau$)}, and with $G(\functau(\lambda),\lambda) = 0$, Eq.~(\ref{eq:dtau_dlambda}) gives
\begin{equation}
\frac{d}{d \lambda}  \functau(\lambda) \leq v_0 \, \left[1 + F(\functau(\lambda),\lambda) \right] \; ,
\label{eq:dtau_dlambda_bound1}
\end{equation}
remembering that $F > 0$ by definition. One moreover has
\begin{equation}
\label{eq:Fcontents_bound}
\Theta_{\mu \nu} \, L^\mu L^\nu =  (b_{\mu \rho} L^\rho) \, \Theta^\mu_{\; \nu} \,  (b^\nu_{\, \sigma} L^\sigma) \leq \boundTheta \left( b_{\mu \nu} L^\mu L^\nu \right)  =  \boundTheta  \; ,
\end{equation}
where we have used again the orthogonality of $\VecField$ and its expansion tensor, implying $\Theta_{\mu \nu} = b_\mu^{\; \rho} \, \Theta_{\rho \nu} = \Theta_{\mu \sigma} b^\sigma_{\, \nu}$. Injecting the above inequality into Eq.~(\ref{eq:defF}) implies $F(\functau(\lambda),\lambda) \leq \exp \! \left( \boundTheta\, \functau(\lambda) \right)$ for any $\lambda \in [0,\length]$, given that $\functau(\lambda) \geq 0$. Eq.~(\ref{eq:dtau_dlambda_bound1}) then becomes,
\begin{equation}
\label{eq:dtau_dlambda_bound2}
\! \! \! \! \frac{d}{d \lambda}  \functau(\lambda) \leq v_0 \, \left[1 + \exp \! \left( \boundTheta \, \functau(\lambda) \right) \right]  , \; i.e. , \; \frac{d}{d \lambda} \! \left( \ln \!\left[1 + \exp \! \left( - \boundTheta \, \functau(\lambda) \right) \right] \right) \geq -v_0 \boundTheta .
\end{equation}

This may then be integrated between $\lambda_0$, where $\functau(\lambda_0) = 0$, and any $\lambda \geq \lambda_0$, to give
\begin{equation}
\ln \!\left[1 + \exp \! \left( - \boundTheta \, \functau(\lambda) \right) \right] \geq \ln(2) - v_0 \boundTheta (\lambda - \lambda_0) \; ,
\end{equation}
that is,
\begin{equation}
\label{eq:tau0_invexpbound}
\exp \! \left( -\boundTheta \, \functau(\lambda) \right) \geq 2 \, \exp \! \left( - v_0 \boundTheta (\lambda - \lambda_0) \right) - 1 \, .
\end{equation}
If the above right-hand side is nonnegative, which is ensured for all $\lambda \in [\lambda_0, \length]$ provided $v_0$ is small enough such that $v_0 \boundTheta (\length - \lambda_0) < \ln(2)$, we then obtain the following bound on $\functau(X^i) = 
\functau(\lambda = \length)$ as a function of the tilt velocity $v_0$:
\begin{equation}
0 \leq \functau(X^i) \leq -\frac{1}{\boundTheta} \ln \! \left[ 2 \, \exp \! \left(-v_0 \boundTheta (\length - \lambda_0) \right) - 1 \right] \; .
\end{equation}
As the integration domain $\Sigma_0 \cap \Tube$ is compact, $v_0 \boundTheta (\length - \lambda_0)$ admits a maximum; we can thus introduce a small parameter $\eta$ (with still $\eta < \ln(2)$ to satisfy the above assumption) such that $v_0 \boundTheta (\length - \lambda_0) \leq \eta$ for any $P$. We can then use the convexity of the function $x \mapsto - \ln \! \left( 2 e^{-x} - 1 \right)$ for $0 \leq x < \ln(2)$ to write
$ - \ln \! \left[ 2 \, \exp \! \left(-v_0 \boundTheta (\length - \lambda_0) \right) - 1 \right] \leq \alpha \, v_0 \boundTheta (\length - \lambda_0)$, with $\alpha \equiv -   \eta^{-1} \ln \! \left( 2 e^{-\eta} - 1 \right)$. This then gives a bound on $\functau(X^i)$ that is linear in $v_0$ and independent of $\boundTheta$ (apart from setting the above condition on $v_0 \boundTheta (\length - \lambda_0)$):
\begin{equation}
0 \leq \functau(X^i) \leq \alpha \, v_0 \, (\length - \lambda_0) \; .
\end{equation}
The (positive) numerical factor $\alpha = -   \eta^{-1} \ln \! \left( 2 e^{-\eta} - 1 \right)$ yields for instance $\alpha \simeq 2.11$ for $\eta = 1/10$, and converges to $\alpha = 2$ for $\eta \rightarrow 0$.

The factor $\length - \lambda_0$ ($\leq \length$) in the above bound and condition, is simply the length of a reduced curve $\tilde{\SC}$, the (still spatially geodesic) part of $\SC$ parametrized by $\lambda \in [\lambda_0, \length]$, \emph{i.e.}, the part of $\SC$ joining the point of coordinate $x^\mu(\lambda_0)$ to $P$. It is in fact clear that this bound may be written in terms of the length of the shortest among all curves on $\Sigma_0$ that join $P$ to a point of $\Sigma_0 \cap \Sigma_0'$. Since this simply amounts to a redefinition (if necessary) of $\SC$ and of its length $\length$, we may simply rewrite the above bound as
\begin{equation}
\label{eq:tau0_positivebound}
0 \leq \functau(X^i) \leq \alpha \, v_0 \, \length \; ,
\end{equation}
where the previous assumed bound on $v_0 \, (\ll 1)$ and $\length$ becomes:
\begin{equation}
v_0 \boundTheta \length \leq \eta \; , \quad \text{with} \quad \eta < \ln(2) \; ,
\label{eq:v0HL_assumption}
\end{equation}
where $\length$ is now interpreted as the length of the shortest curve on $\Sigma_0$ joining $P$ to $\Sigma_0 \cap \Sigma_0'$ as above. This length may be significantly smaller than that of a geodesic on $\Sigma_0$ joining $P$ to a given arbitrary point $P_0 \in \Sigma_0 \cap \Sigma_0'$ in cases where the $\Sigma_0$ and $\Sigma_0'$ slices intersect multiple times (\emph{e.g.}, periodically). 

The symmetric case $\functau(X^i) < 0$, corresponding to $\Sigma_0'$ lying to the past of the $\VecField$-comoving observer at $P$, can be handled in a similar way. One may first assume that $\functau(\lambda) < 0 \, \forall \lambda \in ]0, \length]$ with $\functau(\lambda=0) =0$ up to a suitable redefinition of $\SC$, $\length$ (and the parameter $\lambda$) as above. The bound $v_0$ on $\left| (\VecField \cdot \bm{K})_\lambda \right|$ and on {$\sqrt{1 + (\VecField \cdot \bm K)^2_\lambda}   \, \left| \bm L \cdot \bm{v'} \right|_{(\tau,X^i(\lambda))}$}, along with Eq.~(\ref{eq:Fcontents_bound}) (writing this time $\Theta_{\mu \nu} \, L^\mu L^\nu \geq { - \boundTheta } $), can then again be used to bound the derivative of $\functau(\lambda)$ {as given by Eq.~\eqref{eq:dtau_dlambda}}, this time from below (\emph{cf.} Eqs.~(\ref{eq:dtau_dlambda_bound1}) and (\ref{eq:dtau_dlambda_bound2})):
\begin{equation}
\frac{d}{d \lambda}  \functau(\lambda) \geq - v_0 \, \left[1 + F(\functau(\lambda),\lambda) \right]
 \geq - v_0 \, \left[1 + \exp \! \left( - \boundTheta \, \functau(\lambda) \right) \right]  \;  .
\end{equation}
Proceeding similarly to the $\functau(X^i) > 0$ case above, this results in
\begin{equation}
0 \geq \functau(X^i) \geq - \alpha \, v_0 \, \length \; ,
 \label{eq:tau0_negativebound}
\end{equation} 
still assuming $v_0 \boundTheta \length \leq \eta < \ln(2)$, with the same numerical factor $\alpha = - \eta^{-1} \ln \! \left(2 e^{-\eta}-1 \right)$ as above. 
In this bound, $\length$ is to be interpreted in the same way as discussed for Eqs.~(\ref{eq:tau0_positivebound})--(\ref{eq:v0HL_assumption}) above.

Combining Eqs.~(\ref{eq:tau0_positivebound}), for the $\functau(X^i) > 0$ case, and~(\ref{eq:tau0_negativebound}), for the $\functau(X^i) < 0$ case, implies that in all cases\footnote{%
This bound remains of course valid in the $\functau(X^i) = 0$ case, even considering that $\length$ may then be set to $0$.%
},
\begin{equation}
\left| \functau(X^i) \right| \leq \alpha \, v_0 \, \length \; ,
\label{eq:tau0_absbound}
\end{equation}
under the condition~(\ref{eq:v0HL_assumption}), with $\alpha = \alpha(\eta)$ as above and still the same interpretation for $\length$ as discussed above.
$\left| \functau(X^i) \right|$ corresponds to the distance between the slices $\Sigma_0$ and $\Sigma_0'$ as measured along the flow line of $\VecField$ passing through $P$. The spatial curve length $\length$ depends on the point $P \in \Sigma_0 \cap \Tube$ considered, but it may itself be globally bounded for all points $P$ in this domain by a finite length $\lengthg$. The latter may for instance be defined as the diameter of the domain within $\Sigma_0$,
\begin{equation}
\lengthg = \lengthg_1 \equiv \max_{P_1,P_2 \, \in \, \Sigma_0 \cap \Tube} \mathrm{d} (P_1, P_2) \; ,
\label{eq:diameter_def}
\end{equation}
or as the maximal distance along $\Sigma_0$ of a point in the domain to the intersection of the two slices within the domain,
\begin{equation}
\lengthg = \lengthg_2 \equiv \max_{P \, \in \, \Sigma_0 \cap \Tube, \, P_0 \, \in \, \Sigma_0 \cap \Tube \cap \Sigma_0'} \; \mathrm{d} (P, P_0) \; .
\end{equation}
Above, $\mathrm{d}(P,P')$ is the spatial distance along $\Sigma_0$ between the points $P$ and $P'$, that is, the proper length of the shortest curve on $\Sigma_0$ joining $P$ and $P'$. The existence of the above maxima is guaranteed by the compactness of the spatial domain $\Sigma_0 \cap \Tube$ (and consequently of $\Sigma_0 \cap \Tube \cap \Sigma_0'$, which is non-empty). $\lengthg_2$ is always a smaller (or equal) length than $\lengthg_1$, and may be much smaller, but it could be harder to determine in practice, and it generally depends on $\Sigma_0'$.

We assume that the bounds on the tilt, $v_0 \ll 1$, on the expansion rate, $\boundTheta$, and on the distances on the reference slice as defined by $\lengthg$, can be taken as small enough to ensure a relatively small $v_0 \boundTheta \lengthg$. We denote as $\eta$ the best upper limit that may be set \emph{a priori} on this value, and as above we assume that this limit is strictly smaller than $\ln(2)$. We thus have
\begin{equation}
v_0 \boundTheta \lengthg \leq \eta \; , \quad \text{with} \quad \eta < \ln(2) \; ,
\label{eq:v0HL_assumption_global}
\end{equation}
and consequently the condition~(\ref{eq:v0HL_assumption}) is satisfied with this same $\eta$ for all points $P \in \Sigma_0 \cap \Tube$. Eq.~(\ref{eq:tau0_absbound}) then implies that for all such points, of coordinates $(\tau=0,X^i)$,
\begin{equation}
\left| \functau(X^i) \right| \leq \alpha \, v_0 \, \lengthg \; .
\label{eq:tau0_absbound_global}
\end{equation}
This global bound means in particular that, within the spacetime tube $\Tube$ delineating the domain of interest, the distance between the slices $\Sigma_0$ and $\Sigma_0'$ (along $\VecField$) is everywhere much smaller than the spatial size of the domain as measured in the reference slice $\Sigma_0$, if $v_0 \ll 1$.  

\paragraph{Case of a nonvanishing $4$-acceleration / velocity dispersion.}
In the more general case, the $4$-acceleration $\bm a$ of $\VecField$ cannot be everywhere neglected. When for instance $\VecField$ represents the $4$-velocity of a source fluid, $\bm a$ can correspond to non-gravitational accelerations, which we would consider negligible apart from extremely small fractions of the volume of the domain considered, but it can also arise from the effective modelling of velocity dispersion within the source fluid by effective pressure terms.
We can then assume some global bound $\bounda$ (with dimension time${}^{-1}$ or length${}^{-1}$) on $\sqrt{\bm a \cdot \bm a} = \sqrt{b_{\mu \nu} a^\mu a^\nu}$ and follow a similar derivation as above from Eq.~(\ref{eq:dtau_dlambda}), noting that $|G(\functau(\lambda),\lambda)| \leq \bounda \, \int_0^{\functau(\lambda)} \exp{(\boundTheta \tau)} \, d\tau = \bounda / \boundTheta \left[ \exp{(\boundTheta \functau(\lambda))} - 1 \right]$. For a small enough $\bounda$ so that $\bounda \lengthg$ is at most of order unity and under a possibly more stringent, $\bounda$--dependent constraint on $v_0 \boundTheta \length$ than previously, bounds on $| \functau |$ now involving $\bounda \lengthg$ can be similarly obtained. 

For instance, let us assume that $\zeta(\bounda \lengthg - v_0 \boundTheta \lengthg) \leq K$ for some constant $K>0$, where $\zeta$ is the nonnegative, nondecreasing function defined by $\zeta(x) \equiv (e^x - 1)/x$. The inverse function $\zeta^{-1}$ can be defined and is nondecreasing as well, and the above condition is equivalently rewritten as $\bounda \lengthg \leq v_0 \boundTheta \lengthg + \zeta^{-1}(K)$. Assuming additionally that the bound $\eta$ on $v_0 \boundTheta \lengthg$ is small enough such that $v_0 \boundTheta \lengthg \leq \eta < 1/ (2 K)$, one can show that
\begin{equation}
    \left| \functau(X^i) \right| \leq \left( \frac{-\ln{(1- 2 K \eta)}}{K \eta} \right) \, \times K \times v_0 \, \lengthg  \; .
\end{equation}
Note that the above requirements imply again that $\eta < \ln{2}$, and thus $K > 1/(2 \ln{2})$. 
The prefactor $[ -\ln{(1- 2 K \eta)} ]/[K \eta]$ in the above expression is always larger than $2$ (corresponding to its $K \eta \ll 1$ limit), but remains of order unity if $2 K \eta$ is not too close to $1$. 
Consequently, for the above bound on $| \functau |$ to be significant, $K$ should not be very large, and $\bounda \lengthg$ should accordingly be at most of order unity (with $\zeta^{-1}(K) \sim \ln{K}$ for large $K$). For instance, $\bounda \lengthg \simeq 1$ would still allow for taking $K \lesssim 2$.
Note that one can take \emph{e.g.} $K= 1$ if the $4$-acceleration is small enough that $\bounda \lengthg \leq v_0 \boundTheta \lengthg \leq \eta < 1/2$ (as $\zeta^{-1}(1) = 0$).

In practice, the above requirements may be too restrictive, or the bound too large, to be used directly.\footnote{%
Considering for instance, in the late Universe, the natural case of $\VecField$ representing the $4$-velocity field of a non-relativistic (nearly dust) matter fluid, the main contribution to its $4$-acceleration over most of the volume is expected to arise from velocity dispersion within the fluid acting as effective pressure forces which contribute to oppose gravitational collapse in bound structures. The magnitude of such a $4$-acceleration from effective pressure can then be estimated as being {at most} of the same order as the (Newtonian) gravitational acceleration within those virialized domains. A typical present-day value for these accelerations in the outskirts of galaxies or within galaxy cluster haloes, for instance, would then be of the order of $10^{-10} \, \mathrm{m}.\mathrm{s}^{-2}$ (\emph{e.g.}, \cite{Gaiaacc,eckert2022}), or a little smaller. This value of $\bounda$ accumulated over a large cosmological domain with a diameter $\lengthg$ of a Hubble length, would correspond to $\bounda \lengthg$ of about $10^{-1}$. A more pessimistic estimate of the maximum $\bounda$ of the $4$-acceleration local amplitude, \emph{e.g.} accounting for the central regions of dense clusters or for the haloes of very massive elliptic galaxies, and still with $\lengthg \simeq 1/H_0$, may then lead to an $\bounda \lengthg$ of order unity or beyond. The regions where this may occur, however, would occupy a very small volume fraction at such scales and may accordingly be avoided by the spatial path and/or neglected in the volume-weighted spatial integrals considered.%
} This can be considered as being due to accumulating proper-time differences along $\Sigma_0$ when $4$-accelerations have a specific consistent spatial orientation (corresponding to making the Cauchy-Schwarz bound $| \bm L \cdot \bm a | \leq { \sqrt{\bm a \cdot \bm a} \, \sqrt{b_{\mu \nu} L^\mu L^\nu} \; \left( = \sqrt{\bm a \cdot \bm a} \right) }$ in $G$ into an equality, with no changes of sign). However, especially when $\VecField$ models a physical fluid $4$-velocity, and regardless of $\bm a$ corresponding to non-gravitational forces or to effective velocity dispersion, a residual $4$-acceleration is only expected around localized overdensities, hence small parts of the domain. The acceleration vector is moreover expected to have radial orientations around the centers of those overdensities, leading to compensating signs in $\bm L \cdot \bm a$ over paths crossing such regions. In such a case, we may thus assume that the fluid's $4$-acceleration only contributes small corrections to the above ($\bm a = \bm 0$) bounds. It can moreover be argued that the small spatial regions where $\bm a$ may be non-negligible could be avoided by the spatial path $\mathscr{C}$ up to a small increase in its total length, while the cases where the endpoint $P$ of the path, spanning the whole integration domain, falls within such a dense region may be neglected for their small contribution to the volume-weighted spatial integrals $I(\Scalar)$ in either slice. When $\VecField$ arises from a different, more geometric construction, we shall simply assume it to be a geodesic vector field for simplicity. Accordingly, in the following, we will adopt either of these simplifying assumptions and use the bounds obtained in the vanishing-acceleration case above --- with small corrections being \emph{e.g.} encompassed into taking a slightly pessimistic value for $\eta$ or $v_0$, if necessary.

\subsubsection{Resulting bounds on the norm of $\Delta I(\Scalar)$}
\label{sec:deltaIbounds}

For any given $X^i$ in the domain considered, the above bound on $|\functau(X^i)|$, Eq.~(\ref{eq:tau0_absbound}), results in a constraint on the volume ratios $(\sqrt{b})_{(\tau=\tau_1,X^i)} / (\sqrt{b})_{(\tau=0,X^i)} = \exp \left( \int_{\tau=0}^{\tau_1} (\nabla_\mu \Vec^\mu)_{(\tau,X^i)} \, d\tau \right)$ that appear in $\psi(X^i)$, Eq.~(\ref{eq:psi_result}), for $\tau_1 \equiv \functau(X^i)$. Together with the global bound $\left| \nabla_\mu \Vec^\mu \right| \leq 3 \boundTheta$, Eq.~(\ref{eq:tau0_absbound}) implies
\begin{equation}
\left| \int_{\tau=0}^{\tau_1} (\nabla_\mu \Vec^\mu)_{(\tau,X^i)} \, d\tau \right| \leq 3 \boundTheta |\tau_1| \leq 3 \boundTheta |\functau(X^i) | \leq 3 \, \alpha \, v_0 \boundTheta \length \; ,
\end{equation}
still assuming the condition~(\ref{eq:v0HL_assumption}) to hold.
Noting that $| \exp(x) - 1 | \leq \exp(|x|) -1 $ for any real number $x$, and then successively using the monotonicity and the convexity of the exponential function, the above relation gives
\begin{equation}
\left| \, \exp \left( \int_{\tau=0}^{\tau_1} (\nabla_\mu \Vec^\mu)_{(\tau,X^i)} \, d\tau \right) - 1 \, \right|  \leq e^{3 \boundTheta | \functau(X^i) |} - 1 \leq e^{3 \, \alpha \, v_0 \boundTheta \length} - 1 \leq \tilde\alpha \, v_0 \boundTheta \length \, .
 \label{eq:tau0_exp_bound}
\end{equation}
 The numerical factor $\tilde\alpha$ defined as {$\tilde\alpha \equiv \eta^{-1} (e^{3 \;\! \alpha \;\! \eta} - 1) = 2 \eta^{-1} (e^\eta - 1) (e^{2 \eta} - 2 e^\eta + 4) / (2 - e^\eta)^3$, yields for instance $\tilde\alpha \simeq 8.8$} for $\eta = 1/10$, and converges to $\tilde\alpha = 6$ for $\eta \rightarrow 0$.

Taking $\tau_1 = \functau(X^i)$ and applying this inequality to $\psi$ in Eq.~(\ref{eq:psi_result}) then gives:
\begin{align}
| \psi(X^i) | & \leq \tilde\alpha \, v_0 \boundTheta \length \left| \Scalar_{(\tau=0,X^i)} \right| + (1 + \tilde\alpha \, v_0 \boundTheta \length) \left| \functau(X^i) \right| \! \times \!\! \max_{\tau \, \in \, [0, \, \functau(X^i)]} \left| \left( \Vec^\mu \partial_\mu \Scalar \right)_{(\tau,X^i)} \right|  \nonumber \\
{} & \leq \tilde\alpha \, v_0 \boundTheta \length \left| \Scalar_{(\tau=0,X^i)} \right| + \alpha (1 +  \eta \tilde\alpha) \, v_0 \, \length \times \max_{\tau \, \in \, [0, \, \functau(X^i)]} \left| \left( \Vec^\mu \partial_\mu \Scalar \right)_{(\tau,X^i)} \right| \; .
\label{eq:psi_bound_1}
\end{align}
The numerical factor $\alpha (1+\eta \tilde\alpha)$ in the second term yields for instance $\alpha (1+\eta \tilde\alpha) \simeq 4.0$
for $\eta = 1/10$ and converges to $\alpha (1+\eta \tilde\alpha) = 2$ for $\eta \rightarrow 0$.

The range $\tau \, \in [0 ,\, \functau(X^i)]$ for the maximum simply corresponds geometrically to taking a maximum over the segment of a flow line of $\VecField$ that joins the two slices $\Sigma_0$ and $\Sigma_0'$. The interval $[0, \functau(X^i)]$ (which should be read as $[\functau(X^i),0]$ in case $\functau(X^i) < 0$) is part of the larger interval $[ - \alpha \, v_0 \, \length, \, \alpha \, v_0 \, \length]$ from Eq.~(\ref{eq:tau0_absbound}), and we may then use
\begin{equation}
\max_{\tau \, \in \, [0, \, \functau(X^i)]} \left| \left( \Vec^\mu \partial_\mu \Scalar \right)_{(\tau,X^i)} \right| \leq \max_{\tau \, \in \, [ - \alpha \, v_0 \, \length, \, \alpha \, v_0 \, \length]} \left| \left( \Vec^\mu \partial_\mu \Scalar \right)_{(\tau,X^i)} \right| \; ,
\label{eq:intermediate_bound_on_max}
\end{equation}
to get rid of the remaining dependence of Eq.~(\ref{eq:psi_bound_1}) on the specific foliation $\CF'$. Setting again as $\eta$ the best upper threshold that may be set on the global $v_0 \boundTheta \lengthg$, and assuming that it can obey the constraint~(\ref{eq:v0HL_assumption_global}), $\eta < \ln(2)$, we can moreover replace the two $X^i$-dependent factors $\length$ in Eq.~(\ref{eq:psi_bound_1}) by the global length bound $\lengthg$. With these two remarks, inserting the above Eq.~(\ref{eq:psi_bound_1}) into Eq.~(\ref{eq:delta_I_as_I_psi}) provides the following bound on the variation of the spatial integral of the scalar $\Scalar$ between the two slices under the condition~(\ref{eq:v0HL_assumption_global}):
\begin{align}
& \left| \Delta I(\Scalar) \right| \leq \spatint{\left| \psi(X^i) \right|} \nonumber \\
{} & \;\; \leq \tilde\alpha \, v_0 \boundTheta \lengthg \; \spatint{\left| \Scalar \right|} + \alpha (1 + \eta \tilde\alpha) \, v_0 \, \lengthg \; \spatint{\max_{\tau \, \in \, [ - \alpha \, v_0 \, \length, \, \alpha \, v_0 \, \length]} \left| \left( \Vec^\mu \partial_\mu \Scalar \right)_{(\tau,X^i)} \right|} \, .
\label{eq:delta_I_bound1}
\end{align}
Above, we have kept the $X^i$-dependent distance $\length$ in the second spatial integral, but it may as well be replaced by the global size $\lengthg$ (since $[ - \alpha \, v_0 \, \length, \, \alpha \, v_0 \, \length] \subset [ - \alpha \, v_0 \, \lengthg, \, \alpha \, v_0 \, \lengthg]$) to give a simpler, though \emph{a priori} weaker bound.
We also note that, from Eq.~(\ref{eq:psi_result}), we could instead have written
\begin{equation}
\left| \int_{\tau=0}^{\functau(X^i)} \left( \Vec^\mu \partial_\mu \Scalar \right)_{(\tau,X^i)} \, d\tau \right| \leq \int_{\tau=- \alpha \, v_0 \, \length}^{\alpha \, v_0 \, \length} \left| \Vec^\mu \partial_\mu \Scalar \right|_{(\tau,X^i)} \, d\tau \; ,
\end{equation}
and thus the second spatial integral in Eq.~(\ref{eq:delta_I_bound1}) could alternatively be replaced by:
\begin{equation}
\spatint{\int_{\tau=- \alpha \, v_0 \, \length}^{\alpha \, v_0 \, \length} \left| \Vec^\mu \partial_\mu \Scalar \right|_{(\tau,X^i)} \, d\tau} \; , \quad \text{ or by:} \quad \;\; \spatint{\int_{\tau=- \alpha \, v_0 \, \lengthg}^{\alpha \, v_0 \, \lengthg} \left| \Vec^\mu \partial_\mu \Scalar \right|_{(\tau,X^i)} \, d\tau} \, ,
\label{eq:delta_I_bound1b}
\end{equation}
where we have omitted the prefactor $\alpha (1 + \eta \tilde\alpha) \, v_0 \, \lengthg$ in front of this integral.

Another variant of the above bound, Eq.~\eqref{eq:delta_I_bound1}, is provided in Appendix~\ref{app:psi_form2} as Eq.~\eqref{eq:delta_I_bound2}. It is based on an alternative writing of $\psi(X^i)$ and requires some knowledge about the local variable $\nabla_\mu \left(\Scalar V^\mu \right)$, rather than $\Scalar$ and $\Vec^\mu \partial_\mu \Scalar$ as in the above.

The bound obtained here, Eq.~\eqref{eq:delta_I_bound1}, ensures that $|\Delta I(\Scalar)| / I(|\Scalar|)^\Sc_{\Sc_0} \ll 1$ provided the `tilt-weighted' domain size $v_0 \lengthg$ ($\ll \lengthg$) can be considered much smaller than the characteristic lengths (or times) associated with $\boundTheta^{-1}$ and with $\big| \Scalar_{(0,X^i)} \big| \, / \, | (\Vec^\mu \nabla_\mu \Scalar)_{(\tau,X^i)} |$. At least in the cases where $\Scalar$ does not change sign on $\Sigma_0$ (\emph{e.g.}, for an energy or mass density, or for $\Scalar = 1$, or as a known property of $\Sigma_0$), this expresses a direct constraint on the relative variation of its integral, $| \Delta I(\Scalar) / I(\Scalar)_{\Sc_0}^\Sc |$.

In general, this bound, as well as its alternative form in Eq.~\eqref{eq:delta_I_bound2}, require that the global upper limit $\boundTheta$ on the local directional expansion rates exists over the appropriate spacetime region, and that it and/or the domain size and tilt velocity be sufficiently small for the condition~\eqref{eq:v0HL_assumption_global} to be satisfied. As mentioned, the bounds become more stringent if one can even ensure that $v_0 \boundTheta \lengthg \ll 1$ (\emph{i.e.}, $\eta \ll 1$). In a cosmological setup, and in the case where $\VecField$ can be associated with the $4$-velocity of a matter fluid source component, one can expect the typical expansion rate within the domain to be of the order of the Hubble parameter at a characteristic time of $\Sigma_0$. If local fluctuations of the expansion rate do not substantially exceed this value, then $v_0 \boundTheta \lengthg \ll 1$ is ensured up to a domain size of the order of the associated Hubble length (and possibly larger for $v_0$ small enough). This assumption might be violated in rapidly collapsing, overdense regions. However, as discussed earlier about the $4$-acceleration, which is also expected to only have potentially non-negligible values in strong overdensities, such regions typically occupy very small spatial volumes and may accordingly be avoided and neglected in the spatial integrals, allowing for a tighter $\boundTheta$ bound in the remaining domain. This may as well be applied to the tilt velocities (interpreted as peculiar velocities of the source fluid in each slice), considering that a tighter threshold may be imposed on the tilts everywhere outside small overdense regions. This would allow for an even smaller $v_0$ to be picked while neglecting the contributions of the remaining overdensities as small corrections to the result. 

\subsubsection{Consequences for the foliation dependence of averages}
\label{sec:averagesbound}

We can now derive a bound on the difference of scalar averages between slices of two foliations, while making use of the above results on bounds on the difference of scalar integrals.  

We first obtain a bound on the modulus of the volume difference $\Delta \Vol \equiv \Vol_{\Sc_0}^{\Scp} - \Vol_{\Sc_0}^{\Sc}$, by taking $\Scalar = 1$ in Eq.~(\ref{eq:delta_I_bound1}), from which we have:
\begin{equation}
\label{eq:volumebound}
   |\Delta \Vol| = |\Delta I(1)| \leq \tilde{\alpha} \, v_0 \boundTheta \lengthg \, \Vol_{\Sc_0}^{\Sc} \leq \tilde\alpha \eta \, \Vol_{\Sc_0}^{\Sc} \; .
\end{equation}
Here, we do assume non-weighted averages, \emph{i.e.}, that $\VecField$ is indeed normalized without a need to include its normalization into a redefinition of $\Scalar$.

Considering again any given foliation-independent scalar $\Scalar$, the difference of its average between slices of the two foliations, $\Delta \langle \Scalar \rangle \equiv \averagep{\Scalar} - \average{\Scalar}$, is given by
\begin{align}
   \Delta \langle \Scalar \rangle & = \frac{\spatintp{\Scalar}}{\Vol_{\Sc_0}^{\Scp}} - \frac{\spatint{\Scalar}}{\Vol_{\Sc_0}^{\Sc}} =  \frac{\Delta I(\Scalar)}{\Vol_{\Sc_0}^{\Scp}} - \spatint{\Scalar} \; \frac{\Delta \Vol}{\Vol_{\Sc_0}^{\Sc} \; \Vol_{\Sc_0}^{\Scp}} \nonumber \\
   & {} = \frac{\Delta I(\Scalar) - \Scalar_0 \, \Delta \Vol}{\Vol_{\Sc_0}^{\Scp}} = \frac{ \Delta I(\Scalar - \Scalar_0)}{\Vol_{\Sc_0}^{\Sc} + \Delta \Vol} \; ,
   \label{eq:delta_average_S_firstexpr}
\end{align} 
where we introduced the short-hand notation $\Scalar_0 \equiv \average{\Scalar}$ for the average of $\Scalar$ within the reference slice $\Sigma_0$, seen as a constant number. We can then apply the integral bound in Eq.~(\ref{eq:delta_I_bound1}) to the shifted scalar $\Scalar - \Scalar_0$ to obtain the following bound:
\begin{equation}
    | \Delta I(\Scalar - \Scalar_0) | \leq \tilde\alpha \, v_0 \boundTheta \lengthg \; \spatint{\left| \Scalar - \Scalar_0 \right|}  + \alpha (1 + \eta \tilde\alpha) \, v_0 \, \lengthg \; \spatint{\max_{\tau \, \in \, [ - \alpha \, v_0 \, \length, \, \alpha \, v_0 \, \length]} \left| \left( \Vec^\mu \partial_\mu \Scalar \right)_{(\tau,X^i)} \right|} \, .
\end{equation}
Injecting this into Eq.~\eqref{eq:delta_average_S_firstexpr} and using  Eq.~\eqref{eq:volumebound}, we thus obtain the following bound on the variation of the average of $\Scalar$ between the slices:
\begin{align}
   | \Delta \langle \Scalar \rangle| & \leq \frac{1}{1 - \tilde\alpha \eta} \left[ \tilde\alpha \, v_0 \boundTheta \lengthg \; \average{\left| \Scalar - \average{\Scalar} \right|} \right. \nonumber \\
   & \qquad \qquad \qquad \left. {} + \alpha (1 + \eta \tilde\alpha) \, v_0 \, \lengthg \; \average{\max_{\tau \, \in \, [ - \alpha \, v_0 \, \length, \, \alpha \, v_0 \, \length]} \left| \left( \Vec^\mu \partial_\mu \Scalar \right)_{(\tau,X^i)} \right|} \right] \; ,
\label{eq:boundaverages1}   
\end{align}
provided $\tilde\alpha \eta < 1$, which is ensured for $\eta$ small enough since $\tilde\alpha \eta \sim 6 \eta$ when $\eta \rightarrow 0$.

In the same way as for the bounds on $\Delta I(\Scalar)$ in Eqs.~\eqref{eq:delta_I_bound1}--\eqref{eq:delta_I_bound1b}, we could have replaced in the above the maximum of $\left| V^\mu \partial_\mu \Scalar \right|$ by its time integral (along the worldline of $\VecField$) over the same interval, \emph{i.e.},
\begin{equation}
    \average{\max_{\tau \, \in \, [ - \alpha \, v_0 \, \length, \, \alpha \, v_0 \, \length]} \left| \left( \Vec^\mu \partial_\mu \Scalar \right)_{(\tau,X^i)} \right|} \quad \mapsto \quad
    \average{\int_{\tau=- \alpha \, v_0 \, \length}^{\alpha \, v_0 \, \length} \left| \Vec^\mu \partial_\mu \Scalar \right|_{(\tau,X^i)} \, d\tau}   \; ,
\end{equation}
and $\length$ (appearing in the interval bounds $\pm \, \alpha \, v_0 \, \length$) could be replaced by the global $\lengthg$ in either expression.

In Eq.~\eqref{eq:boundaverages2} of Appendix~\ref{app:psi_form2}, an alternative form for the bound on $\Delta \langle \Scalar \rangle$ is presented. This form is rather based on the local quantities $\nabla_\mu \left( \Scalar \Vec^\mu \right)$ and $\nabla_\mu V^\mu$ thanks to a rewriting of $\psi(X^i)$, following the same logic as for the alternative form of the bound on $\Delta I(\Scalar)$ mentioned hereabove and also presented in Appendix~\ref{app:psi_form2}.

Eq.~\eqref{eq:boundaverages1} and its variants provide upper limits for the possible changes in the spatial {average} of the scalar $\Scalar$ over the domain of interest when going from $\Sigma_0$ to another, intersecting slice that may belong to any other foliation {$\CF'$}, of normal vector $\bm{n'}$, provided it (like $\CF$) obeys the global small tilt condition $(\gvnp^2 - 1)^{1/2} \leq v_0$. This independence in $\Sigma_0'$ is fully achieved if the result is expressed in terms of $\lengthg$ instead of $\length$ and $\lengthg$ is for instance defined as the diameter of the {averaging} domain in $\Sigma_0$ (see Eq.~(\ref{eq:diameter_def})). It is also achieved if some intersection point $P_0$ between $\Sigma_0$ and any possible choice of $\Sigma_0'$ is kept fixed \emph{via} the requirements on the parametrization of $\Scp$, and $\length$ is replaced by the (\emph{a priori} larger) value of the spatial distance along $\Sigma_0$, $\mathrm{d} (P,P_0)$, between the current point $P$ (of coordinates $(\tau=0,X^i)$) and $P_0$.

As for the variation of integrals $\Delta I(\Scalar)$ in Sec.~\ref{sec:deltaIbounds} above, these bounds are expressed as a function of $v_0 \boundTheta \lengthg$ and require this factor to be small enough to obey the constraint~\eqref{eq:v0HL_assumption_global}; hence, similar remarks on setting $\boundTheta$ and/or $v_0$ apply. The bounds on the variation $\Delta \langle \Scalar \rangle$ still depend on the local evolution rate $\Vec^\mu \nabla_\mu \Scalar$ or non-conserved current $\nabla_\mu (\Scalar \Vec^\mu )$; but we note in this case that their additional dependence is on the local fluctuations of $\Scalar$ on the reference slice, rather than just $\average{\Scalar}$.

\subsection{Bounds for constant--proper time foliations}
\label{subsec:constant-tau}

In cosmology, the age of the Universe is typically measured with respect to the proper time of fundamental observers. However, the choice of proper time function is associated with a calibration freedom, and a family of proper time functions in general exist for a given 4-velocity field. In the present section, we shall investigate bounds of integrals formulated within classes of proper time foliations. In particular, we give as an example proper time foliations that are calibrated within the epoch of last scattering. 

\subsubsection{The family of constant--proper time foliations}
\label{gauge_propertime}
We consider proper time foliations of a single 4-velocity field, as an example of a set of foliations that are natural to compare.  
Let us consider the class of proper time functions of a given 4-velocity field. We shall use this (unit) $4$-velocity field as our volume measure vector $\VecField$, and as above use this $\VecField$ to define the propagation of the boundaries of the tube $\Tube$.
We say that $\tau$ is a proper time function of $\VecField$ if it satisfies the equation 
\begin{equation}
\label{eq:Tproper}
\Vec^{\mu} \nabla_{\mu} \tau = 1 . 
\end{equation}
Let $\Sigma_{\text{init}}$ be an initial reference hypersurface chosen at convenience, and let $\tau=\tau_{\text{ref}}$ be the proper time function satisfying $\tau_{\text{ref}}(\Sigma_{\text{init}}) =$ constant. 
All solutions to Eq.~\eqref{eq:Tproper} can be expressed as 
\begin{equation}
\label{eq:Tpropersol}
\tau = \tau_{\text{ref}} + \xi \, , 
\end{equation}
where $\xi$ is a given function satisfying the transport rule $\Vec^{\mu} \nabla_{\mu} \xi = 0$. 
If the function $\xi$, describing the distance of a given solution $\tau$ to the reference time function $\tau_{\text{ref}}$, can be bounded on a single hypersurface, then it can be bounded throughout the tube $\Tube$. 
In the following, we shall consider bounds that are relevant for when we can assume that we can bound $\xi$ on the initial surface $\Sigma_{\text{init}}$ such that $| \xi | \leq \delta T$ everywhere on $\Sigma_{\text{init}}$. It then follows immediately from the above transport rule that $|\xi | \leq \delta T$ globally within $\Tube$. This scenario is thus substantially simpler than the setup considered in subsection~\ref{sec:smalltiltsbounds} above where the main difficulty was bounding the proper time distance between the two slices considered. With such a bound already ensured, we do not need to assume small tilts between the slices of the constant--proper time foliations.

\subsubsection{Resulting bounds on the norm of $\Delta I(\Scalar)$}

We now consider an arbitrary scalar function $\Scalar$, and two intersecting leaves $\Sigma_0$ and $\Sigma_0'$ of the foliations $\mathcal{F}$ and $\mathcal{F}'$ corresponding to the level sets of $\tau_{\text{ref}}$ and $\tau'$, where $\tau'$ is an arbitrary member of the class of solutions (\ref{eq:Tpropersol}).
We may choose $\tau_{\text{ref}}(\Sigma_0) = 0 = \tau'(\Sigma_0')$ without loss of generality, using the gauge freedom of shifting $\tau'$ by an additive constant if necessary. 
Let $(X^i)$ again be a set of three $\VecField$-comoving spatial coordinates. The values $\functau(X^i)$ taken by $\tau_{\text{ref}}$ on the $\Sigma_0'$ hypersurface are equivalently given by the values of $-\xi$ on that same surface since $\tau_\mathrm{ref} = \tau' - \xi$; and it follows that $|\functau(X^i)| \leq \delta T$ {within $\Tube$}. 
From this global upper bound on $|\functau(X^i)|$ we can formulate an upper bound on $| \psi(X^i) |$ from Eq.~\eqref{eq:psi_result}, which reads, in the $(\tau_\mathrm{ref},X^i)$ coordinate system:
\begin{align}
\left| \psi(X^i) \right| & \leq  \left(  \exp \left( \max_{\pm \delta T} \left| \int_{\tau=0}^{\pm \delta T} \left| \nabla_\mu \Vec^\mu \right|_{(\tau,X^i)} \; d\tau \right| \right) - 1 \right) \left| \Scalar_{(\tau_\mathrm{ref}=0,X^i)} \right| \nonumber \\
& \qquad \quad +  \exp \left( \max_{\pm \delta T} \left| \int_{\tau=0}^{\pm \delta T } \left| \nabla_\mu \Vec^\mu \right|_{(\tau,X^i)} \; d\tau \right| \right) \times \max_{\pm \delta T} \left| \int_{\tau=0}^{\pm \delta T} \left| \Vec^\mu \partial_\mu \Scalar \right|_{(\tau,X^i)}  \; d \tau \right| \; .
\label{eq:psi_result_bound_tau}
\end{align}
Assume that we can define a global upper bound $3 \boundTheta$ on the volume expansion rate, time-averaged along $\VecField$ for $\tau_\mathrm{ref}$ spanning $[-\delta T, 0 ]$ or $[0, \delta T]$:
\begin{equation}
\label{eq:proptimebound}
\frac{1}{\delta T} \left| \int_{\tau=0}^{\pm \delta T} \, \left| \nabla_\mu \Vec^\mu\right|_{(\tau,X^i)}  \; d\tau  \right| \leq  3 \boundTheta  \; . 
\end{equation}
This may arise still as a consequence of $3 \boundTheta$ holding as a bound on the local expansion rate everywhere over the spacetime range considered, as in subsection~\ref{sec:smalltiltsbounds} above; but in the present case only the less restrictive time-averaged bound is required. With this assumption, Eq.~\eqref{eq:psi_result_bound_tau} becomes
\begin{equation}
\left| \psi(X^i) \right| \leq  \left(  e^{3 \boundTheta \, \delta T} - 1 \right) \left| \Scalar_{(\tau_\mathrm{ref}=0,X^i)} \right| +  e^{3 \boundTheta \, \delta T} \times \max_{\pm \delta T} \left| \int_{\tau=0}^{\pm \delta T} \left| \Vec^\mu \partial_\mu \Scalar \right|_{(\tau,X^i)}  \; d \tau \right| \; .
\label{eq:psi_result_bound_tau_2}
\end{equation}
Similarly to subsection~\ref{sec:smalltiltsbounds}, the exponential terms above may as well be replaced by affine expressions in $\boundTheta \, \delta T$ as $e^{3 \boundTheta\,\delta T} \leq 1 + \eta^{-1} (e^{3 \eta} - 1) \, \boundTheta \, \delta T$ by considering an upper threshold value $\eta$ that can be set on $\boundTheta\,\delta T$. The above local constraints on $|\psi(X^i)|$ then bound the variation of the integral of $\Scalar$, $\Delta I(\Scalar) = I(\psi(X^i))$, as $| \Delta I(\Scalar) | \leq I(| \psi(X^i) |)$. This translates as well into a bound on the variation of averages, $\Delta \langle \Scalar \rangle$, along the same lines as in Sec.~\ref{sec:averagesbound} above. 
These bounds are useful when it is possible to constrain $(\Vec^\mu \partial_\mu \Scalar)_{(\tau,X^i)}$ over the bounding time span $\pm \delta T$ between the leaves.
When $\delta T$ is much smaller than typical values of the time scales set by $\left| (\nabla_\mu \Vec^\mu)_{(\tau,X^i)} \right|^{-1}$ and $ \left| \Scalar_{(\tau=0,X^i)} \right| / \left| (\Vec^\mu \partial_\mu \Scalar)_{(\tau,X^i)} \right| $ within the subdomain of $\Tube$ defined by $- \delta T \leq \tau_{\text{ref}} \leq \delta T$, then $| \Delta I(\Scalar) | / I(| \Scalar| )_{\Sc_0}^\Sc$ is much smaller than 1. This directly constrains the relative variation $| \Delta I(\Scalar) / I( \Scalar )_{\Sc_0}^\Sc |$ at least when $\Scalar$ has a constant sign along the reference slice.

\subsubsection{Example: proper time foliations synchronised near the last scattering epoch} 

We consider a class of proper time foliation scalars (\ref{eq:Tpropersol}) of a physical matter congruence $\VecField = \bm u$ which are synchronised at the epoch of \emph{last scattering}. 
The epoch of last scattering defines a natural initialisation epoch in cosmology, and this epoch may thus naturally be used for synchronizing the proper time of physical observers. This epoch does however cover a spacetime region with a finite width in cosmic time.
In the physics of recombination, the visibility function quantifies the probability of the streaming of photons as protons and electrons combine into hydrogen. The visibility function might then be used to quantify an epoch of last scattering, after which most photons are freely streaming. 
The full width at half maximum of the visibility function in $\Lambda$CDM cosmology is $\sim 10^{5}$ years, and can be used to define the duration of the last scattering epoch. 
Consider the sub-class of proper time functions (\ref{eq:Tpropersol}) of $\bm u$ {which can all be initialized within the epoch of last scattering. That is, we set $\tau_{\text{ref}}$ as defined from an initialization hypersurface $\Sigma_\mathrm{init}$} that is fully contained within the spacetime region of last scattering, and all remaining proper time functions considered are also required to have an (``initial'') level set contained within the region of last scattering. These proper time functions accordingly satisfy the global constraint $\left| \xi \right| \leq \delta T$ with $\delta T \sim 10^{5}$ years.
We might think of this class of synchronisation hypersurfaces as defining a set of equally preferred calibrations of a cosmic age function. 

Let us investigate how integral quantities computed at the present epoch, defined as the $\tau = t_0$ slice in each foliation for the fixed present-day age $t_0$, differ between foliations of this set. 
Let typical values of the matter fluid's expansion rate $| \nabla_\mu u^\mu |$ around the present-epoch Universe be of the order of $3 H_0$ with the Hubble constant $H_0 \sim 10^{-10}$ years${}^{-1}$, and let its local fluctuations reach at most a factor of a few times this value (say, at most $\sim 10 H_0$), such that $\left| \int_{\tau=0}^{\pm \delta T} \left| \nabla_\mu u^\mu \right|_{(\tau,X^i)}   d\tau \right| \leq 3 \boundTheta \, \delta T \lesssim {10^{-4}}$, with $3 \boundTheta \lesssim 10 H_0$. 
Let us see for instance how the present-epoch volume of the Universe is bounded within members of this set of proper time foliations. 
We accordingly take $\Scalar(\tau,X^i) = 1$, and the bound (\ref{eq:psi_result_bound_tau}) becomes 
\begin{equation}
 \left| \psi(X^i) \right| \leq   \exp \left( \max_{\pm \delta T} \left| \int_{\tau=0}^{\pm \delta T} \left| \nabla_\mu \Vec^\mu \right|_{(\tau,X^i)} \; d\tau \right| \right) - 1 \lesssim  \exp \left( 3 \boundTheta \, \delta T \right) - 1 \approx 3 \boundTheta \, \delta T \lesssim 10^{-4} \; . 
\label{eq:psi_result_bound_tau_Volume}
\end{equation} 
Thus, the difference in the present-epoch volume $\mathcal{V}_{0,\tau} \equiv I(1)^{\tau}_{t_0} $, as measured in two members of the class of proper time foliations synchronised at the last scattering epoch, is bounded as
\begin{equation}
\left| \Delta \mathcal{V}_0 \right| = \left| \Delta I(1) \right| \leq I(|\psi(X^i)|)^{\tau_{\text{ref}} }_{t_0}  \lesssim 3 \boundTheta \, \delta T \,  I(1)^{\tau_{\text{ref}} }_{t_0} = 3 \boundTheta \, \delta T \, \mathcal{V}_{0,\tau_\mathrm{ref}} \lesssim 10^{-4} \times \mathcal{V}_{0,\tau_\mathrm{ref}} \; . 
\label{eq:delta_I_boundVolume}
\end{equation}

Hence, the volume of the present-epoch Universe is well-defined at the level of ${}\sim 10^{-4}$ in the class of proper time foliations calibrated within the region of last scattering. This is due to the time scale set by the duration of last scattering through the width of the visibility function being much smaller than the typical time scale associated with volume expansion in the present-epoch Universe.

\section{Summary and discussion} 
\label{sec:discussion}

The $3+1$ foliation of spacetime in general relativity is a powerful tool for casting Einstein's equations into an initial value problem and for constructing coarse-grained variables by integration operations defined on the leaves. This is in particular used in approaches for cosmological averaging, as in \cite{Zalaletdinov:1992cg,Zalaletdinov:1996aj,Buchert:1999er,Buchert:2001sa,Gasperini:2009wp,Gasperini:2009mu,Green:2010qy,Buchert:2019mvq,Buchert:2022zaa}. 
It has been noted that average quantities that appear in such formalisms have dependence on the foliation within which they are formulated  \cite{Li:2007ci,Clarkson:2010uz,Brown:2012fx,Bolejko:2017wfy,Adamek:2017mzb,Buchert:2018yhd,Verweg:2023fov}. 

In the present paper, we go beyond these previous investigations and consider the $3+1$ foliation problem in a broad context. We are hence not addressing a particular metric model nor a perturbative setting, but are treating the problem of foliation dependence in relativistic integrals and averages of scalar variables over arbitrary bounded regions of 3-dimensional hypersurfaces, in presence of a generic nonsingular spacetime metric. This setup also encompasses the case of integrals and averages over the entire hypersurfaces when those have a closed topology, such as that of a $3$-torus or a $3$-sphere. We view the integral and averaged quantities as functionals of the general scalar function defining the foliation. 
Our systematic analysis of these functionals considering infinitesimal variations of foliations revealed that globally foliation-independent functionals do exist but must be associated with a locally-conserved current. Thus, the only physically relevant exactly preserved quantities are total rest masses within a bounded domain or other quantities that are per construction preserved within the individual volume elements.    
This is not in contradiction with the gauge-independent scheme recently proposed in~\cite{Verweg:2023fov} in the context of perturbation theory, since there, the coordinate transformations considered are not affecting the spacetime foliation, and thus actual foliation changes are by construction absent in the authors' scheme.

We additionally examined choosing foliations specifically to leave a certain functional invariant under infinitesimal deformations, \emph{i.e.}, foliations or individual slices which extremize a given functional, as in the well-known case of extremal-volume slices. Such extremals provide examples of ways to uniquely specify a foliation and thus to eliminate the ambiguity of foliation choice. We briefly discussed, as specific applications, the extremization of entropy functionals and the selection of slices with a minimal average tilt with respect to a given vortical (hence not hypersurface-orthogonal) fluid flow.

Since strictly foliation-independent integral/average quantities are rare, as might have been expected, in the second half of our paper we investigated the bounding of foliation dependence of integrals and averages under finite changes of foliations. 
There we showed that we can in some cases set upper limits on this dependence if we consider foliations with space-like leaves that can be bounded in terms of their relative distance. 
In particular we have considered classes of space-like foliations with associated normal vectors that have a small relative tilt. We have also considered families of constant--proper time foliations (for a given family of observers defining a $4$-velocity field) that are bounded in terms of distance between their respective initial synchronization slices. 
In both cases, we derived bounds on the relative variation of integrals of scalar quantities between different foliations, under the assumption that the local volume expansion rate also remains bounded within a certain spacetime region. These bounds depend on the ability to set constraints on the evolution rate of the scalar quantity considered or on the associated non-conserved current. This implies, as a special case, bounds on the volume within the integration domain. The results on integral functionals of scalars also directly imply bounds on the variation of the associated averages, which we explicited in the small-tilt case, and which depend on the local fluctuations of the corresponding scalar. These bounds are consistent with the qualitative discussion for similarly restricted classes of space-like foliations in~\cite{Buchert:2018yhd}.

The foliation dependence of cosmological backreaction terms as defined in~\cite{Buchert:1999er,Buchert:2001sa} and extended to arbitrary spatial foliations in~\cite{Buchert:2018yhd,Buchert:2019mvq} turns out to be complicated to bound rigorously in general.
This is due to the appearance of the threading lapse $\mathscr{M} \equiv (\VecField \cdot \bm{\nabla} \Sc)^{-1} = (\bm{\nabla} \Sc \cdot \bm{\nabla} \Sc)^{-1/2} \; \gvn^{-1}$, or of $\mathscr{M}^2$, as a factor in the corresponding integrands, inducing a dependence on the local foliation scalar's gradient\footnote{%
The dependence on the slicing in the scheme of~\cite{Buchert:2018yhd,Buchert:2019mvq} remains however limited to this factor in addition to spatial integration itself, thanks to the use of a fixed vector field as a volume measure and spatial boundary propagation vector.%
}.
We have accordingly not derived general analytical expressions for the bounding of variations of backreaction terms. We note, however, that our results in the case of constant--proper time foliations ($\VecField \cdot \bm{\nabla} \Sc = 1$) are directly applicable to the averages and backreaction terms appearing in~\cite{Buchert:2018yhd,Buchert:2019mvq} when the same class of foliations is considered there.
Our results in the case of space-like foliations with small relative tilts also remain applicable to such backreaction terms under the additional assumption of geodesic slicings, for which the slicing lapse $N \equiv (\bm{\nabla} \Sc \cdot \bm{\nabla} \Sc)^{-1/2}$ can be set to $1$ --- up to the small extra corrections induced by the presence of the Lorentz factor $\gvn \sim 1$.
Moreover, since we have succeeded in bounding the \emph{volume} of spatial sections, and since the important implication of backreaction is precisely its impact on the growth of cosmic volume over time, our results still provide implicit bounds on the foliation-dependent contributions to backreaction. In a setting where the foliation dependence of the volume is tightly bounded at all times, while the foliation dependence of a particular backreaction term might be larger, the backreaction effect on the volume (as the combined impact of all backreaction terms accumulated over some time evolution) also has to remain tightly bounded in this scenario.

Our results indicate that while the averaged properties of a given region of spacetime are generally going to depend on the reference time-slicing, there are nevertheless tight bounds that can be constructed within physically motivated classes of foliations that are close to each other by some suitable measure. 
In particular, for cosmological purposes where most of the matter in the Universe is thought to have non-relativistic relative speeds, physically meaningful space-like foliation frames that approximately trace the matter of the Universe will have a relative tilt velocity much smaller than unity.
There are also epochs in the early Universe that constitute natural choices for setting a synchronisation of relevant cosmological foliations. Since, for instance, the time duration of the epoch of last scattering is very short relative to the present-day Hubble time scale, a synchronisation within this epoch provides a class of natural foliations which have small separations, compared to the characteristic time scale of expansion. 
The results that we have obtained are thus directly applicable to the averaging problem in cosmology.

\acknowledgments

The authors are grateful to L\'eo Brunswick, David Wiltshire and Thomas Buchert for helpful discussions, and to an anonymous referee for insightful comments on the draft.

This work is part of a project that has received funding from the European Research Council (ERC) under the European Union's Horizon 2020 research and innovation program (grant agreement ERC advanced grant 740021--ARTHUS, PI: Thomas Buchert) in its earlier stages. It was also supported in early stages by Catalyst grant CSG--UOC1603 administered by the Royal Society of New Zealand.  

AH is funded by the Carlsberg foundation. PM acknowledges support and hospitality for a visit to the University of Canterbury within the context of the above Catalyst grant. In the later stages of this work, PM has been supported by the Universitat de les Illes Balears (UIB), Spain; the Spanish Agencia Estatal de Investigaci\'on grants PID2022-138626NB-I00, PID2019-106416GB-I00, RED2022-134204-E, RED2022-134411-T, funded by MCIN/AEI/10.13039/501100011033; the MCIN with funding from the European Union NextGenerationEU/PRTR (PRTR-C17.I1); Comunitat Auton\`oma de les Illes Balears through the Direcci\'o General de Recerca, Innovaci\'o I Transformaci\'o Digital with funds from the Tourist Stay Tax Law (PDR2020/11 - ITS2017-006), the Conselleria d'Economia, Hisenda i Innovaci\'o grant number SINCO2022/6719 (PI: Alicia Sintes, UIB), co-financed by the European Union and FEDER Operational Program 2021-2027 of the Balearic Islands; the ``ERDF A way of making Europe''; and EU COST Action CA18108.



\begin{appendix}

\section{Complements to the derivation and alternative forms of the bounds on finite variations of spatial integrals and averages}

\subsection{General time coordinate and relation between determinants}
\label{app:volumeelement}

Compared to the derivation above in Sec.~\ref{sec:dif}, Eq.~(\ref{eq:delta_I_as_sqrtb_S_difference}) can as well be obtained along the same lines in the more general coordinates $(T,X^i)$, with any time coordinate $T$ that is nondecreasing along any flow line of $\VecField$. In such coordinates, the resulting equivalent of Eq.~(\ref{eq:delta_I_as_sqrtb_S_difference}) would then feature the factor $\sqrt{g} \, \Vec^0$ instead of $\sqrt{g}$, where $\Vec^\mu = (\Vec^0,0,0,0)$, arising from a factor $\sqrt{g} \left( \Vec^\mu \nabla_\mu T \right)$ in Eq.~(\ref{eq:delta_I_Znablatau}) once re-expressed in terms of the arbitrary time $T$.

As in Eq.~(\ref{eq:delta_I_as_sqrtb_S_difference}) (with $\Vec^0 = 1$ in that case), this factor $\sqrt{g} \, \Vec^0$ --- which can also directly be seen to remain invariant under the change of the time coordinate --- corresponds to $\sqrt{b}$. 
This can be shown by applying Cramer's rule to the inverse metric tensor $\mathbf{g}^{-1}$, expressing the metric itself (as the inverse of $\mathbf{g}^{-1}$) in terms of the determinant and the adjugate matrix of $\mathbf{g}^{-1}$. The $(00)$-component of this relation gives $g_{00} = \det(g^{ij}) / \det(g^{\mu \nu}) = (-g) \det(g^{ij})$. Noting that $\Vec^\mu \Vec_\mu = -1 = g_{00} (\Vec^0)^2$ so that $g_{00} = -1/(\Vec^0)^2$, and that $(b_{ij})$ is the inverse matrix to $(g^{ij})$: $g^{ik} b_{kj} = \delta^i_{\,j}$ (using $\Vec^i = 0$ and $\Vec^\mu \Vec_\mu = \Vec^0 \Vec_0 = -1$) so that $\det(g^{ij}) = 1/ b$, the above relation $g_{00} = (-g) \det(g^{ij})$ becomes $g \, (\Vec^0)^2 = b$. Within the choice $T = \tau$ as used in Sec.~\ref{sec:dif}, the component $\Vec^0 \, ( {} = \Vec^\mu \nabla_\mu \tau)$ reduces to $1$, hence $g$ reduces to $b$ in the coordinates $(\tau,X^i)$.

\subsection{Alternative form of the local integrand and of the global bounds on the variation of spatial integrals and averages}
\label{app:psi_form2}

The integrand of $\Delta I(\Scalar)$ in Eq.~\eqref{eq:delta_I_as_sqrtb_S_difference}, normalized by the reference volume element,
\begin{equation}
    \psi(X^i) \equiv \left[ \big( \sqrt{b} \, \Scalar \big)_{(\tau = \functau(X^i), X^i)} - \big(( \sqrt{b} \, \Scalar \big)_{(\tau=0, X^i)} \right] \, \Big/ \, \big( \sqrt{b} \big)_{(\tau=0, X^i)} \; ,
\end{equation}
can be alternatively written as
\begin{equation}
\psi(X^i) = \left(\sqrt{b} \right)_{(\tau=0,X^i)}^{-1} \, \int_{\tau=0}^{\functau(X^i)} \Vec^\mu \partial_\mu \left( \sqrt{b} \, \Scalar \right)_{(\tau, X^i)} \, d\tau \; ,
\label{eq:int_dot_sqrt_b_S_dtau}
\end{equation}
where, from Eq.~(\ref{eq:evol_sqrt_b_0}),
\begin{equation}
\label{eq:evol_sqrt_b_S_1}
 \Vec^\mu \partial_\mu \! \left( \sqrt{b} \, \Scalar \right)
= \sqrt{b} \, \nabla_\mu \! \left( \Scalar \Vec^\mu \right) \, .
\end{equation}
Substituting expression~(\ref{eq:solution_sqrt_b}) for $\sqrt{b}$ in the above, Eq.~(\ref{eq:int_dot_sqrt_b_S_dtau}) becomes:
\begin{equation}
\psi(X^i) = \int_{\tau=0}^{\functau(X^i)}  \left( \nabla_\mu ( \Scalar \Vec^\mu) \right)_{(\tau,X^i)} \, \exp \left( \int_{\bar{\tau}=0}^{\tau} (\nabla_\mu \Vec^\mu)_{(\bar{\tau},X^i)} \; d\bar{\tau} \right) d\tau \, .
\label{eq:psi_result_alt}
\end{equation}

This alternative form of $\psi$ shows more explicitly the exact foliation-independence of $\spatint{\Scalar}$, $\Delta I(\Scalar) = 0$, in case $\nabla_\mu ( \Scalar \Vec^\mu) = 0$, in agreement with our results from section~\ref{StationarityConditionsIntegral} (in particular Eq.~(\ref{eq:IntegralConstraintsAllReduced})) under the additional simplifying assumptions considered in section~\ref{sec:foliationdependence_restricted}. This form of $\psi$ would have been the one naturally arising under a similar derivation as the one in Sec.~\ref{sec:dif} leading to Eqs.~(\ref{eq:diff_sqrt_b_S_1})--(\ref{eq:delta_I_as_I_psi}), had one started from the alternative form of $\spatint{\Scalar}$ and $\spatintp{\Scalar}$ obtained by integrating by parts, that is, $\Delta I (\Scalar) = \int_{\mathcal{M}} d^4x \, \sqrt{g} \,  \left[\heavi(\Sc - \ScZ) - \heavi(\Scp - \ScZ) \right] \nabla_\mu ( \Scalar \Vec^\mu  )\, \BoundH$.

Both forms for $\psi$, Eqs.~(\ref{eq:psi_result}) and~(\ref{eq:psi_result_alt}), can be of interest, as they can give rise to different upper bounds on $\Delta I(\Scalar) = \spatint{\bar{\psi}}$ and on $\Delta \langle \Scalar \rangle$ (see below), expressed with different local variables --- $| \Scalar |$ (on $\Sigma_0$) and $|\Vec^\mu \partial_\mu \Scalar |$ in the former case, $|\nabla_\mu (\Scalar \Vec^\mu) |$ in the latter, in addition to $|\nabla_\mu\Vec^\mu |$ in both cases. Depending on the scalar $\Scalar$ under consideration, on the choice of $\VecField$, and on the requirements on $\CF'$, available physical or mathematical constraints on the above local quantities would then determine the most suitable of both expressions to derive upper limits on $\Delta I(\Scalar)$ or $\Delta \langle \Scalar \rangle$.

\paragraph{Application to the bounds on the variation of spatial integrals.}
Using this alternative form of $\psi$, in the derivation of the bound on $\Delta I(\Scalar)$ (as in Sec.~\ref{sec:deltaIbounds}), we may write, for $\functau(X^i) \geq 0$,
\begin{align}
 |\psi(X^i)| & \leq \left( \max_{\tau \, \in \, [0,\, \functau(X^i)]} \left|\nabla_\mu \left( \Scalar \Vec^\mu \right) \right|_{(\tau,X^i)} \right) \times \int_{\tau=0}^{\functau(X^i)} \exp \left( \int_{\tau'=0}^{\tau} \left(\nabla_\mu \Vec^\mu \right)_{(\tau',X^i)} \, d\tau' \right) d\tau \nonumber \\
& {} \leq  \left( \max_{\tau \, \in \, [-\alpha \, v_0 \, \length , \, \alpha \, v_0 \, \length]} \left|\nabla_\mu \left( \Scalar \Vec^\mu \right) \right|_{(\tau,X^i)} \right) \times \int_{\tau=0}^{\functau(X^i)} e^{3 \boundTheta \tau} d\tau \nonumber \\
& {} =  \left( \max_{\tau \, \in \, [-\alpha \, v_0 \, \length , \, \alpha \, v_0 \, \length]} \left|\nabla_\mu \left( \Scalar \Vec^\mu \right) \right|_{(\tau,X^i)} \right) \times \frac{1}{3 \boundTheta} \left(e^{3 \boundTheta \functau(X^i)}-1 \right) \nonumber \\
& {} \leq  \left( \max_{\tau \, \in \, [-\alpha \, v_0 \, \length , \, \alpha \, v_0 \, \length]} \left|\nabla_\mu \left( \Scalar \Vec^\mu \right) \right|_{(\tau,X^i)} \right) \times \frac{\tilde\alpha}{3} v_0 \, \length \; ,
\label{eq:psi_bound_2a}
\end{align}
still under the condition~(\ref{eq:v0HL_assumption}) ($v_0 \boundTheta \length \leq \eta < \ln{2}$), and where we have used part of Eq.~(\ref{eq:tau0_exp_bound}) for the last inequality.
For $\functau(X^i) \leq 0$, the first line of the above Eq.~(\ref{eq:psi_bound_2a}) holds with the bounds $\tau=0$ and $\tau=\functau(X^i)$ of the first integral (and of the maximum) reversed, so that the next two lines become
\begin{align}
|\psi(X^i)| & \leq  \max_{\tau \, \in \, [-\alpha \, v_0 \, \length , \, \alpha \, v_0 \, \length]} \left|\nabla_\mu \left( \Scalar \Vec^\mu \right) \right|_{(\tau,X^i)} \times \int_{\tau=\functau(X^i)}^{0} e^{3 \boundTheta (-\tau)} d\tau \nonumber \\
& {} =  \max_{\tau \, \in \, [-\alpha \, v_0 \, \length , \, \alpha \, v_0 \, \length]} \left|\nabla_\mu \left( \Scalar \Vec^\mu \right) \right|_{(\tau,X^i)} \times \frac{1}{3 \boundTheta} \left(e^{3 \boundTheta (-\functau(X^i))}-1 \right) \; ,
\label{eq:psi_bound_2}
\end{align}
and the last line of Eq.~(\ref{eq:psi_bound_2a}) thus holds regardless of the sign of $\functau(X^i)$.
Setting again the global upper threshold $\eta$ on $v_0 \boundTheta \lengthg$, Eq.~(\ref{eq:v0HL_assumption_global}), \emph{i.e.}, $v_0 \boundTheta \lengthg \leq \eta$, we can as previously replace the factor $\length$ appearing above by the global $\lengthg$, and insert the corresponding inequality on $|\psi(X^i)|$ into Eq.~(\ref{eq:delta_I_as_I_psi}) to get the following alternative to Eq.~\eqref{eq:delta_I_bound1} for the bound on $\Delta I(\Scalar)$:
\begin{equation}
\left| \Delta I(\Scalar) \right| \leq \spatint{|\psi(X^i)|} \leq \frac{\tilde\alpha}{3} v_0 \, \lengthg \; \spatint{\max_{\tau \, \in \, [-\alpha \, v_0 \, \length , \, \alpha \, v_0 \, \length]} \left|\nabla_\mu \left( \Scalar \Vec^\mu \right) \right|_{(\tau,X^i)}} \; .
\label{eq:delta_I_bound2}
\end{equation}
As for Eq.~\eqref{eq:delta_I_bound1} earlier, one may either keep the $X^i$-dependent length $\length$ appearing within the spatial integrand, or replace it by the global value $\lengthg$ by computing the corresponding maximum over the range $[-\alpha \, v_0 \, \lengthg , \, \alpha \, v_0 \, \lengthg]$ instead of $[-\alpha \, v_0 \, \length , \:\! \alpha \, v_0 \, \length]$.

\paragraph{Application to the bounds on the variation of averages.}
The alternative form to Eq.~\eqref{eq:boundaverages1} of the bound on $\Delta \langle \Scalar \rangle$ corresponding to the above alternative form of $\psi$, is provided by applying to $\Scalar - \Scalar_0$ (with $\Scalar_0 = \average{\Scalar}$) the bound on $\Delta I(\Scalar)$ above, Eq.~\eqref{eq:delta_I_bound2}. Following the same procedure as in Sec.~\ref{sec:averagesbound}, this gives 
\begin{equation}
    | \Delta \langle \Scalar \rangle| \leq \frac{\tilde\alpha \, v_0 \,\lengthg}{3 \, (1 - \tilde\alpha \eta)} \; \average{\max_{\tau \, \in \, [ - \alpha \, v_0 \, \length, \, \alpha \, v_0 \, \length]} \; \left| \left( \partial_\mu \left( \Scalar \Vec^\mu \right) \right)_{(\tau,X^i)} - \average{\Scalar} \left( \nabla_\mu \Vec^\mu \right)_{(\tau,X^i)} \right| } \; ,
\label{eq:boundaverages2}
\end{equation}
still assuming $\tilde\alpha \eta < 1$.

Again, the use of either Eq.~(\ref{eq:boundaverages1}) or Eq.~(\ref{eq:boundaverages2}) depends on the knowledge of the corresponding local quantities. This holds to a lesser extent as in the case of $\Delta I(\Scalar)$ (with either Eq.~\eqref{eq:delta_I_bound1} or Eq.~\eqref{eq:delta_I_bound2}) though, since the above Eq.~(\ref{eq:boundaverages2}) still involves the local $\Scalar$ through its reference average $\average{\Scalar}$, as well as a contribution from $\nabla_\mu \Vec^\mu$, rather than only a $\partial_\mu \left( \Scalar \Vec^\mu \right)$ term.

\subsection{The proper-time distance between the slices as an integral along a spatial curve}
\label{app:derivation_dtaudlambda}

Let us consider, as in Sec.~\ref{sec:slicedistancebound}, a curve $\mathscr{C}$ within $\Sigma_0$, connecting any two points $P_0$ and $P$ within the integration domain on $\Sigma_0$, and parametrized by its unit space-like $\bm{n}$--orthogonal tangent vector $\bm{K}$ and the associated affine parameter $\lambda$. In the $x^\mu=(\tau,X^i)$ coordinate system, which we will be using throughout this section, the point at parameter $\lambda$ along $\mathscr{C}$ has coordinates $x^\mu (\lambda) = (0, X^i(\lambda))$.

\begin{figure}[h]
    \centering
    \includegraphics[width=0.65\textwidth]{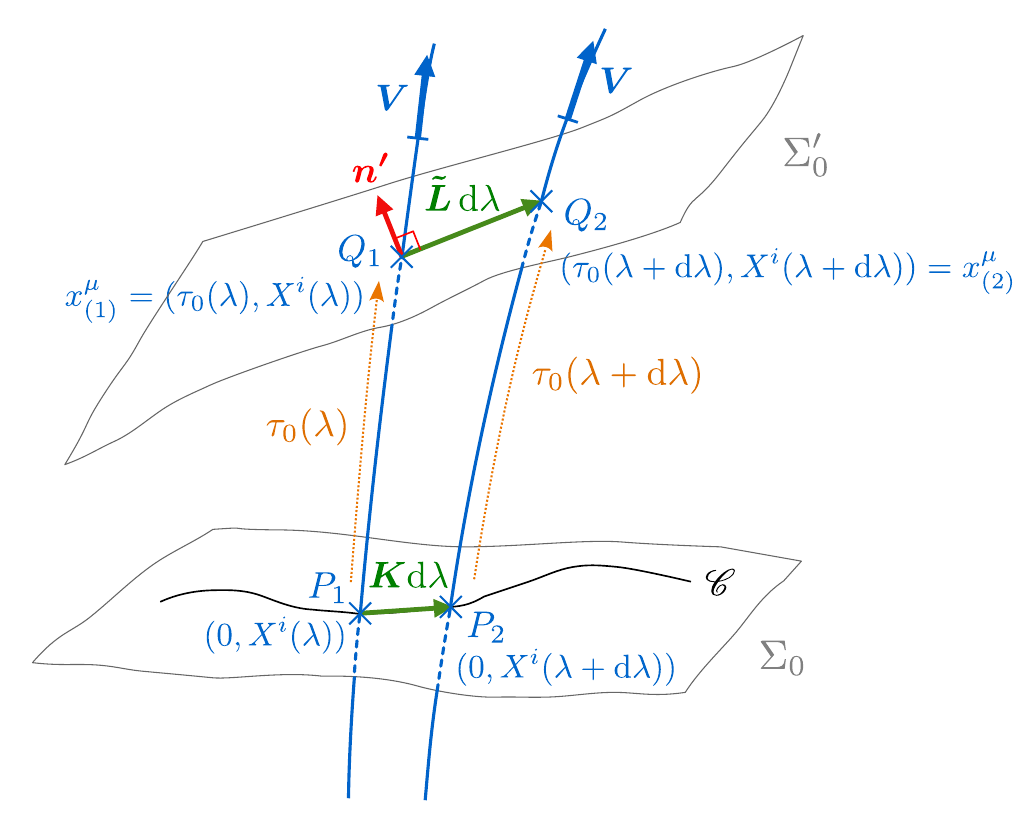}
    \caption{Schematic representation of the geometric configuration and main points ($P_1$, $P_2$, $Q_1$, $Q_2$) and vectors of interest for the derivation of $\mathrm{d}\functau / \dl$. Coordinate labels are included for each named point. We use a Riemannian picture of orthogonality for easier visualisation.}
    \label{fig:dtau_dlambda_deriv}
\end{figure}

In order to determine the evolution along the path $\mathscr{C}$ of the distance $\functau(\lambda) \equiv \functau(X^i(\lambda))$ between $\Sigma_0$ and $\Sigma_0'$, we consider two infinitesimally close points $P_1$ and $P_2$ along $\mathscr{C}$, respectively at parameter $\lambda$ and $\lambda + \mathrm{d}\lambda$, \emph{i.e.} at coordinates $x^\mu(\lambda)=(0,X^i(\lambda))$ and $x^\mu(\lambda+\dl)=(0,X^i(\lambda+\dl))$. The coordinates of these two points are related through the tangent vector $\bm K$ of $\mathscr{C}$ at $P_1$, as $x^\mu(\lambda+\dl) = x^\mu(\lambda) + K^\mu \, \dl$; \emph{i.e.}, $K^0 = 0$ and $X^i(\lambda+\dl) = X^i(\lambda) + K^i \, \dl$.
The flow line of $\VecField$ going through $P_1$ intersects the (Cauchy) slice $\Sigma_0'$ at a single point $Q_1$. The coordinates of $Q_1$ read $x^\mu_{(1)} \equiv (\functau(\lambda),X^i(\lambda))$: they follow from the definition of $\functau$, and by construction of the coordinate system $(\tau,X^i)$ as comoving and synchronous with respect to $\VecField$ with $\tau=0$ on $\Sigma_0$.
Similarly, the flow line of $\VecField$ going through $P_2$ intersects $\Sigma_0'$ at the unique point $Q_2$ of coordinates $x^\mu_{(2)} \equiv (\functau(\lambda+\dl), X^i(\lambda+\dl))$. The geometric framework and points of $\Sigma_0$ and $\Sigma_0'$ under consideration are illustrated on Fig.~\ref{fig:dtau_dlambda_deriv}. 

The fact that the two infinitesimally close points $Q_1$ and $Q_2$ both belong to the $\bm{n'}$--orthogonal slice $\Sigma_0'$ can be expressed \emph{via} the following constraint at $Q_1$: 
\begin{equation}
    0 = n'_\mu \left(x^\mu_{(2)} - x^\mu_{(1)} \right) = n'_0 \, \frac{\mathrm{d}\functau}{\dl} \, \dl + n'_i \left[X^i(\lambda+\dl) - X^i(\lambda) \right] \; .
\label{eq:dtautlambda_deriv_1}
\end{equation}
At $P_1$, \emph{i.e.} on $\mathscr{C}$, the spatial coordinate difference $X^i(\lambda+\dl) - X^i(\lambda)$ corresponds to the spatial components $K^i \, \dl$ of $\bm K \, \dl$. At $Q_1$, it can be expressed in terms of the vector field $\bm{\tilde L}$ extending $\bm{K}$ by Lie dragging along the wordlines of $\VecField$ that intersect $\mathscr{C}$: 
\begin{equation}
    0 = \big( \mathscr{L}_{\VecField} \bm{\tilde{L}} \big)^\nu = (\mathrm{d}/\mathrm{d}\tau) \tilde L^\nu |_{X^i} - \tilde L^\mu \big(\partial_\mu \Vec^\nu \big) = (\mathrm{d}/\mathrm{d}\tau) \tilde L^\nu |_{X^i} \, .
\end{equation}
Hence, $(\tilde L^\mu )_{(\tau,X^i)} = (\tilde L^\mu )_{(0,X^i)} = (K^\mu )_{(0,X^i)} = (0, K^i)_{(0,X^i)} \; \forall \tau$, and at $Q_1$, the following holds:
\begin{equation}
n'_i \left[X^i(\lambda+\dl) - X^i(\lambda) \right] = \dl \, (n'_i \tilde L^i)_{(\functau(\lambda),X^i(\lambda))} = \dl \, (n'_\mu \tilde L^\mu)_{(\functau(\lambda),X^i(\lambda))} \; .
\end{equation}
Injecting the above into Eq.~\eqref{eq:dtautlambda_deriv_1}, and using that, at $Q_1$, one has $n'_0 = n'_\mu \Vec^\mu = - \gvnp$, leads to {the first key expression for $\mathrm{d}\tau_0 / \mathrm{d}\lambda$}:
\begin{equation}
\frac{\mathrm{d} \functau }{d \lambda} = \left( \frac{\bm{n'} \cdot \bm{\tilde L} }{\gvnp}\right)_{(\functau(\lambda),X^i(\lambda))}  \quad .
\end{equation}

Using the local decomposition of $\bm{n'}$ with respect to $\VecField$, Eq.~\eqref{eq:decomp_np}: $\bm{n'} = \gvnp (\VecField + \bm{v'} )$ with $\bm{v'} \cdot \VecField = 0$, the above expression is rewritten as
\begin{equation}
    \frac{\mathrm{d} \functau }{d \lambda} = (\VecField \cdot \bm{\tilde L})_{(\functau(\lambda),X^i(\lambda))} + (\bm{v'} \cdot \bm{\tilde L})_{(\functau(\lambda),X^i(\lambda))} \; .
\label{eq:dtaudlambda_deriv_2}
\end{equation}
Introducing the projected norm $\tilde L_b \equiv (b_{\mu \nu} \tilde L^\mu \tilde L^\nu )^{1/2}$ and the normalized $\VecField$--orthogonal projection $\bm L$ of $\bm{\tilde L}$, defined by $L^\mu \equiv b^\mu_{\ \nu} \tilde L^\nu / \tilde L_b$, the second term in the right-hand side above can be rewritten as $\bm{v'} \cdot \bm{\tilde L} = (\bm{v'} \cdot \bm{L}) \, \tilde L_b$, since $\bm{v'} \cdot \VecField = 0$.

We can now compute both of the right-hand-side terms above from their initial value at $P_1$ (on $\Sigma_0$) and an evolution equation along the flow lines of $\VecField$, making use of the vanishing Lie derivative of $\bm{\tilde L}$ along $\VecField$, this time written as $\Vec^\mu \nabla_\mu \tilde L^\nu - \tilde L^\mu \nabla_\mu \Vec^\nu = 0$.
Projecting this expression onto $\VecField$ gives $\Vec_\nu \Vec^\mu \nabla_\mu \tilde L^\nu = \Vec_\nu \tilde L^\mu \nabla_\mu \Vec^\nu = 0$, hence
\begin{equation}
\left. \frac{\mathrm{d}}{\mathrm{d}\tau} \right|_{X^i} \left(\VecField \cdot \bm{\tilde L} \right) = \Vec^\mu \nabla_\mu \big( \Vec_\nu \tilde L^\nu \big) = a_\nu \tilde L^\nu = \left(\bm a \cdot \bm L \right) \, \tilde L_b \; ,
\label{eq:dtau_dlambda_deriv_ddtau_Vec_Ltilde_term}
\end{equation}
with the $\VecField$--orthogonal $4$-acceleration $\bm a$ of $\VecField$: $a^\nu = \Vec^\mu \nabla_\mu \Vec^\nu$. One thus obtains
\begin{equation}
    \left(\VecField \cdot \bm{\tilde L} \right)_{(\functau(\lambda),X^i(\lambda))} = (\VecField \cdot \bm{K})_{(0,X^i(\lambda))} + \int_{\tau = 0}^{\functau(\lambda)}{(\bm L \cdot \bm a)_{(\tau,X^i(\lambda))} \, (\tilde L_b)_{(\tau,X^i(\lambda))} \; d\tau} \; .
\label{eq:dtau_dlambda_deriv_Vec_Ltilde_term_1}
\end{equation}

Now projecting the vanishing Lie derivative expression above onto $\bm{\tilde L}$ instead, and using the kinematic decomposition of the covariant derivative of $\underline{\VecField}$ in Eq.~\eqref{eq:decomp_grad_Vec}, results in the following evolution equation for $\bm{\tilde L} \cdot \bm{\tilde L}$:
\begin{equation}
\Vec^\mu \nabla_\mu (\tilde L_\nu \tilde L^\nu) = 2 \, \tilde L_\nu \Vec^\mu \nabla_\mu \tilde L^\nu = 2 \, \tilde L^\mu \tilde L^\nu \nabla_\mu \Vec_\nu = - 2 \, (\tilde L^\mu \Vec_\mu) (\tilde L^\nu a_\nu) + 2 \, \Theta_{\mu \nu} \tilde L^\mu \tilde L^\nu \; .
\end{equation}

Using the evolution equation~\eqref{eq:dtau_dlambda_deriv_ddtau_Vec_Ltilde_term} for $\VecField \cdot \bm{\tilde L}$, as well as $\tilde L_b ^2 = \tilde L_\mu \tilde L^\mu + (\VecField \cdot \bm{\tilde L})^2$, leads to an evolution equation for $\tilde L_b$:
\begin{equation}
    \frac{\mathrm{d}}{\mathrm{d}\tau} \big(\tilde L_b^2 \big) \Big|_{X^i} = \Vec^\mu \nabla_\mu \big(\tilde L_\nu \tilde L^\nu \big) + 2 \, \big(\tilde L^\mu \Vec_\mu \big) \, \frac{\mathrm{d}}{\mathrm{d}\tau} \big(\tilde L^\nu \Vec_\nu \big)\Big|_{X^i} = 2 \, \Theta_{\mu \nu} \tilde L^\mu \tilde L^\nu = 2 \, \tilde L_b^2 \; \Theta_{\mu \nu} L^\mu L^\nu \; ,
\label{eq:dtau_dlambda_deriv_evol_L_b_norm}
\end{equation}
where the last equality uses the orthogonality of $\Theta_{\mu \nu}$ to $\VecField$.
At $\tau = 0$, \emph{i.e.}, at $P_1$, we have: 
\begin{equation}
\tilde L_b = \sqrt{b_{\mu \nu} K^\mu K^\nu} = \sqrt{ 1 + (\bm K \cdot \VecField)^2_{(0,X^i(\lambda))} }\ \; ;
\end{equation}
and the above evolution equation for $\tilde L_b$, Eq.~\eqref{eq:dtau_dlambda_deriv_evol_L_b_norm}, is solved as
\begin{align}
\label{eq:dtau_dlambda_deriv_L_b_norm}
    & (\tilde L_b)_{(\tau,X^i(\lambda))} = \sqrt{1 + (\bm K \cdot \VecField)_{(0,X^i(\lambda))}^2} \;\; F(\tau,\lambda) \;\; , \\
    \mathrm{with} \quad & F(\tau,\lambda) \equiv \exp \left[ \int_{\tau_1=0}^{\tau} {\, (\Theta_{\mu \nu} \, L^\mu L^\nu)_{(\tau_1,X^i(\lambda))} \; d\tau_1} \right] \; .
\end{align}

Injecting this expression for $\tilde L_b$ into Eq.~\eqref{eq:dtau_dlambda_deriv_Vec_Ltilde_term_1}
for $\VecField \cdot \bm{\tilde L}$, while noticing that the $\sqrt{1+(\VecField \cdot \bm K)^2}$ prefactor is time-independent, then gives the following:
\begin{align}
\label{eq:dtau_dlambda_deriv_Vec_Ltilde_term_2}
    & (\VecField \cdot \bm{\tilde L})_{(\functau(\lambda),X^i(\lambda))} = (\VecField \cdot \bm{K})_{(0,X^i(\lambda))} + \sqrt{1+(\VecField \cdot \bm K)_{(0,X^i(\lambda))}^2} \;\; G(\functau(\lambda),\lambda) \;\; , \\
    & \qquad \mathrm{with} \quad G(\tau,\lambda) \equiv \int_{\tau_1=0}^{\tau}{ \, (\bm L \cdot \bm a)_{(\tau_1,X^i(\lambda))} \; F(\tau_1,\lambda) \; d\tau_1} \;\; .
\end{align}

Finally, by injecting Eq.~\eqref{eq:dtau_dlambda_deriv_Vec_Ltilde_term_2} above for $\VecField \cdot \bm{\tilde L}$ and Eq.~\eqref{eq:dtau_dlambda_deriv_L_b_norm} for $\tilde L_b$ into Eq.~\eqref{eq:dtaudlambda_deriv_2}, with $\bm{v'} \cdot \bm{\tilde L} = (\bm{v'} \cdot \bm L) \, \tilde L_b$, one recovers the main result for $\mathrm{d}\functau / \dl$:
\begin{equation}
    \frac{\mathrm{d} \functau }{\dl} = (\VecField \!\cdot \bm K)_{(0,X^i(\lambda))} + \sqrt{1+(\VecField \!\cdot \bm K)_{(0,X^i(\lambda))}^2}  \left[ (\bm{v'} \cdot \bm L)_{(\functau(\lambda),X^i(\lambda)) }\,  F(\functau(\lambda),\lambda) + G(\functau(\lambda),\lambda) \right] \, .
\label{eq:dtau_dlambda_appendix}
\end{equation}

One can then bound the distance $| \functau(\lambda)|$ between the slices, as per the derivation of section~\ref{sec:slicedistancebound}, from writing $\functau(\lambda)$ as an integral of the above expression of $\mathrm{d}\functau / \dl$ along $\mathscr{C}$, setting the starting point $P_0$ as belonging to $\Sigma_0 \cap \Sigma_0'$ --- where $\functau  = 0$ by construction.

\end{appendix}

\bibliographystyle{mnras_unsrt}
\bibliography{Foliation_v9}

\end{document}